\documentclass[12pt,oneside,english]{book}

\usepackage[T1]{fontenc}
\usepackage[latin9]{inputenc}
\usepackage[a4paper]{geometry}
\usepackage{array}
\usepackage{float}
\usepackage{textcomp}
\usepackage{amsmath}
\usepackage{amssymb}
\usepackage{graphicx}
\usepackage{setspace}
\usepackage{esint}
\makeatletter

\newcommand{\noun}[1]{\textsc{#1}}
\providecommand{\tabularnewline}{\\}

\usepackage{tocbibind}

\makeatother

\usepackage{babel}
\begin{document}
\noindent \begin{center}
{\LARGE{}Lee-Yang Model in Presence of Defects}
\par\end{center}{\LARGE \par}

\bigskip{}

\begin{center}
{\Large{}PhD Thesis}
\par\end{center}{\Large \par}

\vspace{2cm}

\noindent \begin{center}
{\Large{}Omar K. El Deeb}
\par\end{center}{\Large \par}

\noindent \begin{center}
{\Large{}}%
\footnote{Physics department, Beirut Arab Univeristy, o.deeb@bau.edu.lb%
}
\par\end{center}{\Large \par}

\bigskip{}

\noindent \begin{center}
{\large{}July, 2012}
\par\end{center}{\large \par}

\vspace{3.5cm}

\begin{singlespace}
\noindent \noun{\large{}Supervisor:}{\large{} }\emph{\large{}Zoltan
Bajnok }{\large \par}

\noindent \noun{\large{}Doctoral School leader:}{\large{} }\emph{\large{}Ferenc
Csikor}{\large \par}

\noindent \noun{\large{}Program leader:}{\large{} }\emph{\large{}Ferenc
Csikor}\vspace{3.5cm}

\noindent {\large{}Eotvos Lorand University, Budapest}{\large \par}

\noindent {\large{}Physics Doctoral School}{\large \par}

\noindent {\large{}Particle Physics and Astronomy Program}{\large \par}

\noindent {\large{}Theoretical Physics Group of the Hungarian Academy
of Sciences}{\large \par}
\end{singlespace}

\pagebreak{}

\chapter*{Acknowledgments}

I would like to pay my deep gratitude to Zoltan Bajnok, my adviser
who played a significant role in my academic development and then
in guiding me throughout the whole period of my work in this thesis
in particular. Dr Bajnok saved no effort in putting me on the right
track of work for the past four years. He is a mentor whom I praise
and thank.

I also thank Dr Paul Pearce for his helpful discussions and directions
during the work on the lattice model of the Lee-Yang theory. His humbleness
and acceptance to work together on that project is highly appreciated.

The university of Eotvos Lorand was a helpful environment to develop
this work, with its superb teachers and researchers.

Finally, my family: Father, mother and brother were stubborn enough
to carry out all of my expenses during this work, sometimes under
hard periods of need. I hereby thank them for their endless support.

No hard work could have been achieved without the moral support of
friends and colleagues.

It is the time to acknowledge all of those people, and their help
and support.

\begin{onehalfspace}
\noindent \pagebreak{}

\noindent \tableofcontents{}

\noindent \pagebreak{}
\end{onehalfspace}

\chapter{Introduction}

The study of the Lee-Yang model is important  for a general understanding
of two dimensional integrable models. The main motivation behind that
comes from $AdS/CFT$ duality. In our quest for understanding realistic
but very complicated models like the Super Yang-Mills gauge theories,
there is an important conjecture called the AdS/CFT duality and it
\cite{MGKPW} states the equivalence of $\mathcal{N}=4$ Super Yang-Mills
gauge theory with superstrings on $AdS_{5}\times S^{5}$ . The correspondence
is extremely interesting as it links the very difficult non-perturbative
physics of gauge theory to (semi) classical string/supergravity theory. 

As such it allows to gain new insight into various gauge theoretical
phenomena but at the same time makes it very difficult to test and
prove. A real breakthrough in this respect is the discovery of integrability
on both sides of the duality \cite{MinZa,BKS,BStau,BDKM,BPR,DNW}.
On the string theory side it means that the light-cone quantized worldsheet
sigma model is an integrable quantum field theory, while on the gauge
theory side it manifests itself in the appearance of spin chains.

It becomes obvious that we need to have a deeper understanding of
two dimensional integrable quantum field theoretical models as a starting
step to understand the more complicated theories leading to realistic
models.

In this thesis I choose the Lee-Yang model and go through different
approaches to analyze the model using the form factor approach and
the bootstrap program, the lattice model and the TBA equations from
the lattice as different approaches that lead to a full picture about
the model.

The bootstrap program aims to classify and explicitly solve 1+1 dimensional
integrable quantum field theories by constructing all of their Wightman
functions ( for a recent review \cite{FFBootstrap} and references
\cite{chew,STW,KarT,KarT2,KarW,ZamZ,BKW}). In the first step, called
the S-matrix bootstrap, the scattering matrix, connecting asymptotic
in and out states, is determined from its properties such as factorizability,
unitarity, crossing symmetry and Yang-Baxter equation (YBE) supplemented
by the maximal analyticity assumption \cite{exactS}. In the second
step, called the form factor bootstrap, matrix elements of local operators
between asymptotic states are computed using their analytical properties
originating from the already computed S-matrix. Supposing maximal
analycity leads to a set of solutions each of which corresponds to
a local operator of the theory \cite{KW}. In the third step these
\emph{bulk form factors} are used to build up the correlation (Wightman)
functions via their spectral representations and describe the theory
completely off mass shell. This program has been implemented for a
wide range of theories as in \cite{BFK1,BFK2,BK1,Bab1,Bab2,BF1,FR1,SV1}
. 

The analogous bootstrap program for 1+1 dimensional integrable \emph{boundary}
quantum field theories has been already developed. The first step
is called the R-matrix bootstrap \cite{GZ}: In boundary theories
the asymptotic states are connected by the R-matrix, which, as a consequence
of integrability factorizes and satisfies the unitarity, boundary
crossing unitarity and the boundary YBE (BYBE) requirements. These
equations supplemented by the maximal analytical assumptions makes
it possible to determine the reflection matrices and provide the complete
information about the theory on mass shell. In the second step we
are interested in the matrix elements of local operators localized
both in the bulk and also at the boundary. Due to the absence of translational
invariance the \emph{bulk operators}' one point functions acquire
nontrivial space dependence which can be calculated in the crossed
channel using the knowledge of the boundary state together with the
bulk form factors \cite{1ptf}. 

For the matrix elements of local \emph{boundary operators} axioms
can be derived from their analytical properties originating from the
already computed R-matrix \cite{BFF}. Supposing maximal analytical
leads to a set of solutions each of which corresponds to a local boundary
operator of the theory and is uniquely related to a vector in the
ultraviolet Hilbert space. The explicit form of the boundary form
factors via the spectral representation of the boundary correlation
functions provides a partial description of the theory off the mass
shell. A full description would include correlation functions of operators
localized in bulk as well, but this complicated problem has not been
addressed yet.

Since any two dimensional \emph{defect} theory can be mapped to a
boundary theory \cite{DefBound} the development of a separate bootstrap
program for their solution seems to be redundant. However, integrable
defect theories are severely restricted and one can go much beyond
the boundary bootstrap program explained above: We can determine the
form factors of both types of operators, those localized in the bulk
and also the ones localized on the defect. With the help of these
form factors we are able to derive spectral representation for any
correlation function and in principle fully solve the theory off the
mass shell as we show in \cite{BaOmFF}.

In developing a defect form factor program the first step is the T-matrix
bootstrap. Interacting integrable defect theories are purely transmitting
\cite{DMS1,DMS2,Defboot} and topological. As a consequence a momentum
like quantity is conserved \cite{ClDef,DAT} and the location of the
defect can be changed without affecting the spectrum of the theory
\cite{BS,XXZ}. This fact, together with integrability lead to the
factorization of scattering amplitudes into the product of pairwise
scattering and individual transmission and enable one to determine
the transmission factors from defect YBE (DYBE), unitarity and defect
crossing unitarity \cite{KL,QAT,sgdef}. The second step is the defect
form factor bootstrap: Once the transmission factors are known we
can formulate the axioms that have to be satisfied by the matrix elements
of local defect operators. We will analyze both operators localized
in the bulk and also on the defect. By finding their solutions the
spectral representation of any correlator can be determined and theory
can be solved completely.

In \cite{YangYang,Zam90,Zam91a,Zam91b,Zam91c}, Thermodynamic Bethe
Ansatz (TBA) equations have been introduced as an important tool in
the study of both massive and massless integrable quantum field theories.
Extensive studies have been carried out on scaling energies of vacuum
or ground states. However only relatively few excited states~\cite{Martins,Rav,BLZ,Dorey,Fendley}
were possible by TBA analysis and these are primarily restricted to
massive and diagonal scattering theories. So despite considerable
successes, the application of TBA methods was limited. The primary
obstacle is that there is no systematic and unified derivation of
excited state TBA equations \cite{PCAI}. 

The Lee-Yang model was studied from the TBA approach. The periodic
Lee-Yang was analyzed in \cite{Zam90} for the groundstate, and in
\cite{BLZ} and \cite{DPTW} for the excited states. The equations
were solved based on assumptions about the analytic structure of the
model, and were also supported by numerical results from the TCSA.

The TBA approach was also used to study the boundary Lee-Yang in groundstate
\cite{LMSS} and also in excited states, \cite{DPTW} where as the
defect groundstate was analyzed in \cite{BS}.

However, the lattice approach is far more reaching. It is a systematic
approach that allows to obtain both massive and massless excited TBA
equations by studying the continuum scaling limit of the associated
integrable lattice models. The most important input from the lattice
approach is an insight into the analytic structure of the excited
state solutions of the TBA equations. Previously this structure had
to be guessed. In contrast, in the lattice approach, the analytical
structure can be probed by direct numerical calculations of finite
size transfer matrices.

The lattice model is very general and was used to study several models
like the tricritical Ising model \cite{PCAI,PCAII}, by considering
the massive tricritical Ising model ${\cal M}(4,5)$ perturbed by
the thermal operator $\varphi_{1,3}$. 

There have been many relevant studies of the $A_{4}$ lattice model
and the more general $A_{L}$ models from the lattice viewpoint. For
the $A_{4}$ model, the off-critical TBA functional equation for periodic
boundary conditions has been derived and solved~\cite{BaxP82,BaxP83}
for the bulk properties and correlation lengths. The off-critical
TBA functional equations for the $A_{L}$ models were derived by Klümper
and Pearce~\cite{PearK91,KlumP91,KlumP92}. But only the critical
or ``conformal TBA\textquotedbl{} equations were derived and solved
in the critical scaling limit for the central charges and conformal
weights. The very same off-critical TBA functional equations for $A_{L}$
models were subsequently derived~\cite{BPO96} in the presence of
integrable boundaries showing that the TBA functional equations are
universal in the sense that they are independent of the boundary conditions. 

In this thesis we turn our attention to the simplest example of a
non-unitary minimal theory ${\cal M}(p,p')$ with $|p-p'|\ne1$, namely,
the Lee-Yang minimal model ${\cal M}(2,5)$ \cite{BaOmL}.

Here we study the Lee-Yang model on the lattice. We analyze the periodic,
boundary and the seam cases in both massive and massless regimes.
We derive their ground state TBA and analyze the flows from the $(r,s)=(1,1)$
to the $(r,s)=(1,2)$ sectors.

The thesis is organized as follows:

In chapter 2, I present an introduction to the basic conformal field
theory and define the Lee-Yang model, and introduce the necessary
methods to be used later. 

In chapter 3 I introduce asymptotic states in defect theories and
the notion of the transmission matrix. Then I determine the coordinate
dependence of defect form factors. By specifying the boundary form
factor axioms we postulate the axioms for diagonal defect theories.
Using an analogy between defects and standing particles we subject
our axioms to a consistency check. Then we determine the form factors
of any bulk operator in terms of the transmission factor and the already
calculated bulk form factors and outline the procedure to calculate
the general solution for operators localized on the defects. 

Afterward we apply this technology to determine the defect form factors
of the Lee-Yang model. By calculating the dimension of the operators
we can map them to the UV Hilbert space of the model. Finally I introduce
a method to derive the different boundary conditions via the defects.

In chapter 4, we start with the definition of the lattice theory.
We define the face weights and the periodic, defect and boundary raw
transfer matrices. We derive the functional relations they satisfy.
Then we analyze the analytical structure of the transfer matrices
to turn the functional relations into integral equations, obtaining
the TBA equations in the three models in the massive and the massless
cases.

We start with the trigonometric/conformal case: First we make correspondence
between the UV Hilbert space in terms of Virasoro modes and the zeros
of the transfer matrix and the paths. Then we analyze the lattice
flow in the parameter $\xi$ and describe our findings in the three
languages: paths, zeros, modes. We repeat this analysis for the periodic
and defect cases. We derive the integral TBA equations for the massless
cases. 

Switching into the massive case, and using the elliptic theta functions
for the face weights instead of the trigonometric ones, we get the
same analytic structure and the same paths, zeroes , modes. We derive
the massive TBA equations for the ground state and obtain the source
terms.

Then in chapter 5, I present the basic results which we obtained \cite{BaOm}
for the Luscher correction terms to the asymptotic Bethe Ansatz energy
of an $su(2)$ particle state in finite volume by vacuum polarization
effects due to wrapping interactions.

I present the conclusions in chapter 6. 

Finally the appendix presents how to calculate the finite size correction
which originates from virtual particles propagating around the circle.

Those Luscher correction terms which were presented in the outline
are calculated by the weak coupling expansion and shown in the appendix.

\vspace{3.5cm}

The material presented in this thesis is based on the papers \cite{BaOmFF,BaOmL,BaOm}:
\begin{itemize}
\item Zoltan Bajnok, Omar el Deeb: \emph{Form factors in the presence of
integrable defects}, Nuclear Physics B 832, p. 500-519, 2010, arXiv:0909.3200
{[}hep-th{]}
\item Zoltan Bajnok, Omar el Deeb, Paul Pearce: Lattice, In Progress
\item Zoltan Bajnok, Omar el Deeb: \emph{6-loop anomalous dimension of a
single impurity operator from AdS/CFT and multiple zeta values}, JHEP
1101:054,2011, arXiv:1010.5606 {[}hep-th{]}
\end{itemize}
\pagebreak{}

\chapter{The Lee-Yang Model}

We start by considering a massive scattering theory with $n$ types
of particles with masses $m_{a}$. One-particle states are denoted
by $|\theta\rangle_{a}$, where $\theta$ is the particle rapidity.
The Lorentz invariant normalization implies that $_{a}\langle\theta_{a}|\theta_{b}\rangle_{b}=2\pi\delta_{ab}\delta(\theta_{a}-\theta_{b})$ 

Asymptotic states of the theory are defined as tensor products of
one-particle states and are denoted by $|\theta_{1},\theta_{2},...,\theta_{n}\rangle_{a1a2...an}$,
where $\theta_{1}>\theta_{2}>....>\theta_{n}$ for in states and $\theta_{1}<\theta_{2}<....<\theta_{n}$
for out states.

\section{Conformal Field Theories}

Conformal Field Theories are two-dimensional Euclidean field theories,
which possess invariance under conformal transformations, including
scale-invariance. A general introduction to conformal field theories
was presented in \cite{Ginsparg,Cardy}. CFTs are a special set among
the QFT's in the sense that they represent fixed points under the
renormalization group flow. In statistical physics, they describe
fluctuations of critical systems in the continuum limit \cite{Polyakov}.
Conformal invariance highly constrains the behavior of the correlation
functions, and even the operator content of the theory \cite{BPZ84}. 

Due to scale invariance the energy-momentum tensor $T_{\mu\nu}(x,y)$
is constrained to be traceless. In complex coordinates $z=x+iy$ and
$\bar{z}=x-iy$ this condition is expressed as $T_{z\bar{z}}=T_{\bar{z}z}=0$.
Conservation of the energy-momentum tensor means that $\partial_{\bar{z}}T_{zz}=\partial_{z}T_{\bar{z}\bar{z}}=0$.
Therefore, being holomorphic and anti-holomorphic quantities $T(z)\equiv T_{zz}(z)$
and $\bar{T}(\bar{z})\equiv T_{\bar{z}\bar{z}}(\bar{z})$ are chiral
and anti-chiral tensor components respectively.

$T(z)$ may be expanded into its Laurent-series by summing over its
modes around $z=0$ as 
\[
T(z)=\sum_{n}z^{-n-2}L_{n}
\]
 The algebraic operators $L_{n}$ satisfy the Virasoro algebra: 
\[
[L_{n},L_{m}]=(n-m)L_{n+m}+\frac{c}{12}(n^{3}-n)\delta_{n+m,0}
\]
 which is a an extension of the symmetry algebra of the classical
conformal group. The new element $c$ that appears in this expression
is the central charge of the theory.

The value of $c$ restricts the possible representations of the Virasoro-algebra,
the operator content and the spectrum of the theory. The simplest
theories are the minimal models \cite{PBZstat}, which contain a finite
number of primary fields and possess no additional symmetries. For
those models the central charge is determined from two coprime integers
$p$ and $q$: 
\[
c=1-\frac{6(p-q)^{2}}{pq}
\]
 Theories with $q=p+1$ are unitary theories having no negative-norm
states. However some non-unitary models are also interesting models
to study like the Lee-Yang due to their simplicity, and the possibility
of generalizing the outcomes of their study to physical models of
higher complexity.

The $L_{n}$'s generate the conformal transformations associated with
$z=x+iy$ with $c$ denoting the central charge of the conformal theory.
The same algebra holds with $\bar{L}_{n}$ associated with $\bar{z}=x-iy$,
the complex conjugate of $z$. The $L$ and $\bar{L}$ commute. Each
operator family is formed of a primary operator $\Phi$ and its descendants
formed by the repeated action of $L_{n}$ and $\bar{L}_{n}$ with
negative $n.$ Positive values of $n$ annihilate $\Phi.$

The descendant operators are of the form:

\[
(L_{-i_{1}}....\ L_{-i_{I}}\bar{L}_{-j_{1}}....\ \bar{L}_{-j_{J}})\Phi
\]

with

\[
0<i_{1}\leq....\leq i_{I}
\]
and 
\[
0<j_{1}\leq....\leq j_{J}
\]

The descendant operators will have the levels $(l,\bar{l})$ where
$l=\sum_{n=1}^{I}i_{n}$ and $\bar{l}=\sum_{n=1}^{J}j_{n}$. The conformal
dimensions of the descendant operators will be $(\Delta,\bar{\Delta})=(\Delta_{\Phi}+l,\bar{\Delta}_{\Phi}+\bar{l})$.

A scaling operator $\Phi$ has the conformal dimension $(\Delta_{\Phi},\bar{\Delta}_{\Phi})$
which determines the scaling dimension and the euclidean spin by:

\begin{equation}
X_{\Phi}=\Delta_{\Phi}+\bar{\Delta}_{\Phi}
\end{equation}
and 

\begin{equation}
s_{\Phi}=\Delta_{\Phi}-\bar{\Delta}_{\Phi}
\end{equation}

One has to pay attention that not all descendants are independent
due to the presence of degenerate representations that possess vanishing
linear combinations of descendant operators.

\section{CFTs on the cylinder}

The conformal transformations 
\begin{equation}
z=\exp\Big(\frac{2\pi}{L}(\tau-ix)\Big)\quad\quad\bar{z}=\exp\Big(\frac{2\pi}{L}(\tau+ix)\Big)\label{eq:exponentialmap}
\end{equation}

can be used to map the complex plane into a cylinder of spatial circumference
$L$ , where $x$ and $\tau$ are the spatial and the imaginary time
coordinates, respectively. With this transformation the Hamiltonian
operator will be defined as: 
\begin{equation}
H=\int_{0}^{L}dx\ T_{\tau\tau}(x,\tau=0)=\frac{2\pi}{L}\Big(L_{0}+\bar{L}_{0}-\frac{c}{12}\Big)\label{conform_H}
\end{equation}
 In minimal models the Hilbert-space is given by 
\begin{equation}
\mathcal{H}=\bigoplus_{h}V_{h}\otimes\bar{V}_{h}\label{conformal_Hilbert_space}
\end{equation}
 where $V$ and $\bar{V}$ denote the irreducible representation of
the left and right Virasoro algebras with highest weight $h$.

\section{Perturbing CFT's}

Conformal field theories represent statistical physical or quantum
systems at criticality. However, they can be also used to approach
noncritical models.

If there is only one perturbation present, which only brakes a subset
of the conformal symmetries, the theory may still possess an infinite
number of conservation laws, and it may remain integrable \cite{Zam89,Zam87,Zam87R}.
We assume that the theory defined by the action 
\begin{equation}
A=A_{CFT}+g\int d^{2}x\ \Phi(x)\label{integrable_action}
\end{equation}
 is integrable. Scale-invariance is broken by the perturbation. This
action can define a massless or a massive perturbation depending on
the original $CFT$ and the nature of the perturbation.

The correspondence between perturbed $CFT$s and scattering theories
can be inspected by several ways. 
\begin{itemize}
\item Zamolodchikov's counting argument \cite{Zam89}. 
\item Thermodynamic Bethe Ansatz \cite{Zam90tba,KlMe}, which determines
the finite size dependence of the vacuum energy in terms of the S-matrix
of the theory. It allows to predict the central charge of the CFT
and the scaling dimension of the perturbing field.
\item Truncated Conformal Space Approach (TCSA) which can be used to numerically
determine the low-lying energy levels of the finite size spectrum.
One can identify multi-particle states and test the phase shifts $S_{ab}(\theta)$. 
\end{itemize}

\section{The Lee-Yang model}

The non-unitary minimal model $M_{2,5}$ has central charge $c=-22/5$
and a unique nontrivial primary field $\Phi$ with scaling weights
$\Delta=\bar{\Delta}=-1/5$. The field $\Phi$ is normalized so that
it has the following operator product expansion:
\begin{equation}
\Phi(z,\bar{z})\Phi(0,0)=\mathcal{C}(z\bar{z})^{1/5}\Phi(0,0)+(z\bar{z})^{2/5}\mathbb{I}+\dots\label{eq:lyconfope}
\end{equation}
 where $\mathbb{I}$ is the identity operator and the only nontrivial
structure constant is 
\[
\mathcal{C}=1.911312699\dots\times i
\]
 The Hilbert space of the conformal model is given by
\[
\mathcal{H}_{LY}=\bigoplus_{h=0,-1/5}\mathcal{V}_{h}\otimes\bar{\mathcal{V}}_{h}
\]
 where $\mathcal{V}_{h}$ ($\bar{\mathcal{V}}_{h}$) denotes the irreducible
representation of the left (right) Virasoro algebra with highest weight
$h$.

The off-critical Lee-Yang model is defined by the Hamiltonian 
\begin{equation}
H^{SLY}=H_{CFT}+i\lambda\int_{0}^{L}dx\Phi(0,x)\label{eq:lypcftham}
\end{equation}
 where $H_{CFT}$ is the conformal Hamiltonian. 

When $\lambda>0$ the theory above has a single particle in its spectrum
with mass $m$.

The $S$-matrix reads \cite{CaMu}
\begin{equation}
S_{LY}(\theta)=\frac{\sinh\theta+i\sin\frac{2\pi}{3}}{\sinh\theta-i\sin\frac{2\pi}{3}}\label{eq:Smatly}
\end{equation}
 and the particle occurs as a bound state of itself at $\theta=2\pi i/3$
with the three-particle coupling given by
\[
\Gamma^{2}=-2\sqrt{3}
\]
 where the negative sign is due to the non-unitarity of the model. 

\pagebreak{}

\chapter{Form Factors in Presence of Defects}

\section{Defect form factors }

In this section we present the axioms for the matrix elements of local
operators between asymptotic states. To shorten the discussion we
introduce Zamolodchikov-Faddeev (ZF) operators in order to describe
both the multiparticle transmission process as well as the properties
of the form factors.

\subsection{Asymptotic states and transmission matrix}

Multi-particle asymptotic states in integrable \emph{bulk} theories
can be formulated in terms of the ZF operators as 
\[
\vert\theta_{1},\dots,\theta_{n}\rangle=A^{+}(\theta_{1})\dots A^{+}(\theta_{n})\vert0\rangle
\]
All particles have different momenta $p_{i}=m\sinh\theta_{i}$, thus
in the remote past they are not interacting and form an initial state
$\theta_{1}<\dots<\theta_{n}$. When time evolves they approach each
other and after the consecutive scatterings they rearrange themselves
into the opposite order:

\[
\vert\theta_{1},\dots,\theta_{n}\rangle=\prod_{i<j}S(\theta_{i}-\theta_{j})\vert\theta_{n},\dots,\theta_{1}\rangle
\]
Here $S(\theta)$ is the two particle scattering matrix which satisfies
unitarity and crossing symmetry

\[
S(-\theta)=S^{-1}(\theta)\quad;\qquad S(i\pi-\theta)=S(\theta)
\]
This multi-particle scattering process is easily formulated with the
ZF algebra: 

\begin{equation}
A^{+}(\theta_{1})A^{+}(\theta_{2})=S(\theta_{1}-\theta_{2})A^{+}(\theta_{2})A^{+}(\theta_{1})+2\pi\delta(\theta_{1}-\theta_{2}-i\pi)\label{eq:ZFalgebra}
\end{equation}
where we extended their definition for imaginary $\theta$ by postulating
the crossing property \cite{BFF}: 
\begin{equation}
A(\theta)=A^{+}(\theta+i\pi)\label{eq:crossing}
\end{equation}

Once defects are introduced we have to make a distinction whether
the particle arrives from the left ($A$) or from the right ($B$)
to the defect. These particles can be even different from each other
as they live in different subsystems. A multiparticle state is then
described by 
\[
\vert\theta_{1},\dots,\theta_{n};\theta_{n+1},\dots,\theta_{m}\rangle=A^{+}(\theta_{1})\dots A^{+}(\theta_{n})D^{+}B^{+}(\theta_{n+1})\dots B^{+}(\theta_{m})\vert0\rangle
\]
where the ZF operators $B^{+}$ create particles on the right of the
defect and satisfy similar defining relations to (\ref{eq:ZFalgebra})
with possibly different scattering matrix. Yet, for simplicity, we
restrict our discussion to the case when the two subsystems are identical
with the same scattering matrix. Observe however, that this does not
imply space parity invariance, since the defect may break it. In the
initial state rapidities are ordered as $\theta_{1}>\dots>\theta_{n}>0>\theta_{n+1}>\dots>\theta_{m}$.
The final state, in which all scatterings and transmissions are already
terminated, can be expressed in terms of the initial state via the
multiparticle transmission matrix. 
\[
\vert\theta_{1},\dots,\theta_{n};\theta_{n+1},\dots,\theta_{m}\rangle=\prod_{i<j}S(\theta_{i}-\theta_{j})\prod_{i=1}^{n}T_{-}(\theta_{i})\prod_{i=n+1}^{m}T_{+}(-\theta_{i})\vert\theta_{m},\dots,\theta_{n+1};\theta_{n},\dots,\theta_{1}\rangle
\]
Due to integrability it factorizes into pairwise scatterings and individual
transmissions: $T_{-}(\theta)$ and $T_{+}(-\theta)$. We parametrize
$T_{+}$ in such a way that for its physical domain ($\theta<0$)
its argument is always positive. Transmission factors satisfy unitarity
and defect crossing symmetry \cite{DefBound}
\begin{equation}
T_{+}(-\theta)=T_{-}^{-1}(\theta)\quad;\qquad T_{-}(\theta)=T_{+}(i\pi-\theta)\label{Defprop}
\end{equation}
The multiparticle transition amplitude can be derived by introducing
the defect operator $D^{+}$ and the following relations in the ZF
algebra: 
\[
A^{+}(\theta)D^{+}=T_{-}(\theta)D^{+}B^{+}(\theta)\qquad;\quad D^{+}B^{+}(-\theta)=T_{+}(\theta)A^{+}(-\theta)D^{+}
\]
A defect is parity symmetric if $T_{-}(\theta)=T_{+}(\theta)$. Clearly
$A^{+}(\theta=0)$ satisfies the properties of $D^{+}$ with $T_{-}(\theta)=S(\theta)=T_{+}(\theta)$
. Thus a standing particle can be considered as the prototype of a
parity symmetric defect.

\subsection{Coordinate dependence of the form factors}

The form factor of a local operator $\mathcal{O}(x,t)$ is its matrix
element between asymptotic states:
\[
\langle\theta_{m^{'}}^{'},\dots,\theta_{n^{'}+1}^{'};\theta_{n^{'}}^{'},\dots,\theta_{1}^{'}\vert\mathcal{O}(x,t)\vert\theta_{1},\dots,\theta_{n};\theta_{n+1},\dots,\theta_{m}\rangle
\]
where the adjoint state is defined to be 
\[
\langle\theta_{m^{'}}^{'},\dots,\theta_{n^{'}+1}^{'};\theta_{n^{'}}^{'},\dots,\theta_{1}^{'}\vert=\langle0\vert B(\theta_{m^{'}}^{'})\dots B(\theta_{n^{'}+1}^{'})DA(\theta_{n^{'}}^{'})\dots A(\theta_{1}^{'})
\]
Strictly speaking the form factor is defined only for initial/final
states (i.e. for decreasingly/increasingly ordered arguments) but
using the ZF algebra we can generalize them for any values and orders
of the rapidities. 

The multiparticle asymptotic states are eigenstates of the conserved
energy. This fact can be formulated in the language of the ZF algebra
as 
\[
[H,A^{+}(\theta)]=m\cosh\theta\, A^{+}(\theta)\qquad;\qquad[H,D^{+}]=e_{D}D^{+}
\]
In the second equation we supposed that the vacuum containing the
defect has energy $e_{D}$. Classical considerations together with
the topological nature of the defect suggest the existence of a conserved
momentum with properties
\[
[P,A^{+}(\theta)]=m\sinh\theta\, A^{+}(\theta)\qquad;\qquad[P,D^{+}]=p_{D}D^{+}
\]
Thus, opposed to a general boundary theory, the defect breaks translation
invariance by having a nonzero momentum eigenvalue $p_{D}$ and not
by destroying the existence of the momentum itself. As a consequence
the time and space dependence of the form factor can be obtained as
\begin{eqnarray*}
\langle\theta_{m^{\prime}}^{\prime},\dots,\theta_{n^{\prime}+1}^{\prime};\theta_{n^{\prime}}^{\prime},\dots,\theta_{1}^{\prime}\vert\mathcal{O}(x,t)\vert\theta_{1},\dots,\theta_{n};\theta_{n+1},\dots,\theta_{m}\rangle=\\
e^{it\Delta E-ix\Delta P}F_{(n^{\prime},m^{\prime})(n,m)}^{\mathcal{O}}(\theta_{n\prime+m\prime}^{\prime},...,\theta_{n\prime+1}^{\prime};\theta_{n\prime}^{\prime},...,\theta_{1}^{\prime}\vert\theta_{1},\dots,\theta_{n};\theta_{n+1},...,\theta_{n+m})
\end{eqnarray*}
 where $\Delta E=m(\sum_{j}\cosh\theta_{j}-\sum_{j^{\prime}}\cosh\theta_{j^{\prime}}^{\prime})$
and $\Delta P=m(\sum_{j}\sinh\theta_{j}-\sum_{j^{\prime}}\sinh\theta_{j^{\prime}}^{\prime})$.
The very same simple space and time dependence can be seen in a theory
without the defect and it is substantially different from what we
would expect from a general boundary theory where even the one point
function has a nontrivial space-dependence. These considerations remain
valid for operators inserted at the defect $\mathcal{O}(t)=\mathcal{O}(0,t)$,
too.

\subsection{Crossing transformation of defect form factors}

The properties and analytical structure of the form factor $F_{(n^{\prime}m^{\prime}),(nm)}$
can be derived via the reduction formula from the correlation functions
similarly to the boundary case \cite{BFF}. Instead of going to the
details of the calculation of \cite{BFF} we note that all equations
follow from the defining relations of the ZF algebra and the locality
of the operator $[\mathcal{O}(0,0),A^{+}(\theta)]=0$ except the crossing
relation. It reads as 
\begin{eqnarray*}
F_{(n\prime,m\prime)(n,m)}^{\mathcal{O}}(\theta_{n\prime+m\prime}^{\prime},...,\theta_{n\prime+1}^{\prime};\theta_{n\prime}^{\prime},...,\theta_{1}^{\prime}\vert\theta_{1},\dots,\theta_{n};\theta_{n+1},...,\theta_{n+m})=\\
F_{(n\prime,m\prime+1)(n,m-1)}^{\mathcal{O}}(\theta_{n+m}+i\pi,\theta_{n\prime+m\prime}^{\prime},...,\theta_{n\prime+1}^{\prime};\theta_{n\prime}^{\prime},...,\theta_{1}^{\prime}\vert\theta_{1},\dots,\theta_{n};\theta_{n+1},...,\theta_{n+m-1})
\end{eqnarray*}
and can be obtained as follows: We fold the system to a boundary one:
$B^{+}(\theta)\leftrightarrow\tilde{B}^{+}(-\theta)$, and consider
$A^{+}$ and $\tilde{B}^{+}$ as creation operators of two different
type of particles which scatter trivially on each other. Now we apply
the crossing equation of $\tilde{B}^{+}$ for the resulting boundary
form factor. If we fold back the system to the original defect theory
we obtain the defect crossing equation above. 

By analyzing the crossing equation of the particle $A^{+}$ instead
of $B^{+}$ we obtain 
\begin{eqnarray*}
F_{(n\prime,m\prime)(n,m)}^{\mathcal{O}}(\theta_{n\prime+m\prime}^{\prime},...,\theta_{n\prime+1}^{\prime};\theta_{n\prime}^{\prime},...,\theta_{1}^{\prime}|\theta_{1},\dots,\theta_{n};\theta_{n+1},...,\theta_{n+m}) & =\\
F_{(n\prime+1,m\prime)(n-1,m)}^{\mathcal{O}}(\theta_{n\prime+m\prime}^{\prime},...,\theta_{n\prime+1}^{\prime};\theta_{n\prime}^{\prime},...,\theta_{1}^{\prime},\theta_{1}-i\pi|\theta_{2},\dots,\theta_{n};\theta_{n+1},...,\theta_{n+m})
\end{eqnarray*}
This crossing equation can also be obtained from (\ref{eq:crossing}).
Using any of the crossing equations above we can express all form
factors in terms of the one-sided form factors: 
\[
F_{(n,m)}^{\mathcal{O}}(\theta_{1},\dots,\theta_{n};\theta_{n+1},...,\theta_{n+m}):=F_{(0,0)(n,m)}^{\mathcal{O}}(;\vert\theta_{1},\dots,\theta_{n};\theta_{n+1},...,\theta_{n+m})
\]
The properties of this form factor follows from the ZF algebra relations
and from the crossing relations and we postulate them in the next
subsection as axioms.

\subsection{Defect form factor axioms}

The matrix elements of local operators satisfy the following axioms:

I. Transmission:

\begin{center}
\[
F_{(n,m)}^{\mathcal{O}}(\theta_{1},\dots,\theta_{n};\theta_{n+1},...,\theta_{n+m})=T_{-}(\theta)F_{(n-1,m+1)}^{\mathcal{O}}(\theta_{1},\dots,\theta_{n-1};\theta_{n},\theta_{n+1},...,\theta_{n+m})
\]
\includegraphics{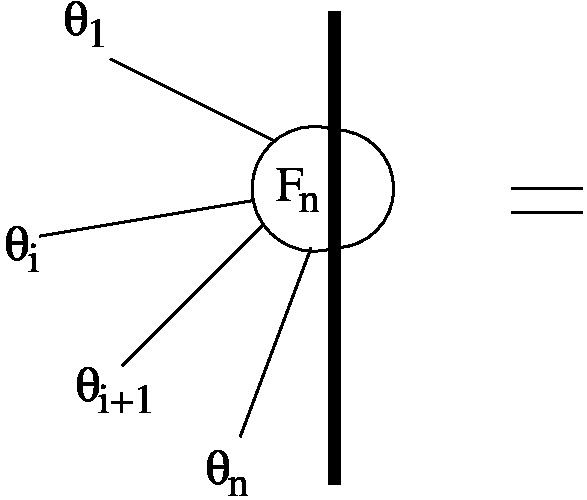}~~~~~~~\includegraphics{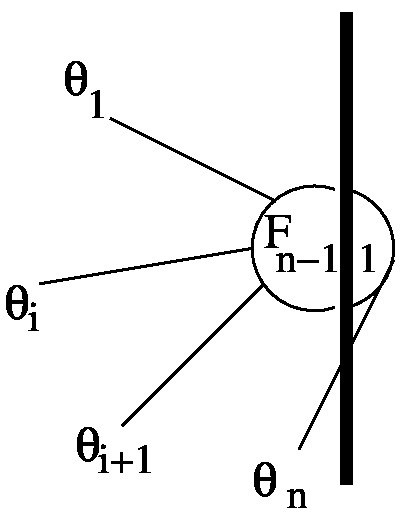}
\par\end{center}

By means of this axiom we can express every form factor in terms of
the elementary one
\[
F_{n}^{\mathcal{O}}(\theta_{1},\dots,\theta_{n})=F_{(n,0)}^{\mathcal{O}}(\theta_{1},\dots,\theta_{n};)
\]
It satisfies the further axioms:

II. Permutation:
\[
F_{n}^{\mathcal{O}}(\theta_{1},\dots\theta_{i},\theta_{i+1},\dots,\theta_{n})=S(\theta_{i}-\theta_{i+1})F_{n}^{\mathcal{O}}(\theta_{1},\dots\theta_{i+1},\theta_{i},\dots,\theta_{n})
\]

\begin{center}
\includegraphics[height=3cm]{bd1}~~~~~~~~~\includegraphics[height=3cm]{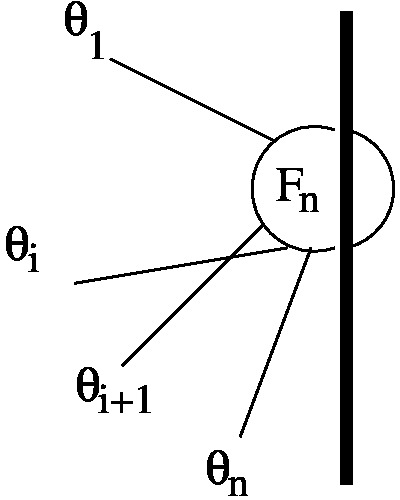}
\par\end{center}

III. Periodicity:
\[
F_{n}^{\mathcal{O}}(\theta_{1},\theta_{2},\dots,\theta_{n})=F_{n}^{\mathcal{O}}(\theta_{2},\dots\theta_{n},\dots,\theta_{1}-2i\pi)
\]

\begin{center}
\includegraphics[height=3cm]{bd1}~~~~~~~~\includegraphics[height=3cm]{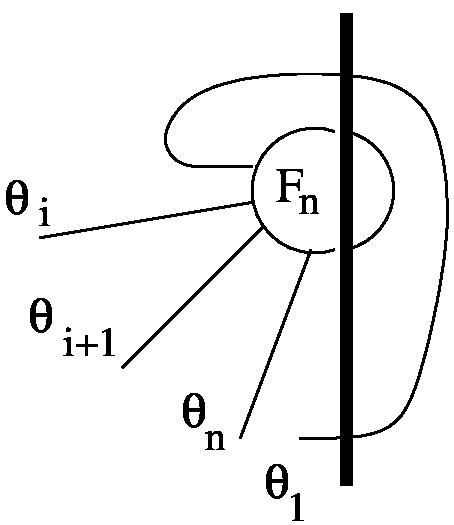}
\par\end{center}

The physical singularities can be formulated as follows. 

IV. Kinematic singularity:
\[
-i\mbox{Res}_{\theta=\theta\prime}F_{n+2}^{\mathcal{O}}(\theta+i\pi,\theta^{\prime},\theta_{1},...,\theta_{n})=\left(1-\prod_{j=1}^{n}S(\theta-\theta_{j})\right)F_{n}^{\mathcal{O}}(\theta_{1},...,\theta_{n})
\]

\begin{center}
\includegraphics[height=3cm]{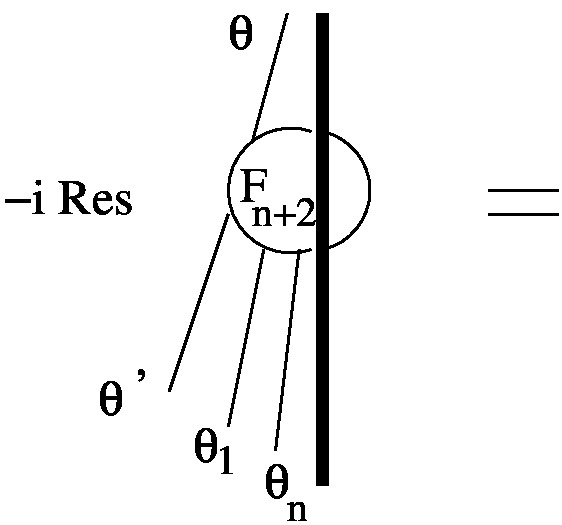}~~~\includegraphics[height=3cm]{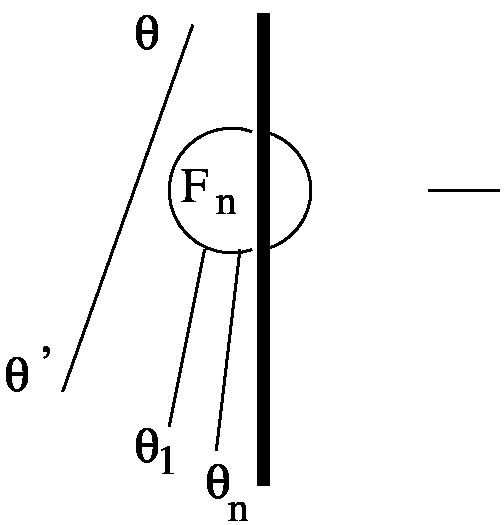}~~~\includegraphics[height=3cm]{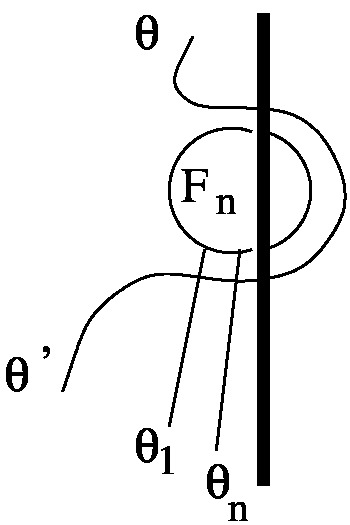}
\par\end{center}

V. Dynamical bulk singularity:
\[
-i\mbox{Res}_{\theta^{\prime}=\theta}F_{n+2}^{\mathcal{O}}(\theta^{\prime}+\frac{i\pi}{3},\theta-\frac{i\pi}{3},\theta_{1},\ldots,\theta_{n})=\Gamma F_{n+1}^{\mathcal{O}}(\theta,\theta_{1},\ldots,\theta_{n})
\]

\begin{center}
\includegraphics[height=3cm]{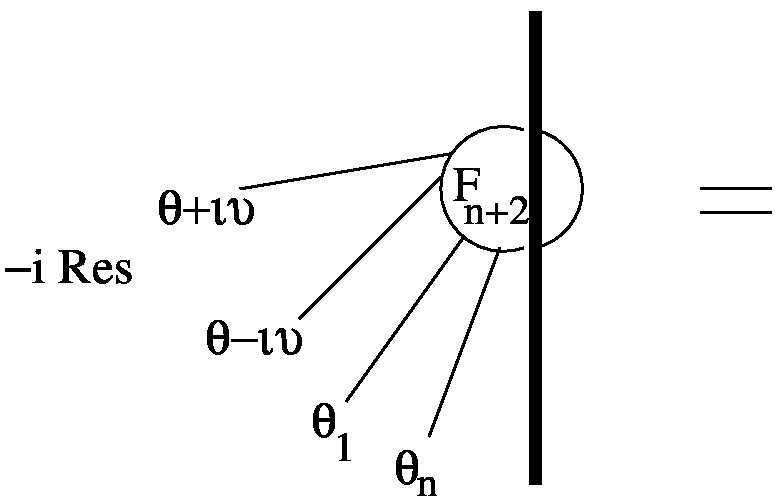}~~~~~~\includegraphics[height=3cm]{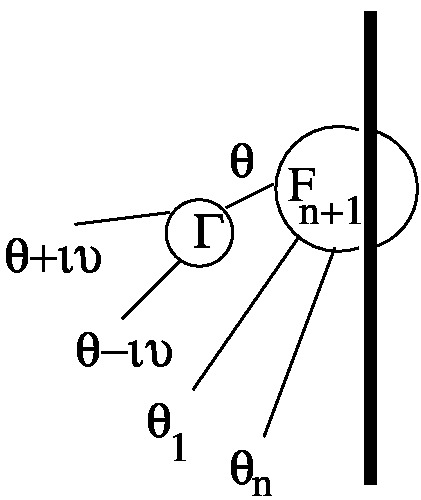}
\par\end{center}

where $\Gamma$ is the 3 particle on-shell coupling.

VI. Dynamical defect singularity:
\[
-i\mbox{Res}_{\theta=iu}F_{n+1}^{\mathcal{O}}(\theta_{1},\ldots,\theta_{n},\theta)=ig\tilde{F}_{n}^{\mathcal{O}}(\theta_{1},\ldots,\theta_{n})
\]

\begin{center}
\includegraphics[height=3cm]{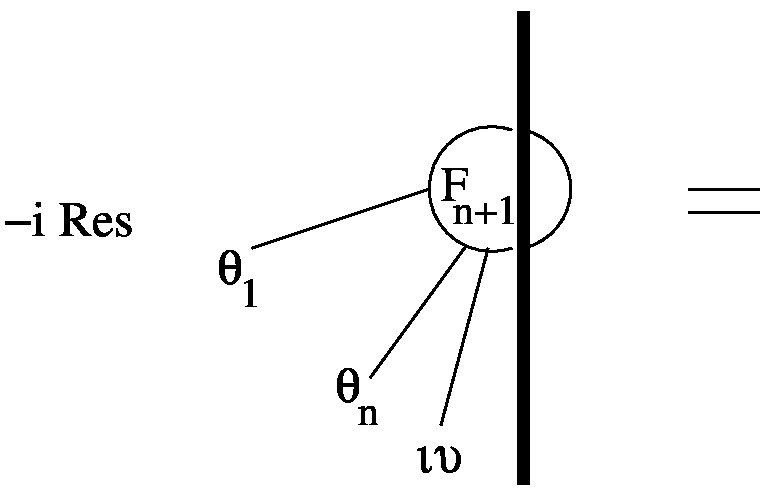}~~~~~~~\includegraphics[height=3cm]{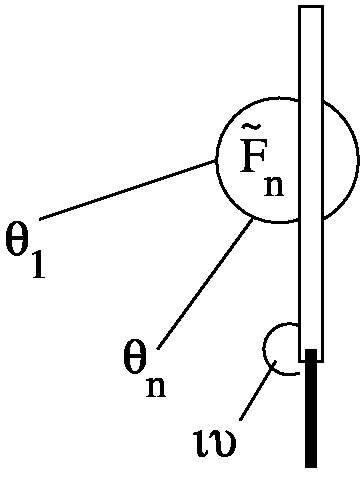}
\par\end{center}

where $g$ is the defect bound-state coupling.

A few remarks are in order: Although the form factor axioms (II-V)
are the same as the axioms of the form factors in a theory without
the defect, the axioms (I,VI) are different and in general we will
have different solutions. An exception is the invisible defect $T_{-}(\theta)=1$
when we recover the usual form factor equation providing a consistency
check for our axioms. Another consistency check can be obtained by
considering a standing particle as the defect. Then $T_{\pm}(\theta)=S(\theta)$
and the two additional axioms become part of the old ones: (I,VI)
will be special cases of (II,V), respectively.

\section{Form factor solutions, two point functions}

In this section we determine the solutions of the form factor equations
for operators localized in the bulk and at the defect. For operators
localized in the bulk the solutions can be built up form the bulk
form factors and from the transmission factors. Using the defect form
factor solution we determine the spectral representation of the two
point function for the situations when the operators are localized
on the same or on the opposite sides of the defect. Finally, for operators
localized on the defect we outline the strategy for the general solution.

\subsection{Bulk operators}

The form factor axioms for $F_{n}$ are the same as in the bulk so
we expect to use the bulk form factor solutions. Clearly we have to
make a distinction whether the operator are localized on the left,
or on the right of the defect. If the operator is localized on the
left then particles arriving from the left can reach the operator
without crossing the defect. Since the defect is topological changing
its location does not alter the form factor (as far as we do not cross
the insertion point of the operator). Shifting the defect faraway
we expect to obtain the form factors of the bulk theory. 

\begin{center}
\includegraphics[height=3cm]{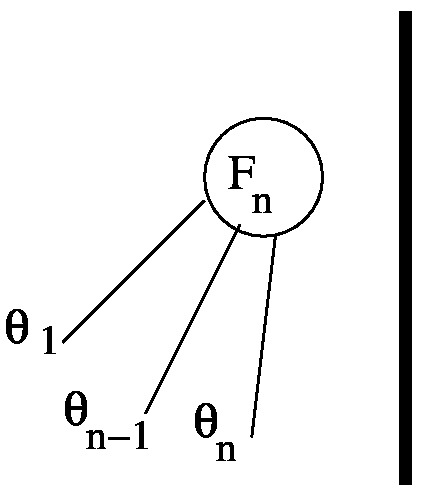}~~~~~\includegraphics[height=3cm]{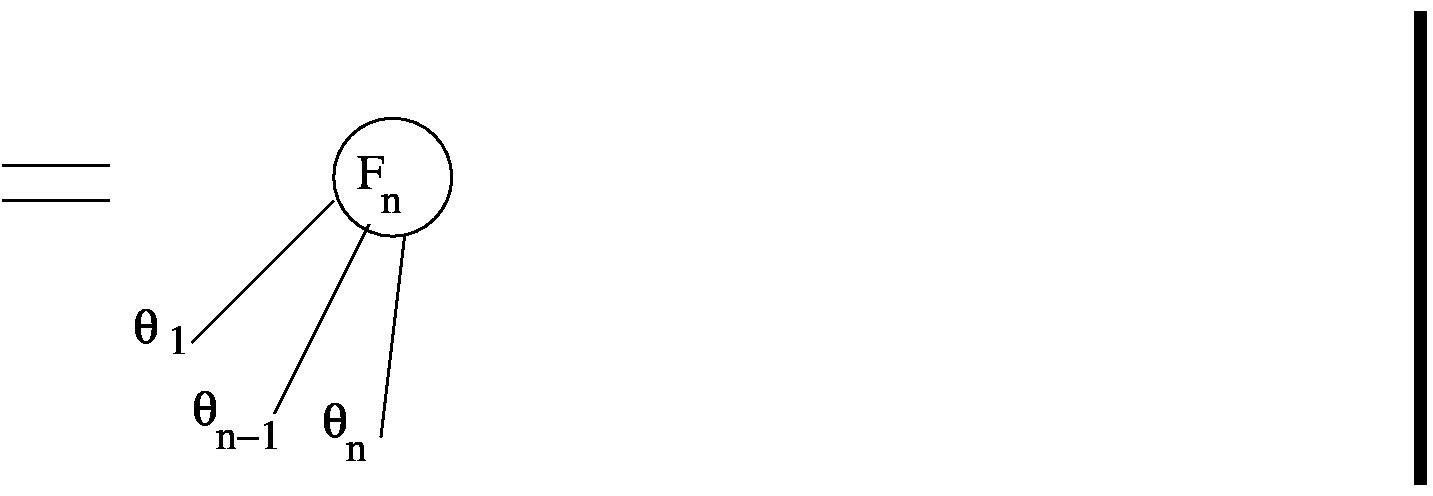}
\par\end{center}

Thus we can conclude that for the initial state $\theta_{1}>\dots>\theta_{n}>0$
the defect form factor coincides with the bulk form factor. Let us
denote the solutions of the bulk form factor equations by $B_{n}(\theta_{1},\dots,\theta_{n})$.
Then we claim that for an operator localized on the left ($\mathcal{O}^{<}$
for short) we have 
\begin{equation}
F_{n}^{<}(\theta_{1},\dots,\theta_{n})=B_{n}(\theta_{1},\dots,\theta_{n})=F_{(n,0)}^{<}(\theta_{1},\dots,\theta_{n};)\label{eq:bulkleft}
\end{equation}
By using the transmission axiom and the crossing relation we can express
all other matrix elements in terms of the bulk matrix element and
the transmission factor. If the operator is localized on the right
of the defect ($\mathcal{O}^{>}$) then, by similar argumentation,
we expect the defect form factor to coincide with the bulk form factor
for particles coming from the right. Those initial states have the
ordering $0>\theta_{1}>\dots>\theta_{n}$ and the form factor is then
\[
F_{(0,n)}^{>}(;\theta_{1},\dots,\theta_{n})=B_{n}(\theta_{1},\dots,\theta_{n})
\]
The solution for the elementary defect form factor for operators localized
on the right thus turns out to be 
\begin{equation}
F_{n}^{>}(\theta_{1},\dots,\theta_{n})=\prod_{i}T_{-}(\theta_{i})B_{n}(\theta_{1},\dots,\theta_{n})\label{eq:bulkright}
\end{equation}
which satisfies all the bulk form factor axioms but \emph{does not
coincide} with the bulk form factor solution. Having calculated the
form factor solutions we use them to construct the two point functions,
which for operators localized on the opposite side of the defect will
be intrinsically different from the one without the defect.

We analyze the following two point function 
\[
\langle;\vert\mathcal{O}_{1}(x_{1},t_{1})\mathcal{O}_{2}(x_{2},t_{2})\vert;\rangle
\]
where we denote by $\vert;\rangle$ the vacuum of the defect theory.
Formally $\vert;\rangle=D^{+}\vert0\rangle$. Now we insert the resolution
of the identity. It can be composed both from initial and from final
states and for definiteness we choose initial states. It is instructive
to list the possible states. If we have no particles we have only
the vacuum: $\vert;\rangle$. One particle states can be of two types,
depending on whether the particle arrives from the left or from the
right: $\vert\theta;\rangle$ for $\theta>0$ and $\vert;\theta\rangle$
for $\theta<0$. A general $N=n+m$ particle state $\vert\theta_{1},\dots,\theta_{n};\theta_{n+1},\dots,\theta_{n+m}\rangle$
with $\theta_{1}>\dots>\theta_{n}>0>\theta_{n+1}>\dots>\theta_{m}$
has to cover all possible cases ranging from $n=0$ to $n=N$ . The
two point function then can be written formally as 
\[
\langle;\vert\mathcal{O}_{1}(x_{1},t_{1})\mathcal{O}_{2}(x_{2},t_{2})\vert;\rangle=\sum_{N=0}^{\infty}\langle;\vert\mathcal{O}_{1}(0,0)\vert N\rangle\langle;\vert\mathcal{O}_{2}^{*}(0,0)\vert N\rangle^{*}e^{iE(N)(t_{1}-t_{2})-iP(N)(x_{1}-x_{2})}
\]
We have to specify the integration ranges for the multiparticle state
$N$. Originally we have to integrate only for the multiparticle momentum
range of the \emph{initial }states. If we exchange the order for $\theta_{1}>\theta_{2}$
to the nonphysical $\theta_{2}<\theta_{1}$ then the form factor of
$\mathcal{O}_{1}$ picks up a factor $S(\theta_{1}-\theta_{2})$ while
that of $\mathcal{O}^{*}$ the inverse factor $S^{*}(\theta_{1}-\theta_{2})$,
so the integrand is a symmetric function. For each integration with
$\theta_{1}>0$ we also have an analogous integration for $-\theta_{1}<0$
. Their contributions differ by a factor $T_{+}(\theta_{1})$ for
the form factor of $\mathcal{O}_{1}$ and by the inverse $T_{+}^{*}(\theta_{1})$
for $\mathcal{O}_{2}$. As a consequence we can express the correlator
in terms of the elementary form factors $F_{n}$ as:

\noindent \begin{center}
{\small{}$\begin{array}{cc}
\langle;\vert\mathcal{O}_{1}(x_{1},t_{1})\mathcal{O}_{2}(x_{2},t_{2})\vert;\rangle=\\
\sum_{n=0}^{\infty}\frac{1}{n!}\int_{-\infty}^{\infty}\frac{d\theta_{1}}{2\pi}\dots\int_{-\infty}^{\infty}\frac{d\theta_{n}}{2\pi}F_{n}^{O_{1}}(\theta_{1},\dots,\theta_{n})F_{n}^{O_{2}^{*}}(\theta_{1},\dots,\theta_{n})^{*}e^{iE(n)(t_{1}-t_{2})-iP(n)(x_{1}-x_{2})}
\end{array}$}
\par\end{center}{\small \par}

Although we transported the operators $\mathcal{O}_{1}$ and $\mathcal{O}_{2}$
into the origin, the form factor solutions remember whether the operators
are localized on the left or on the right of the defect. 

If both operators are localized on the left, ($x_{1}<0,\, x_{2}<0$)
then the elementary form factors are the same as the bulk form factors
(\ref{eq:bulkleft}) and we can conclude that the two point function
is\emph{ exactly the same} as the bulk two point function 

\noindent \begin{center}
{\small{}$\begin{array}{cc}
\langle;\vert\mathcal{O}_{1}(x_{1},t_{1})\mathcal{O}_{2}(x_{2},t_{2})\vert;\rangle=\\
\sum_{n=0}^{\infty}\frac{1}{n!}\int_{-\infty}^{\infty}\frac{d\theta_{1}}{2\pi}\dots\int_{-\infty}^{\infty}\frac{d\theta_{n}}{2\pi}B_{n}^{O_{1}}(\theta_{1},\dots,\theta_{n})B_{n}^{O_{2}^{*}}(\theta_{1},\dots,\theta_{n})^{*}e^{iE(n)(t_{1}-t_{2})-iP(n)(x_{1}-x_{2})}
\end{array}$}
\par\end{center}{\small \par}

This is intuitively clear: we can transport the defect to infinity
without crossing any of the insertion points thus leaving invariant
the two point function. When the defect is at infinity it does not
influences the two point function which then has to be the same as
in the bulk. The same result can be obtained when both operators are
localized on the right of the defect. 

If the operators are localized on different sides of the defect ($x_{1}<0$,
$x_{2}>0$) then additionally to (\ref{eq:bulkleft}) we also have
to use (\ref{eq:bulkright}). As a result the two point function is
expressed in terms of the bulk form factors $B_{n}$ and the transmission
matrix $T_{-}(\theta)$ as 

\noindent \begin{center}
{\footnotesize{}$\begin{array}{cc}
\langle;\vert\mathcal{O}_{1}(x_{1},t_{1})\mathcal{O}_{2}(x_{2},t_{2})\vert;\rangle=\\
\sum_{n=0}^{\infty}\frac{1}{n!}\int_{-\infty}^{\infty}\frac{d\theta_{1}}{2\pi}T_{-}(\theta_{1})\dots\int_{-\infty}^{\infty}\frac{d\theta_{n}}{2\pi}T_{-}(\theta_{n})B_{n}^{O_{1}}(\theta_{1},\dots,\theta_{n})B_{n}^{O_{2}^{*}}(\theta_{1},\dots,\theta_{n})^{*}e^{iE(n)(t_{1}-t_{2})-iP(n)(x_{1}-x_{2})}
\end{array}$}
\par\end{center}{\footnotesize \par}

This is the main result of this section. This formula shows how the
correlation function can be calculated in the presence of an integrable
defect in terms of the transmission factor and the bulk form factors.
It can be generalized to any correlators localized in the bulk using
the resolution of the identity together with the exact form factors
(\ref{eq:bulkleft}) and (\ref{eq:bulkright}). It cannot be applied,
however, for operators localized on the defect, which is the subject
of the next subsection.

\subsection{Defect operators}

We have seen that although the minimal form factors $F_{n}$ are subject
to the same requirement as the bulk form factors they are not necessarily
the same. In this subsection we develop a general methodology to determine
the defect form factors. Let us analyze them for increasing particle
numbers:

The first form factor is the vacuum expectation value of a defect
field 
\[
\langle;\vert\mathcal{O}(t)\vert;\rangle=F_{0}
\]

The one particle form factor is defined to be 
\[
\langle;\vert\mathcal{O}(t)\vert\theta;\rangle=F_{1}(\theta)\quad;\qquad\langle;\vert\mathcal{O}(t)\vert;\theta\rangle=T_{-}(\theta)F_{1}(\theta)
\]
 Contrary to the bulk case it has a nontrivial rapidity dependence:
it is not natural to take $F_{1}$ to be a constant. In a parity invariant
theory for a parity symmetric operators, for example, we have $F_{1}(\theta)=T_{-}(-\theta)F_{1}(-\theta)$.
If parity is broken then $F_{1}(\theta)$ can be an arbitrary defect
condition-dependent $2\pi i$-periodic function. The only restriction
came from the defect bound-state axiom (VI): it must have a pole at
$iv$ whenever $T_{-}(\theta)$ has a pole corresponding to a bound-state.
Let us denote the minimal function which satisfies this requirement
by $d(\theta)$. The general form of the one particle form factor
is then 
\[
\langle;\vert\mathcal{O}(t)\vert\theta;\rangle=d(\theta)Q^{\mathcal{O}}(x)\quad;\qquad x=e^{\theta}
\]
where $d(\theta)$ depends on the defect condition, while $Q(x)$
depends on the operator we are dealing with. 

The two particle form factor must also have a singularity at $i\nu$
and additionally it satisfies the bulk form factor axioms so we expect
it to be written into the form 
\[
F_{2}(\theta_{1},\theta_{2})=d(\theta_{1})d(\theta_{2})f_{min}(\theta_{1}-\theta_{2})
\]
where $f_{min}(\theta)$ is the minimal solution of the bulk two particle
form factor equations
\[
f_{min}(\theta)=S(\theta)f_{min}(-\theta)\quad;\qquad f_{min}(i\pi-\theta)=f_{min}(i\pi+\theta)
\]
 Taking into account the general parametrization of the bulk and boundary
form factors together with the dynamical and kinematical singularity
axioms we parametrize our minimal defect form factors as 
\[
F_{n}(\theta_{1},\dots,\theta_{n})=\prod_{i}d(\theta_{i})\prod_{i<j}\frac{f_{min}(\theta_{i}-\theta_{j})}{x_{i}+x_{j}}Q_{n}(x_{1},\dots,x_{n})\quad;\qquad x_{i}=e^{\theta_{i}}
\]
where $Q(x_{1},\dots,x_{n})$ is a symmetric function expected to
be a polynomial, if there is no bulk dynamical singularity. If there
is such a singularity we have to include the corresponding singularity
into $f_{min}$ . The dependence on the defect condition is contained
in $d(\theta)$ with possible defect bound-state singularities, while
the dependence on the operator is in $Q$. If for instance the defect
is the invisible defect with $T_{\pm}=1$ then $d=1$ and we recover
the solution of the bulk form factor equation as it should be. From
the kinematical singularity equations recursion relations can be obtained
among the polynomials $Q_{n+2}$ and $Q_{n}$.

\section{Model studies}

In this section we analyze the solutions of the defect form factor
axioms for the free boson and for the Lee-Yang models.

\subsection{Free boson }

The purely transmitting free bosonic theory was analyzed in \cite{BS}
as the limiting case of the sinh-Gordon theory. The Lagrangian of
the model reads as 

\begin{eqnarray*}
\mathcal{L} & = & \Theta(-x)\left[\frac{1}{2}(\partial_{\mu}\Phi_{-})^{2}-\frac{m^{2}}{2}\Phi_{-}^{2}\right]+\Theta(x)\left[\frac{1}{2}(\partial_{\mu}\Phi_{+})^{2}-\frac{m^{2}}{2}\Phi_{+}^{2}\right]\\
 &  & -\frac{\delta(x)}{2}\left(\Phi_{+}\dot{\Phi}_{-}-\Phi_{-}\dot{\Phi}_{+}+m\left[(\cosh\mu)\left(\Phi_{+}^{2}+\Phi_{-}^{2}\right)+2(\sinh\mu)\,\Phi_{+}\Phi_{-}\right]\right)
\end{eqnarray*}
where $\Phi_{\pm}$ are the fields living on the right/left part of
the defect and $\mu$ is a free parameter. By varying the action we
obtain the free equation of motion in the bulk 
\[
(\partial_{x}^{2}-\partial_{t}^{2})\Phi_{\pm}=m^{2}\Phi_{\pm}
\]
and the defect conditions: 
\begin{eqnarray*}
\pm\partial_{t}\Phi_{\pm}\mp\partial_{x}\Phi_{\mp} & = & m(\sinh\mu)\,\Phi_{\pm}+m(\cosh\mu)\,\Phi_{\mp}
\end{eqnarray*}
Since $\Phi_{\pm}$ are free fields, they have an expansion in terms
of plane waves and creation/annihilation operators 
\[
\Phi_{\pm}(x,t)=\int_{-\infty}^{\infty}\frac{dk}{2\pi}\frac{1}{2\omega(k)}\left(a_{\pm}(k)e^{ikx-i\omega(k)t}+a_{\pm}^{+}(k)e^{-ikx+i\omega(k)t}\right)\quad;\quad\omega(k)=\sqrt{k^{2}+m^{2}}
\]
 where the $a,a^{+}$ operators are adjoint of each other with commutators:
\[
[a_{\pm}(k),a_{\pm}^{+}(k^{'})]=2\pi2\omega(k)\delta(k-k^{'})
\]
They are not independent, the defect condition connects them as 
\[
a_{\pm}(\pm k)=T_{\mp}(k)a_{\mp}(\pm k)\quad;\qquad T_{\mp}(k)=-\frac{m\sinh\mu\mp i\omega(k)}{m\cosh\mu-ik}\quad;\quad k>0
\]
This shows that the defect is purely transmitting, that is we do not
have any reflected wave. The transmission factor in the rapidity parametrization
($k=m_{\mathrm{cl}}\sinh\theta)$ can be written also in the following
form: 
\[
T_{-}(\theta)=-i\frac{\sinh(\frac{\theta}{2}-\frac{i\pi}{4}+\frac{\mu}{2})}{\sinh(\frac{\theta}{2}+\frac{i\pi}{4}+\frac{\mu}{2})}=\frac{1+wx}{1-wx}
\]
where $w=e^{\mu}$ and $x=e^{\theta}$. Sometimes we also use $\bar{w}=w^{-1}$
and $\bar{x}=x^{-1}$. In the next subsection we will set $m=1$ and
use dimensionless quantities.

\subsubsection{Form factors }

In the Free boson model, we have the advantage that we can explicitly
calculate the form factors of all the operators, and then check that
they satisfy the defect form factor axioms. Additionally we can also
confirm that we have as many polynomial solutions of the axioms as
many local operators exist in the theory. 

We work with the Euclidean version of the theory ($t=iy$) and introduce
complex coordinates $z=y+ix$ , $\bar{z}=y-ix$ . We use the explicit
expressions of $\Phi_{\pm}(z,\bar{z})$ above to calculate the form
factors. The one particle form factors turn out to be: 

\[
F_{1}^{\Phi_{-}}=\langle0|\Phi_{-}(z,\bar{z})|a_{-}(\theta)\rangle=e^{zx+\bar{z}\bar{x}}
\]
\[
F_{1}^{\Phi_{+}}=\langle0|\Phi_{+}(z,\bar{z})|a_{-}(\theta)\rangle=e^{zx+\bar{z}\bar{x}}T_{-}(\theta)
\]
from which it is easy to calculate the defect form factors of the
derivative of the elementary fields:

\[
\langle0|\partial^{n}\Phi_{-}(0)|a_{-}(\theta)\rangle=x^{n}\quad;\qquad\langle0|\bar{\partial}^{n}\Phi_{-}(0)|a_{-}(\theta)\rangle=\bar{x}^{n}
\]
\[
\langle0|\partial^{n}\Phi_{+}(0)|a_{-}(\theta)\rangle=x^{n}T_{-}(\theta)\quad;\qquad\langle0|\bar{\partial}^{n}\Phi_{+}(0)|a_{-}(\theta)\rangle=\bar{x}^{n}T_{-}(\theta)
\]
 We can unify the notation by $\partial^{-n}=\bar{\partial}^{n}$.
It is instructive to see how we can recover these form factors from
the solution of the form factor axioms. Now using the parametrization
of the form factors in terms of $d(\theta)$ , we know that at level
1 the solutions of the form factor axioms have the form:

\[
F_{1}(\theta)=d(\theta).Q_{1}(\theta)
\]
Thus if we choose
\[
d(\theta)=\frac{1}{1-wx}
\]
 we obtain

\[
Q_{1}^{\partial^{n}\Phi_{-}}(\theta)=x^{n}(1-wx)\quad;\qquad Q_{1}^{\partial^{n}\Phi_{+}}(\theta)=x^{n}(1+wx)
\]
Naively it seems we have less polynomial solutions of the form factor
equations as operators: We have extra relations among the form factors
originating from 
\[
\partial\bar{\partial}\Phi_{\pm}=\Phi_{\pm}\quad;\qquad\bar{\partial}\Phi_{-}-\bar{\partial}\Phi_{+}=w(\Phi_{+}+\Phi_{-})\quad;\qquad\partial\Phi_{-}+\partial\Phi_{+}=\bar{w}(\Phi_{+}-\Phi_{-})
\]
However, these relations are satisfied due to the bulk equation of
motion and the defect conditions. Note that the form factor solutions
are even more simple in terms of $\phi=\Phi_{+}+\Phi_{-}$ and $\bar{\phi}=\Phi_{+}-\Phi_{-}$.
Actually $\bar{\phi}$ is not independent since $\bar{\phi}=w\partial\phi$.
Their form factors read as: 
\[
Q_{1}^{\partial^{n}\phi}=x^{n}\qquad;Q_{1}^{\partial^{n}\bar{\phi}}=wx^{n+1}
\]
The general multiparticle form factor as calculated from the explicit
solution of the model reads as 

\noindent \begin{center}
$\begin{array}{cc}
\langle0\vert:\partial^{n_{1}}\Phi_{-}\dots\partial^{n_{k}}\Phi_{-}\partial^{n_{k+1}}\Phi_{+}\dots\partial^{n_{N}}\Phi_{+}:\vert\theta_{1},\dots,\theta_{N}\rangle\\
=x_{1}^{n_{1}}\dots x_{N}^{n_{N}}T(x_{k+1})\dots T(x_{N})+\mbox{permutations}
\end{array}$
\par\end{center}

where $x_{i}=e^{\theta_{i}}$. In the form factor bootstrap the general
parametrization is 
\[
F_{n}(;\theta_{1},\dots,\theta_{N})=\prod_{i=1}^{N}d(\theta_{i})Q_{N}(x_{1},\dots,x_{N})
\]
 Thus we can read off the corresponding form factor solution
\[
Q_{N}=x_{1}^{n_{1}}\dots x_{N}^{n_{N}}(1-wx_{1})\dots(1-wx_{k})(1+wx_{k+1})\dots(1+wx_{N})+\mbox{permutations}
\]
Since the scattering matrix in the free boson theory is trivial $S=1$,
the form factors of different levels are not connected to each other,
and in this way we solved the theory completely. 

In terms of the field $\phi$ the form factor solutions are exactly
the same as in the bulk free bosonic theory: 
\[
Q_{N}=x_{1}^{n_{1}}\dots x_{N}^{n_{N}}+\mbox{permutations}
\]
thus we have exactly the same number of solution of the form factor
axioms as many independent local operators in the theory.

\subsection{Defect scaling Lee-Yang model}

The scaling Lee-Yang model can be defined as a perturbation of the
$\mathcal{M}_{(2,5)}$ conformal minimal model with central charge
$c=-\frac{22}{5}$. It contains two chiral representations of the
Virasoro algebra, $V_{0},V_{1}$ with highest weights $0$ and $-\frac{1}{5}$,
respectively. The fusion rules can be summarized as: $N_{0i}^{i}=N_{i0}^{i}=1$
and $N_{11}^{i}=1$ for $i=0,1$ and all others are zero. The Hilbert
space on the torus corresponds to the (diagonal) modular invariant
partition function and contains modules corresponding to the $Id$
and the $\Phi(z,\bar{z})$ primary fields with weights $(0,0)$ and
$(-\frac{1}{5},-\frac{1}{5})$:
\begin{equation}
\mathcal{H}=V_{0}\otimes\bar{V}_{0}+V_{1}\otimes\bar{V}_{1}\label{eq:bulkH}
\end{equation}
 The only relevant perturbation by the field $\Phi$ results in the
simplest scattering theory with one neutral particle of mass $m$
and scattering matrix \cite{SYLbulk} 
\[
S(\theta)=\frac{\sinh\theta+i\sin\frac{\pi}{3}}{\sinh\theta-i\sin\frac{\pi}{3}}
\]
The pole at $\theta=\frac{i\pi}{3}$ (with residue $\Gamma^{2}$)
shows that the particle can form a bound-state. The relation
\[
S(\theta+i\frac{\pi}{3})S(\theta-i\frac{\pi}{3})=S(\theta)
\]
however, implies that the bound-state is the original particle itself
and the bulk bootstrap is closed.

\subsection{Integrable defects}

Two types of topological defects can be introduced in the $\mathcal{M}_{(2,5)}$
minimal model. They can be considered as operators acting on the bulk
Hilbert space (\ref{eq:bulkH}) commuting with the action of the left
and right Virasoro generators. They have to act diagonally on each
factor in (\ref{eq:bulkH}) and satisfy a Cardy type condition. This
leads to two choices which can be labeled by the same way as the bulk
fields: $(0,0)$ and $(1,1)$. After making a modular transformation
the defect is inserted in space and the corresponding Hilbert space
can be described as 
\[
\mathcal{H}^{(a,a)}=\sum_{i,j}(V_{i}\otimes\bar{V}_{j})^{\oplus(\sum_{c\in\{0,1\}}N_{ia}^{c}N_{cj}^{a})}
\]
For the topological defect labeled by $(0,0)$ the Hilbert space turns
out to be 
\[
\mathcal{H}^{(0.0)}=V_{0}\otimes\bar{V}_{0}+V_{1}\otimes\bar{V}_{1}
\]
and coincides with the bulk Hilbert space. This defect is the trivial
(invisible) defect. 

For the other defect labeled by $(1,1)$ we obtain 
\[
\mathcal{H}^{(1,1)}=V_{0}\otimes\bar{V}_{0}+V_{1}\otimes\bar{V}_{0}+V_{0}\otimes\bar{V}_{1}+2\, V_{1}\otimes\bar{V}_{1}
\]
For each of the representation spaces we associate a primary field
$Id,\varphi(z),\bar{\varphi}(\bar{z}),\Phi_{-}(z,\bar{z})$ and $\Phi_{+}(z,\bar{z})$
with highest weights $(0,0),(-1/5,0),(0,-1/5),(-1/5,-1/5)$, respectively.
The non-chiral fields $\Phi_{\pm}(z,\bar{z})$ can be considered as
the left/right limits of the bulk field $\Phi$ on the defect. 

The bulk perturbation by $\Phi$ in the defect conformal field theory
does not break integrability. In the case of the trivial defect the
transmission factor is simply the identity $T=1.$ In the case of
the defect labeled by $(1,1)$ we can introduce a one parameter family
of defect perturbations by properly harmonizing the coefficients of
the $\varphi(z)$, $\bar{\varphi}(\bar{z})$ and $\Phi(z,\bar{z})$
terms. We plan to analyze this issue in a forthcoming publication.
Related investigations with only defect perturbations can be found
in \cite{KRW}. The resulting theory is integrable and can be solved
by exploiting how the defect acts on integrable boundaries \cite{BS}.
In the calculation the bootstrap relation 

\begin{equation}
T_{-}(\theta+\frac{i\pi}{3})T_{-}(\theta-\frac{i\pi}{3})=T_{-}(\theta)\label{eq:LYTboot}
\end{equation}
was used together with defect unitarity and defect crossing symmetry
(\ref{Defprop}) to fix the transmission factor as 
\begin{equation}
T_{-}(\theta)=[b+1][b-1]\quad;\qquad[x]=i\frac{\sinh(\frac{\theta}{2}+i\frac{\pi x}{12})}{\sinh(\frac{\theta}{2}+i\frac{\pi x}{12}-i\frac{\pi}{2})}\label{eq:YLdefsol}
\end{equation}
Actually the inverse of the solution is also a solution but the two
are related by the $b\to6+b$ transformation.

We also note that the defect with parameter $b=3$ behaves as a standing
particle both from the energy and from the scattering point of view.

\subsection{Defect form factors}

In this subsection we apply the general method developed in Section
3 to determine the form factors of the defect Lee-Yang model. The
form factor can be written as 
\[
F_{n}(\theta_{1},\dots,\theta_{n})=H_{n}\prod_{i}d(\theta_{i})\prod_{i<j}\frac{f_{\mbox{min}}(\theta_{i}-\theta_{j})}{x_{i}+x_{j}}Q_{n}(x_{1},\dots,x_{n})
\]
The minimal solution of the two particle form factor equation is well-known
reads as \cite{LY2pt}: 
\[
f_{\mbox{min}}(\theta)=\frac{x+x^{-1}-2}{x+x^{-1}+1}\, v(i\pi-\theta)\, v(-i\pi+\theta)
\]
where 
\[
v(\theta)=\exp\left\{ 2\int_{0}^{\infty}\frac{dx}{x}e^{\frac{i\theta x}{\pi}}\frac{\sinh\frac{x}{2}\sinh\frac{x}{3}\sinh\frac{x}{6}}{\sinh^{2}x}\right\} 
\]
We also included the pole corresponding to the dynamical singularity
equation by the denominator. We choose the normalization of the form
factors as in the bulk 
\[
H_{n}=-\frac{\pi m^{2}}{4\sqrt{3}}\left(\frac{3^{\frac{1}{4}}}{2^{\frac{1}{2}}v(0)}\right)^{n}
\]
$Q_{n}(x_{1},\dots,x_{n})$ is expected to be a symmetric polynomial
in $x_{i}$ and $\bar{x}_{i}$. 

Let us turn to the determination of $d(\theta)$. Due to the defect
dynamical singularity for $F_{n}(\theta_{1},\dots,\theta_{n})$ the
defect dependent term $d(\theta)$ must have a pole whenever $T_{-}(\theta)$
has a pole. Similar equation is valid for $F_{0,n}(;\theta_{1},\dots,\theta_{n})=\prod T_{-}^{-1}(\theta_{i})F_{n}(\theta_{1},\dots,\theta_{n})$
at the defect bound-states poles of $T_{+}(\theta)$ . We will take
into account that the transformation $b\leftrightarrow6-b$ exchanges
$T_{-}(\theta)$ with $T_{+}(\theta)$ and we expect that it acts
in a similar way on the form factors. The minimal solution with these
requirements turns out to be: {\small{}
\[
d(\theta)=\frac{1}{4\sinh(\frac{\theta}{2}+\frac{i\pi}{12}(b-5))\sinh(\frac{\theta}{2}+\frac{i\pi}{12}(b-7))}=\frac{1}{\sqrt{3}+2\cos(\frac{b\pi}{6}-i\theta)}=\frac{1}{\sqrt{3}+x\nu+x^{-1}\bar{\nu}};
\]
}where we introduced $\nu=e^{i\frac{\pi b}{6}}$ and $\bar{\nu}=\nu^{-1}$.
This function satisfies two relevant relations: 
\[
d(\theta+i\pi)d(\theta)=\frac{1}{1-2\cos(\frac{b\pi}{3}-2i\theta)}=\frac{1}{1-x^{2}\nu^{2}-x^{-2}\bar{\nu}^{2}}
\]
and 
\[
d(\theta+\frac{i\pi}{3})d(\theta-\frac{i\pi}{3})=\frac{1}{2\cos(\frac{b\pi}{6}-i\theta)}d(\theta)=\frac{1}{x\nu+x^{-1}\bar{\nu}}d(\theta)
\]

Singularity axioms generate recursive relations between the polynomials.
The kinematical recursion relation is given by: 
\[
Q_{n+2}(-x,x,x_{1},...,x_{n})=D_{n}(x,x_{1},...,x_{n})Q_{n}(x_{1},...,x_{n})
\]
with{\small{}
\begin{eqnarray*}
D_{n}(x,x_{1},...,x_{n}) & = & (-1)^{n+1}(x^{4}\nu^{2}-x^{2}+\nu^{-2})\\
 &  & \frac{1}{2x(\omega-\bar{\omega})}\left(\prod_{i=1}^{n}(x\omega+x_{i}\bar{\omega})(x\bar{\omega}-x_{i}\omega)-\prod_{i=1}^{n}(x\omega-x_{i}\bar{\omega})(x\bar{\omega}+x_{i}\omega)\right)
\end{eqnarray*}
} where we introduced $\omega=e^{\frac{i\pi}{3}}$, $\bar{\omega}=\omega^{-1}$,
while the the bound state recursion relation is :
\begin{eqnarray*}
Q_{n+1}(x\omega,x\bar{\omega},x_{1},...,x_{n-1}) & = & (\nu x^{2}+\bar{\nu})\prod_{i=1}^{n-1}(x+x_{i})Q_{n}(x,x_{1},...,x_{n-1})
\end{eqnarray*}
Now I proceed to solve these recursions.

\subsubsection{Solutions}

Since $Q_{n}(x_{1},...,x_{n})$ is a symmetric polynomial, it is useful
to introduce the elementary symmetric polynomials $\sigma_{k}^{(n)}(x_{1},...,x_{n})$
defined through the generating function:
\[
\prod_{i=1}^{n}(x+x_{i})=\sum_{k=0}^{n}x^{n-k}\sigma_{k}^{(n)}(x_{1},...,x_{n})
\]
By means of these functions the kinemetical recursive relation for
$Q_{n}$ reads as: {\small{}
\[
(-1)^{n+1}Q_{n+2}(-x,x,x_{1},...,x_{n})=(x^{4}\nu^{2}-x^{2}+\nu^{-2})\tilde{D}_{n}(x,x_{1},...x_{n})Q_{n}(x_{1},...,x_{n})
\]
 
\[
\hat{D}_{n}(x,x_{1},...x_{n})=\sum_{k=1}^{n}\sum_{m=1,odd}^{k}\frac{sin(\frac{2\pi}{3}m)}{sin(\frac{2\pi}{3})}x^{2(n-k)+m}\sigma_{k}^{(n)}\sigma_{k-m}^{(n)}(-1)^{k+1}
\]
}We are going to find the form factors of the operators $\Phi_{\pm}(z,\bar{z})$,
$\varphi(z)$, $\bar{\varphi}(\bar{z})$ and their descendants. We
can choose $\Phi_{\pm}$as the defect limits of the right/left bulk
fields, thus we know already all of their form factors. Taking into
account the explicit form of $d(\theta)$ together with $T_{-}(\theta)$
we find {\small{}
\[
Q_{1}^{\Phi_{-}}=\nu\sigma+\bar{\nu}\bar{\sigma}+\sqrt{3}\quad;\qquad Q_{1}^{\Phi_{+}}=\nu\sigma+\bar{\nu}\bar{\sigma}-\sqrt{3}
\]
For two particle form factors we get 
\[
Q_{2}^{\Phi_{-}}=\sigma_{1}(v^{2}\sigma_{2}+\sqrt{3}v\sigma_{1}+\sigma_{1}\bar{\sigma}_{1}+1+\sqrt{3}\bar{\nu}\bar{\sigma}_{1}+\bar{\nu}^{2}\bar{\sigma}_{2})
\]
 
\[
Q_{2}^{\Phi_{+}}=\sigma_{1}(v^{2}\sigma_{2}-\sqrt{3}v\sigma_{1}+\sigma_{1}\bar{\sigma}_{1}+1-\sqrt{3}\bar{\nu}\bar{\sigma}_{1}+\bar{\nu}^{2}\bar{\sigma}_{2})
\]
}where we used the solution of the bulk form-factor equation $Q_{2}^{\Phi}=\sigma_{1}$.
They satisfy the dynamical recursion relations. The asymptotics of
the solutions for $x\to\pm\infty$ reflect the dimensions of the fields
$(-\frac{1}{5},-\frac{1}{5})$. We would like to describe two more
chiral fields $\varphi(z)$ and $\bar{\varphi}(\bar{z})$ with dimensions
$(-\frac{1}{5},0)$ and $(0,-\frac{1}{5})$. The corresponding solutions
at level one turn out to be{\small{}
\[
Q_{1}^{\varphi}=\sigma_{1}\quad;\qquad Q_{1}^{\bar{\varphi}}=\bar{\sigma}_{1}
\]
}They are related by the $x\leftrightarrow x^{-1}$ transformation.
Using our recursion relations we find the related solutions at level
2 {\small{}
\[
Q_{2}^{\varphi}=\sigma_{1}(v\sigma_{1}^{2}+\bar{\nu})\quad;\qquad Q_{2}^{\bar{\varphi}}=\bar{\sigma}_{1}(\bar{\nu}\bar{\sigma}_{1}^{2}+v)
\]
}We summarize the FF solutions of the primary fields up to level 2
in this table:

\begin{center}
{\small{}}%
\begin{tabular}{|c|c|c|}
\hline 
{\small{}Operator} & {\small{}$Q_{1}$} & {\small{}$Q_{2}$}\tabularnewline
\hline 
\hline 
{\small{}$\Phi_{-}$} & \emph{\small{}$\nu\sigma_{1}+\bar{\nu}\bar{\sigma}_{1}+\sqrt{3}$} & {\small{}$\sigma_{1}(v^{2}\sigma_{2}+\sqrt{3}v\sigma_{1}+\sigma_{1}\bar{\sigma}_{1}+1+\sqrt{3}\bar{\nu}\bar{\sigma}_{1}+\bar{\nu}^{2}\bar{\sigma}_{2})$}\tabularnewline
\hline 
{\small{}$\Phi_{+}$} & \emph{\small{}$\nu\sigma_{1}+\bar{\nu}\bar{\sigma}_{1}-\sqrt{3}$} & {\small{}$\sigma_{1}(v^{2}\sigma_{2}-\sqrt{3}v\sigma_{1}+\sigma_{1}\bar{\sigma}_{1}+1-\sqrt{3}\bar{\nu}\bar{\sigma}_{1}+\bar{\nu}^{2}\bar{\sigma}_{2})$}\tabularnewline
\hline 
{\small{}$\varphi$} & {\small{}$\sigma_{1}$} & {\small{}$\sigma_{1}(v\sigma_{1}^{2}+\bar{\nu})$}\tabularnewline
\hline 
{\small{}$\bar{\varphi}$} & {\small{}$\bar{\sigma}_{1}$} & {\small{}$\bar{\sigma}_{1}(\bar{\nu}\bar{\sigma}_{1}^{2}+v)$}\tabularnewline
\hline 
\end{tabular}
\par\end{center}{\small \par}

Together with the Identity operator, we have 5 independent operators
in this model, and with their descendants they describe the full spectrum
of the solutions. However, the one-to-one matching remains an open
problem due to the large number of operators and their descendants
in this model.

We also list the first and second order descendant operators and their
corresponding solutions at levels 1 and 2 below:
\begin{itemize}
\item First Order Descendants:\vspace{-0.9cm}

\end{itemize}
\noindent \begin{center}
\begin{tabular}{|c|c|c|}
\hline 
{\small{}Operator} & {\small{}$Q_{1}$} & {\small{}$Q_{2}$}\tabularnewline
\hline 
\hline 
$\partial\phi_{+}$ & $\sigma_{1}(v\sigma_{1}+\bar{\nu}\bar{\sigma}_{1}-\sqrt{3})$ & $\sigma_{1}^{2}(v^{2}\sigma_{2}-\sqrt{3}v\sigma_{1}+\sigma_{1}\bar{\sigma}_{1}+1-\sqrt{3}v^{-1}\bar{\sigma}_{1}+v^{-2}\bar{\sigma}_{2})$\tabularnewline
\hline 
$\bar{\partial}\phi_{+}$ & $\bar{\sigma}_{1}(v\sigma_{1}+\bar{\nu}\bar{\sigma}_{1}-\sqrt{3})$ & $\bar{\sigma}_{1}\sigma_{1}(v^{2}\sigma_{2}-\sqrt{3}v\sigma_{1}+\sigma_{1}\bar{\sigma}_{1}+1-\sqrt{3}v^{-1}\bar{\sigma}_{1}+v^{-2}\bar{\sigma}_{2})$\tabularnewline
\hline 
$\partial\phi_{-}$ & $\sigma_{1}(v\sigma_{1}+\bar{\nu}\bar{\sigma_{1}}+\sqrt{3})$ & $\sigma_{1}^{2}(v^{2}\sigma_{2}+\sqrt{3}v\sigma_{1}+\sigma_{1}\bar{\sigma}_{1}+1+\sqrt{3}v^{-1}\bar{\sigma}_{1}+v^{-2}\bar{\sigma}_{2})$\tabularnewline
\hline 
$\bar{\partial}\phi_{-}$ & $\bar{\sigma}_{1}(v\sigma_{1}+\bar{\nu}\bar{\sigma}_{1}+\sqrt{3})$ & $\bar{\sigma}_{1}\sigma_{1}(v^{2}\sigma_{2}+\sqrt{3}v\sigma_{1}+\sigma_{1}\bar{\sigma}_{1}+1+\sqrt{3}v^{-1}\bar{\sigma}_{1}+v^{-2}\bar{\sigma}_{2})$\tabularnewline
\hline 
$\partial\varphi$ & $\sigma_{1}^{2}$ & $\sigma_{1}^{2}(v\sigma_{1}^{2}+v^{-1})$\tabularnewline
\hline 
$\bar{\partial}\bar{\varphi}$ & $\bar{\sigma}_{1}^{2}$ & $\sigma_{1}\bar{\sigma}_{1}(v^{-1}\bar{\sigma}_{1}^{2}+v)$\tabularnewline
\hline 
\end{tabular}\vspace{0.8cm}

\par\end{center}
\begin{itemize}
\item Second Order Descendants:\vspace{-0.9cm}

\end{itemize}
\noindent \begin{center}
{\footnotesize{}}%
\begin{tabular}{|c|c|c|}
\hline 
{\footnotesize{}Operator } & {\footnotesize{}$Q_{1}$} & {\footnotesize{}$Q_{2}$}\tabularnewline
\hline 
\hline 
{\footnotesize{}$\partial^{2}\phi_{+}$} & {\footnotesize{}$\sigma_{1}^{2}(v\sigma_{1}+v^{-1}\bar{\sigma}_{1}-\sqrt{3})$} & {\footnotesize{}$\sigma_{1}^{3}(v^{2}\sigma_{2}-\sqrt{3}v\sigma_{1}+\sigma_{1}\bar{\sigma}_{1}+1-\sqrt{3}v^{-1}\bar{\sigma}_{1}+v^{-2}\bar{\sigma}_{2})$}\tabularnewline
\hline 
{\footnotesize{}$\partial\bar{\partial}\phi_{+}$} & {\footnotesize{}$\sigma_{1}\bar{\sigma}_{1}(v\sigma_{1}+v^{-1}\bar{\sigma}_{1}-\sqrt{3})$} & {\footnotesize{}$\bar{\sigma}_{1}\sigma_{1}^{2}(v^{2}\sigma_{2}-\sqrt{3}v\sigma_{1}+\sigma_{1}\bar{\sigma}_{1}+1-\sqrt{3}v^{-1}\bar{\sigma}_{1}+v^{-2}\bar{\sigma}_{2})$}\tabularnewline
\hline 
{\footnotesize{}$\bar{\partial}^{2}\phi_{+}$} & {\footnotesize{}$\bar{\sigma}_{1}^{2}(v\sigma_{1}+v^{-1}\bar{\sigma}_{1}-\sqrt{3})$} & {\footnotesize{}$\bar{\sigma}_{1}^{2}\sigma_{1}(v^{2}\sigma_{2}-\sqrt{3}v\sigma_{1}+\sigma_{1}\bar{\sigma}_{1}+1-\sqrt{3}v^{-1}\bar{\sigma}_{1}+v^{-2}\bar{\sigma}_{2})$}\tabularnewline
\hline 
{\footnotesize{}$\partial^{2}\phi_{-}$} & {\footnotesize{}$\sigma_{1}^{2}(v\sigma_{1}+v^{-1}\bar{\sigma}_{1}+\sqrt{3})$} & {\footnotesize{}$\sigma_{1}^{3}(v^{2}\sigma_{2}+\sqrt{3}v\sigma_{1}+\sigma_{1}\bar{\sigma}_{1}+1+\sqrt{3}v^{-1}\bar{\sigma}_{1}+v^{-2}\bar{\sigma}_{2})$}\tabularnewline
\hline 
{\footnotesize{}$\partial\bar{\partial}\phi_{-}$} & {\footnotesize{}$\sigma_{1}\bar{\sigma}_{1}(v\sigma_{1}+v^{-1}\bar{\sigma}_{1}+\sqrt{3})$} & {\footnotesize{}$\bar{\sigma}_{1}\sigma_{1}^{2}(v^{2}\sigma_{2}+\sqrt{3}v\sigma_{1}+\sigma_{1}\bar{\sigma}_{1}+1+\sqrt{3}v^{-1}\bar{\sigma}_{1}+v^{-2}\bar{\sigma}_{2})$}\tabularnewline
\hline 
{\footnotesize{}$\bar{\partial}^{2}\phi_{-}$} & {\footnotesize{}$\sigma_{1}^{2}(v\sigma_{1}+v^{-1}\bar{\sigma}_{1}+\sqrt{3})$} & {\footnotesize{}$\bar{\sigma}_{1}^{2}\sigma_{1}(v^{2}\sigma_{2}+\sqrt{3}v\sigma_{1}+\sigma_{1}\bar{\sigma}_{1}+1+\sqrt{3}v^{-1}\bar{\sigma}_{1}+v^{-2}\bar{\sigma}_{2})$}\tabularnewline
\hline 
{\footnotesize{}$\partial^{2}\varphi$} & {\footnotesize{}$\sigma_{1}^{3}$} & {\footnotesize{}$\sigma_{1}^{3}(v\sigma_{1}^{2}+v^{-1})$}\tabularnewline
\hline 
{\footnotesize{}$\bar{\partial}^{2}\bar{\varphi}$} & {\footnotesize{}$\bar{\sigma}_{1}^{3}$} & {\footnotesize{}$\sigma_{1}\bar{\sigma}_{1}^{2}(v^{-1}\bar{\sigma}_{1}^{2}+v)$}\tabularnewline
\hline 
{\footnotesize{}$T$} & {\footnotesize{}$0$} & {\footnotesize{}$\sigma_{1}(\sigma_{1}^{2}+\sigma_{2})(v^{2}\sigma_{2}+\sqrt{3}v\sigma_{1}+\sigma_{1}\bar{\sigma}_{1}+1+\sqrt{3}v^{-1}\bar{\sigma}_{1}+v^{-2}\bar{\sigma}_{2})$}\tabularnewline
\hline 
\end{tabular}
\par\end{center}{\footnotesize \par}

\vspace{1cm}

\subsubsection{Parity Symmetry}

In this part we analyze how the parity transformation acts on the
form factor solutions. The action of the parity operator $P$ on the
operators can be written as 
\[
P\mathcal{O}P^{-1}=\mathcal{O}^{P}
\]
The action on the form factors is 
\[
P\langle0|\mathcal{O}(0)|\theta;\ \rangle=\langle0|\mathcal{O}^{P}|\ ;-\theta\rangle
\]

Checking for $\Phi_{-}$and $\Phi_{+}$up to level 2, we find that
they are parity even with 
\[
P\langle0|\Phi_{-}(0)|\theta;\ \rangle=\langle0|\Phi_{+}(0)|\ ;-\theta\rangle
\]
 while on the contrary, $\varphi$ and $\bar{\varphi}$ are parity
odd with

\[
P\langle0|\varphi(0)|\theta;\ \rangle=-\langle0|\bar{\varphi}(0)|\ ;-\theta\rangle
\]
To confirm these parity properties of the primary fields one should
work out the defect Lee-Yang conformal field theory.

\section{Boundary form factors via defects }

In this section we intend to illustrate how defects can be used to
generate new boundary form factor solutions from old ones. The underlying
fusing idea for the reflection matrices can be explained as follows:
Suppose we place an integrable defect with transmission factor $T_{-}(\theta)$
in front of an integrable boundary with reflection factor $R(\theta)$,
which satisfies unitarity and boundary crossing unitarity: 
\[
R(-\theta)=R^{-1}(\theta)\quad;\qquad R(\frac{i\pi}{2}-\theta)=S(2\theta)R(\frac{i\pi}{2}+\theta)
\]
If we fuse the defect to the boundary the composite boundary system
will be integrable and will have reflection factor 
\[
\bar{R}(\theta)=T_{+}(\theta)R(\theta)T_{-}(\theta)
\]
which, due to the defect unitarity and crossing equations, will satisfy
boundary unitarity and crossing unitarity. This idea has been used
to calculate the transmission factors from the already determined
reflection factors $R,\bar{R}$ in the sinh-Gordon and Lee-Yang models
in \cite{BS}. In contrast, here we would like to use the fusion idea
to generate new form factor solutions from old ones. For this purpose
we suppose that we determined already the boundary form factors $F_{n}^{\mathcal{O}}(\theta_{1},\dots,\theta_{n})$
of a boundary operator $\mathcal{O}$. It satisfies besides some singularity
axioms the following requirements: permutation \emph{
\[
F_{n}^{\mathcal{O}}(\theta_{1},\dots,\theta_{i},\theta_{i+1},\dots,\theta_{n})=S(\theta_{i}-\theta_{i+1})F_{n}^{\mathcal{O}}(\theta_{1},\dots,\theta_{i+1},\theta_{i},\dots,\theta_{n})
\]
}reflection
\[
F_{n}^{\mathcal{O}}(\theta_{1},\dots,\theta_{n-1},\theta_{n})=R(\theta_{n})F_{n}^{\mathcal{O}}(\theta_{1},\dots,\theta_{n-1},-\theta_{n})
\]
and crossing reflection 
\[
F_{n}^{\mathcal{O}}(\theta_{1},\theta_{2},\dots,\theta_{n})=R(i\pi-\theta_{1})F_{n}^{\mathcal{O}}(2i\pi-\theta_{1},\theta_{2},\dots,\theta_{n})
\]
We claim that the fused form factor 
\begin{equation}
\bar{F}_{n}^{\mathcal{O}}(\theta_{1},\dots,\theta_{n})=\prod_{i=1}^{n}T_{-}(\theta_{i})F_{n}^{\mathcal{O}}(\theta_{1},\dots,\theta_{n})\label{eq:barF}
\end{equation}
 satisfies the boundary form factor axioms of the fused boundary corresponding
to the reflection factor $\bar{R}$. Let us analyze them one by one.
Since the extra factor is symmetric in $\theta_{i}$ the permutation
axiom is trivially satisfied. To show the reflection property we use
defect unitarity 
\begin{equation}
\bar{R}(\theta)=T_{+}(\theta)R(\theta)T_{-}(\theta)=T_{-}(-\theta)^{-1}R(\theta)T_{-}(\theta)\label{eq:Funi}
\end{equation}
while for the crossing reflection we use defect crossing symmetry:
\begin{equation}
\bar{R}(i\pi-\theta)=T_{+}(i\pi-\theta)R(i\pi-\theta)T_{-}(i\pi-\theta)=T_{-}(\theta)R(i\pi-\theta)T_{-}(2i\pi-\theta)^{-1}\label{eq:Fcuni}
\end{equation}
Now multiplying both sides of the reflection and crossing reflection
equation by $\prod_{i}T_{-}(\theta_{i})$ and using (\ref{eq:Funi})
and (\ref{eq:Fcuni}) the claim follows. Similarly one can show the
satisfaction of the singularity axioms. 

By this method form factor solution of a given boundary can be used
to generate form factor solutions for the fused boundary. It is practically
useful if we can follow the identification of the operators under
the fusion procedure. This is the case for example if the operator
in the UV limit commutes with the defect. Say for example if in the
Lee-Yang model we determine the form factors of the operators of the
identity module on the trivial boundary, then by the fusion procedure
we can generate the form factors of the same module on the fused $\phi$
boundary.\pagebreak{}

\chapter{Spectrum of Lee-Yang model in finite volume}

\noindent We consider the non-unitary Yang-Lee minimal model ${\cal M}(2,5).$
It is obtained~\cite{Riggs89} as the continuum scaling limit of
the $A_{4}$ lattice model of Forrester-Baxter~\cite{FB85} in Regime~III
with crossing parameter $\lambda=\frac{3\pi}{5}$. We consider it
in three different finite geometries: on the strip with integrable
boundary conditions labeled by the Kac labels $(r,s)=(1,1),(1,2)$,
on cylindrical geometry with either periodic boundary condition or
by including an integrable purely transmitting defect. We then apply
$\varphi_{1,3}$ integrable perturbations both on the strip's boundary
and the defect and describe the flow of the spectrum. Introducing
an additionally $\Phi_{1,3}$ integrable perturbation in the bulk
we can go off-critical and determine the finite size spectrum of the
massive scattering theory in the three geometries, via thermodynamic
Bethe ansatz (TBA) equations. We derive these equations for all excitations
by solving, in the continuum scaling limit, the TBA functional equation
satisfied by the transfer matrices of the associated $A_{4}$ lattice
model of Forrester and Baxter in Regime~III. The excitations are
classified in terms of simple $(m,n)$ systems. The excited state
TBA equations agree with the previously conjectured equations in the
strip and periodic cylindrical geometries, giving novel equations
for the defect case, and confirming also the previously conjectured
transmission factors.

\section{Lee-Yang Lattice Model and Transfer Matrices}

The Lee-Yang lattice model is defined on a square lattice with spins
or heights $a=1,2,3,4$ restricted so that nearest neighbor heights
differ by $\pm1$. The spins thus live on the $A_{4}$ Dynkin diagram.
It is helpful to regard the Lee-Yang model as a special case of the
$A_{L}$ Forrester-Baxter~\cite{FB85} model.

\subsection{Lee-Yang lattice model as the $A_{4}$ BF model}

The $A_{L}$ Forrester-Baxter~\cite{FB85} models with spins $a=1,\ldots,L$
are defined by the Boltzmann weights 
\begin{eqnarray}
W\left(\left.\begin{array}{cc}
a\pm1 & a\\
a & a\mp1
\end{array}\right|u\right) & = & \frac{s(\lambda-u)}{s(\lambda)}\\
W\left(\left.\begin{array}{cc}
a & a\pm1\\
a\mp1 & a
\end{array}\right|u\right) & = & \frac{g_{a\mp1}}{g_{a\pm1}}\left(\frac{s((a\pm1)\lambda)}{s(a\lambda)}\right)\,\frac{s(u)}{s(\lambda)}\\
W\left(\left.\begin{array}{cc}
a & a\pm1\\
a\pm1 & a
\end{array}\right|u\right) & = & \frac{s(a\lambda\pm u)}{s(a\lambda)}.
\end{eqnarray}
 Here we differentiate between the massive and the massless cases
as follows:
\begin{itemize}
\item $s(u)=\sin u$ , the trigonometric sine function for the massless
model
\item $s(u)=\vartheta_{1}(u,q)$ the standard elliptic theta function~for
the massive model where
\end{itemize}
\begin{eqnarray}
\vartheta_{1}(u,q)=2q^{1/4}\sin u\prod_{n=1}^{\infty}(1-2q^{2n}\cos2u+q^{4n})(1-q^{2n})
\end{eqnarray}
 where $u$ is the spectral parameter and the elliptic nome $q$ is
a temperature-like variable corresponding to the $\varphi_{1,3}$
integrable bulk perturbation. 

In our massive calculations for the Lee-Yang model we have found that
$t^{\nu}=\frac{mL}{4\sqrt{3}}\frac{1}{N}$ where 
\[
\mu=\frac{mL}{4\sqrt{3}}=\lim_{N\to\infty,\, q\to0}Nt^{\nu}
\]
 or more precisely 
\[
L=\lim_{N\to\infty,\, a\to0}Na,\qquad m=\lim_{t^{\nu}\to0,\, a\to0}\frac{4\sqrt{3}t^{\nu}}{a}
\]
 where $a$ is the lattice spacing, $m$ is a mass, $L$ is the continuum
length scale, and $t=q^{2}$ is the deviation from critical temperature
variable, and $v=\frac{5}{12}$.

The crossing parameter $\lambda$ is 
\begin{equation}
\lambda=\frac{(p'-p)\pi}{p'}\label{crossing}
\end{equation}
 where $p<p'$. Integrability derives from the fact that these local
face weights satisfy the Yang-Baxter equation.

Here we will only consider the Lee-Yang model with 
\[
p'=5,\quad p=2,\quad\lambda=\frac{3\pi}{5}
\]
 The gauge factors $g_{a}$ are arbitrary but we will take $g_{a}=\left(-1\right)^{a/2}$,
so that $\frac{g_{a\mp1}}{g_{a_{\pm1}}}=-1$. With this choice the
face weights are only symmetric under reflections about one of the
two diagonals. Since the Lee-Yang model is non-unitary ($p'\ne p+1$),
some Boltzmann weights are negative.

The critical theory correspond to $r=0$ and then the function $s(u)$
degenerates simply to $s(u)=\sin(u)$.

\subsection{Transfer matrices and functional relations}

The transfer matrices $T_{1}(u)$ can be built up from the local face
weights. As the local face weights satisfy the Yang-Baxter equations
the transfer matrices will form commuting families $[T_{j}(u),T_{j'}(u')]=0$
from which integrable Hamiltonians can be derived. Due to the underlying
symmetry of the model the transfer matrices will satisfy the functional
relations
\begin{equation}
T_{1}(u)T_{1}(u+\lambda)=\mathbb{I}+T_{2}(u+3\lambda)\label{eq:funcrel}
\end{equation}
where in our case $T_{2}(u)$ is spectrally equivalent to $T_{1}(u)$.
Let us see how this functional relations are realized in the various
circumstances.

\subsubsection{Periodic boundary condition}

The transfer matrix with periodical boundary condition can be defined
on a lattice of $2N$ sites from the local face weights as follows:{\small{}
\[
T_{j}(u)_{\mathbf{a}}^{\mathbf{b}}=W_{j}\left(\left.\begin{array}{cc}
b_{1} & b_{2}\\
a_{1} & a_{2}
\end{array}\right|u\right)W_{j}\left(\left.\begin{array}{cc}
b_{2} & b_{3}\\
a_{2} & a_{3}
\end{array}\right|u\right)\dots W_{j}\left(\left.\begin{array}{cc}
b_{2N-1} & b_{2N}\\
a_{2N-1} & a_{2N}
\end{array}\right|u\right)W_{j}\left(\left.\begin{array}{cc}
b_{2N} & b_{1}\\
a_{2N} & a_{1}
\end{array}\right|u\right)
\]
}and $W=W_{1}$. This transfer matrix is a first in a series of transfer
matrices obtained by fused weights. The transfer matrix $T_{2}(u)$,
for instance, is defined from the fused face weight $W_{2}:$
\[
W_{2}\left(\left.\begin{array}{cc}
d & c\\
a & b
\end{array}\right|u\right)=\frac{s(\lambda)}{s(u)}\sum_{e}W\left(\left.\begin{array}{cc}
e & f\\
a & b
\end{array}\right|u\right)W\left(\left.\begin{array}{cc}
d & c\\
e & f
\end{array}\right|u+\lambda\right)
\]
being non-vanishing only when $|a-d|=\{0,2\}$ and $a+d=\{4,6\}$.
It is not hard to see that the definition is good, i.e. the fused
weights does not depend on $f$. These transfer matrices form a simple
fusion hierarchy:
\[
T_{1}(u)T_{1}(u+\lambda)=\left(\frac{s(u+\lambda)s(\lambda-u)}{s^{2}(\lambda)}\right)^{N}\mathbb{I}+\left(\frac{s(u)}{s(\lambda)}\right)^{N}T_{2}(u)
\]

Here $N$ is taken to be even. The Lee-Yang theory is the simplest
theory as in this case the fused weights $W_{2}$ are trivially related
to $W_{1}$ modulo some $u$-independent gauge factors: 
\[
W_{2}\left(\left.\begin{array}{cc}
d & c\\
a & b
\end{array}\right|u\right)\equiv W\left(\left.\begin{array}{cc}
d & c\\
5-a & 5-b
\end{array}\right|u+3\lambda\right)
\]
This gauge transformation can be dropped at the level of the transfer
matrix, thus by introducing the height reversal matrix $Y$ we conclude
that $T_{1}(u)\equiv YT_{2}(u)=T_{2}(u)Y$. If we renormalize the
transfer matrix as 
\[
T(u)=\left(\frac{s(\lambda)s(u+2\lambda)}{s(u+\lambda)s(u+3\lambda)}\right)^{N}T_{1}(u)
\]
then $T(u)$ will satisfy the functional relation
\[
T(u)T(u+\lambda)=\mathbb{I}+T(u+3\lambda)Y
\]
As the height reversal matrix $Y$ commutes with $T$ we can diagonalize
it in the same basis. Since $Y^{2}=1$ the eigenvalues are $Y=\pm1$.
Restricting the analysis to the $Y=+1$ eigenspace the eigenvalues
of the transfer matrix $t(u)$ will satisfy the relation
\[
t(u)t(u+\lambda)=1+t(u+3\lambda)
\]
Clearly for any eigenvalue $-t(u)$ is an eigenvalue, too. 

The transfer matrix also satisfies the crossing relation: 
\[
T(\lambda-u)=T(u)
\]

and periodicity

\[
T(u+\pi)=T(u)
\]

\subsubsection{Periodic boundary condition with a seam}

The transfer matrix with periodical boundary condition with a seam
of parameter $\xi$ can be defined on a lattice of $N-1$ (with even
$N$) sites from the local face weights as follows:{\footnotesize{}
\[
T_{j}(u)_{\mathbf{a}}^{\mathbf{b}}=W_{j}\left(\left.\begin{array}{cc}
b_{1} & b_{2}\\
a_{1} & a_{2}
\end{array}\right|u+\xi\right)W_{j}\left(\left.\begin{array}{cc}
b_{2} & b_{3}\\
a_{2} & a_{3}
\end{array}\right|u\right)\dots W_{j}\left(\left.\begin{array}{cc}
b_{2N-1} & b_{2N}\\
a_{2N-1} & a_{2N}
\end{array}\right|u\right)W_{j}\left(\left.\begin{array}{cc}
b_{2N} & b_{1}\\
a_{2N} & a_{1}
\end{array}\right|u\right)
\]
}where $n_{seam}(u,\xi)=\left(\frac{s(\lambda)s(u+\xi+2\lambda)}{s(u+\xi+\lambda)s(u+\xi+3\lambda)}\right)\left(\frac{s(\lambda)s(u+2\lambda)}{s(u+\lambda)s(u+3\lambda)}\right)^{N-1}$
is a normaliztion scalar factor which ensures that the transfer matrices
satisfies

\[
T_{1}(u)T_{1}(u+\lambda)=1+T_{2}(u+3\lambda)
\]
where $T_{2}(u)$ is defined from the fused face weights and fused
seam. Just as in the periodical case we have $T_{1}(u)\equiv T_{2}(u)$.
We note that the $\xi\to0$ limit reproduces the periodic result.

\subsubsection{Boundary case: double row transfer matrices}

To ensure integrability of Lee-Yang lattice model in the presence
of a boundary~\cite{BPO96} we need commuting double row transfer
matrices and triangle boundary weights which satisfy the boundary
Yang-Baxter equations. The integrable boundary conditions are labeled
by the Kac labels $(r,s)=(1,s)$ with $s=1,2$. For $s=1$, the non-zero
left and right triangle weights are given by 
\begin{eqnarray}
K_{L}\left(\left.\begin{array}{c}
1\\
1
\end{array}2\right|u\right)=\frac{s(2\lambda)}{s(\lambda)},\qquad\qquad\qquad\quad K_{R}\left(\left.2\begin{array}{c}
1\\
1
\end{array}\right|u\right)=1
\end{eqnarray}
 For $s=2$, the non-zero right boundary weights can be obtained by
placing a seam in front of the $s=1$ boundary {\small{}
\begin{eqnarray}
K_{R}\left(\left.2\begin{array}{c}
1\\
1
\end{array}\right|u,\xi\right) & = & W\left(\left.\begin{array}{cc}
a & 2\\
2 & 1
\end{array}\right|u+\xi\right)W\left(\left.\begin{array}{cc}
2 & 1\\
a & 2
\end{array}\right|u+\xi\right)K_{R}\left(\left.2\begin{array}{c}
1\\
1
\end{array}\right|u\right)\\[6pt]
 & = & \frac{s(u+\xi+(2-a)\lambda)s(u-\xi+(2-a)\lambda)}{s(\lambda)s(2\lambda)},\quad a=1,3\nonumber 
\end{eqnarray}
} The parameter $\xi$ is arbitrary and can be complex. However, to
obtain conformal boundary conditions at the isotropic point $u=\lambda/2$,
we choose $\Re e(\xi)=0$. Integrability in the presence of these
boundaries derives from the fact that these boundary weights satisfy
the left and right boundary Yang-Baxter equations respectively.

The face and triangle boundary weights are used to construct~\cite{BPO96}
a family of commuting double row transfer matrices $D(u)$. For a
lattice of width $N$, the entries of $D(u)$ are given diagrammatically
by {\footnotesize{}
\begin{eqnarray*}
D(u)_{\mathbf{a},\mathbf{b}} & = & \sum_{c_{0},\dots,c_{N}}W\left(\left.\begin{array}{cc}
c_{0} & c_{1}\\
1 & a_{1}
\end{array}\right|u\right)W\left(\left.\begin{array}{cc}
c_{1} & c_{2}\\
a_{1} & a_{2}
\end{array}\right|u\right)\dots W\left(\left.\begin{array}{cc}
c_{N-1} & s\\
a_{N-1} & s
\end{array}\right|u\right)K_{R}\left(\left.c_{N}\begin{array}{c}
s\\
s
\end{array}\right|u\right)\times\\
 &  & K_{L}\left(\left.\begin{array}{c}
1\\
1
\end{array}2\right|\lambda-u\right)W\left(\left.\begin{array}{cc}
1 & b_{1}\\
c_{0} & c_{1}
\end{array}\right|\lambda-u\right)W\left(\left.\begin{array}{cc}
b_{1} & b_{2}\\
c_{1} & c_{2}
\end{array}\right|\lambda-u\right)\dots W\left(\left.\begin{array}{cc}
b_{N-1} & s\\
c_{N-1} & s
\end{array}\right|\lambda-u\right)
\end{eqnarray*}
} This transfer matrix is positive definite and satisfies crossing
symmetry $D(u)=D(\lambda-u)$. Although $D(u)$ is not symmetric or
normal, this one-parameter family of transfer matrices can be diagonalized
because $\tilde{D}(u)=GD(u)=\tilde{D}(u)^{T}$ is symmetric where
the diagonal gauge matrix $G$ is given by 
\begin{eqnarray}
G_{\mathbf{a},\mathbf{b}}=\prod_{j=1}^{N-1}G(a_{j},a_{j+1})\,\delta(a_{j},b_{j})
\end{eqnarray}
 with 
\begin{eqnarray*}
G(a,b)=\begin{cases}
s(\lambda)/s(2\lambda), & b=1,4\\
1, & \mbox{otherwise}
\end{cases}
\end{eqnarray*}
It is convenient to define the normalized transfer matrix 
\begin{equation}
T(u)=S_{s}(u)\,\frac{s^{2}(2u-\lambda)}{s(2u+\lambda)s(2u-3\lambda)}\Bigl[\frac{s(\lambda)s(u+2\lambda)}{s(u+\lambda)s(u+3\lambda)}\Bigr]^{N}D(u)
\end{equation}
 with 
\begin{eqnarray}
S_{s}(u) & = & \begin{cases}
1, & s=1\\
{\displaystyle \frac{s^{2}(\lambda)s(u+\xi+2\lambda)s(u-\xi+2\lambda)}{s(u+\xi+\lambda)s(u-\xi+\lambda)s(u+\xi+3\lambda)s(u-\xi+3\lambda)},} & s=2
\end{cases}
\end{eqnarray}
 It can then be shown~\cite{BPO96} that the normalized transfer
matrix satisfies the universal TBA functional equation 
\begin{equation}
T(u)T(u+\lambda)=1+T(u+3\lambda)\label{e:functional}
\end{equation}
 independent of the boundary conditions. Apart from the change in
the value of the crossing parameter $\lambda$, it is also the same
functional equation that holds for the tricritical hard square and
hard hexagon models~\cite{BaxP82}. However, this change in the crossing
parameter drastically changes the relevant analytical properties.
Since the transfer matrices commute this functional equation also
holds for each eigenvalue $t(u)$ of $T(u)$.

\section{Classification of Excited States}

Here we start the classification of states in the critical case when
$s(u)=\sin(u)$. We will make correspondence with the conformal Lee-Yang
model thus we recall first the description of their Virasoro modules.
The Virasoro algebra contains two relevant modules for $c=-\frac{22}{5}$. 

The identity module, is built over the vacuum $\vert0\rangle$ 
\[
L_{-n_{1}}\dots L_{-n_{m}}\vert0\rangle\quad;\qquad n_{m}>1\qquad;\qquad n_{i}>n_{i+1}+1
\]
Interestingly this basis is linearly independent (no singular vectors).
The representation has the reduced character 
\[
\chi_{0}(q)=\sum_{n=1}^{\infty}\mbox{dim}(V_{n}^{0})q^{n}=\sum_{n=1}^{\infty}\frac{q^{n^{2}+n}}{(1-q)\dots(1-q^{n})}=\prod_{n=1}^{\infty}\frac{1}{(1-q^{5n-3})(1-q^{5n-2})}
\]
where $n$ can be considered as the energy ($L_{0}$ eigenvalue) of
the given state. The sum and the product form is related by the Andrews-Gordon
identity, which is the generalization of the Rogers-Ramanujan identities. 

The other appearing module is built over the highest weight state
$\vert h\rangle$ where $h=-\frac{1}{5}$. The module is generated
by the linearly independent modes 
\[
L_{-n_{1}}\dots L_{-n_{m}}\vert h\rangle\quad;\qquad n_{m}>0\qquad;\qquad n_{i}>n_{i+1}+1
\]
 and has the reduced character: 

\[
\chi_{1}(q)=\sum_{n=1}^{\infty}\mbox{dim}(V_{n}^{1})q^{n}=\sum_{n=1}^{\infty}\frac{q^{n^{2}}}{(1-q)\dots(1-q^{n})}=\prod_{n=1}^{\infty}\frac{1}{(1-q^{5n-4})(1-q^{5n-1})}
\]

\subsection{$(m,n)$ systems, zeros, paths and characters}

The Hilbert space of the lattice model consist of paths. By diagonalizing
the various transfer matrices we can characterize a given eigenvector
by the distribution of the zeros of the transfer matrix on the complex
plane. We will make correspondence between the paths, the distribution
of zeros and the Virasoro descendants in the three cases. We start
with the simplest boundary case as the Hilbert space contains one
single Virasoro module only. Then we turn to the analysis of the periodic
case with and without the seam, where tensor products of Virasoro
modules appear. We also analyze the flows between the Hilbert spaces
induced by $\xi$ going from $0$ to $\infty$.

\subsubsection{Boundary case}

\subsubsection*{$(m,n)$ system, characterization of the eigenvectors by zeros of
the transfer matrix}

Let us consider the sectors with boundary conditions $(r,s)=(1,1),(1,2)$
which we often label simply by $s=1,2$. The excitation energies are
given by the scaling limit of the eigenvalues of the double-row transfer
matrix $D(u)$, or equivalently the normalized transfer matrix $T(u)$,
where $u$ is the spectral parameter. The single relevant analytical
strip in the complex $u$-plane is the full periodicity strip 
\begin{equation}
-\frac{\pi}{5}<\Re e(u)<\frac{4\pi}{5}
\end{equation}
 Additionally, the transfer matrix is symmetric for the real line
thus it is enough to analyze the analytical structure on the upper
half plane. The excitations are classified by the string content in
this analytical strip. There are four kinds of strings which we call
``$1$-strings\textquotedbl{}, ``short $2$-strings\textquotedbl{},
``long $2$-strings\textquotedbl{} and ``real $2$-strings\textquotedbl{}.
See the next figure for two typical configurations in the two sectors. 

\begin{figure}[H]
\begin{centering}
\includegraphics[height=7cm]{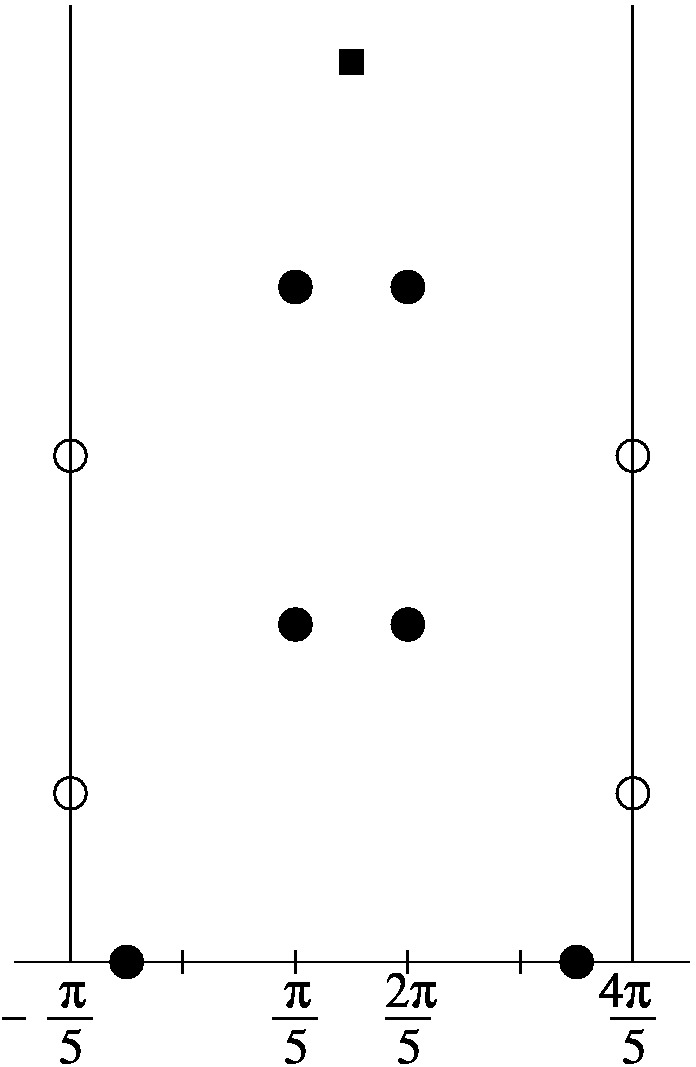}~~~~~~~\includegraphics[height=7cm]{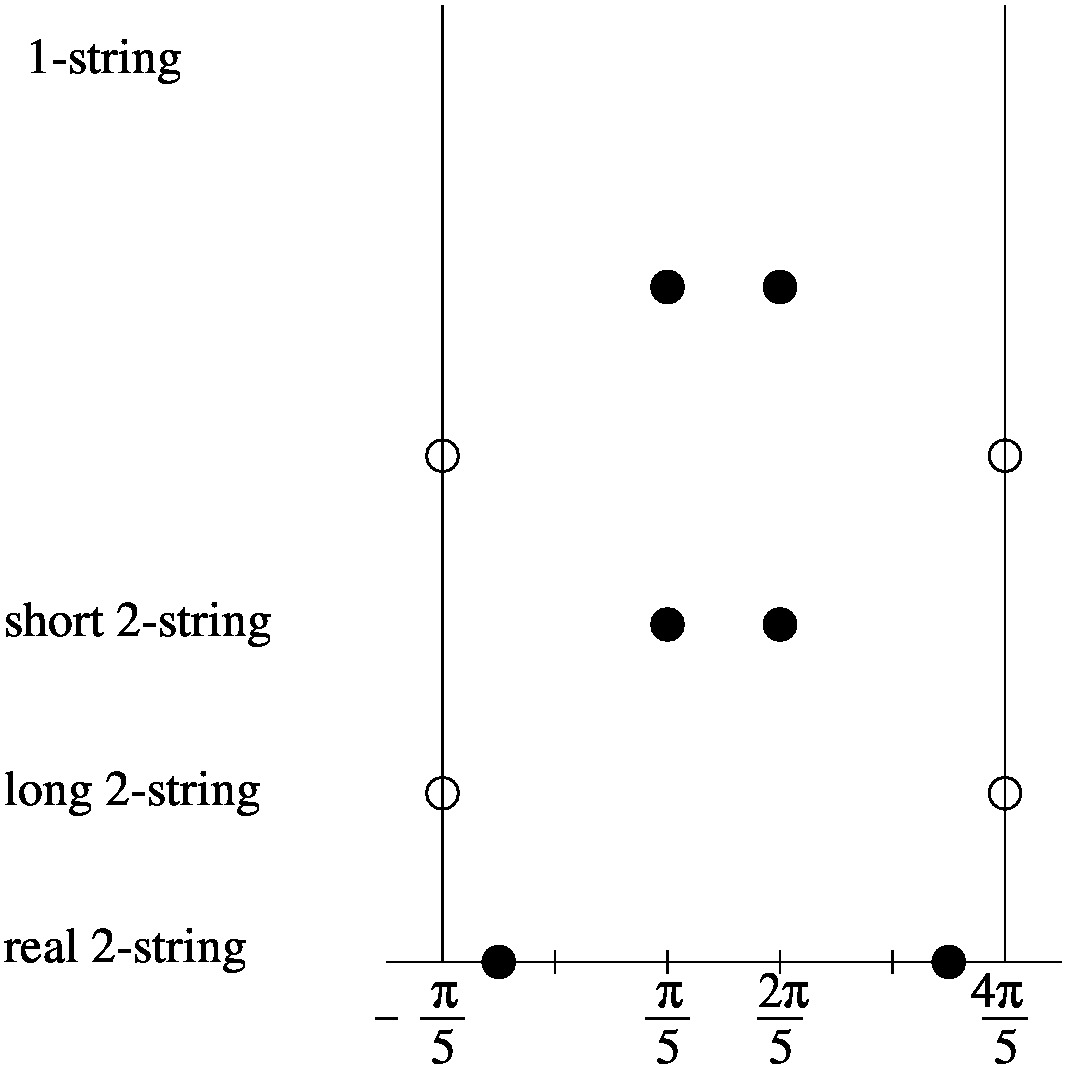}
\par\end{centering}

\protect\caption{Typical configuration of zeros of the transfer matrix eigenvalue.
The boundary condition $(1,1)$ is on the left, while $(1,2)$ is
on the right.}
\end{figure}

The $1$-string $u_{j}=3\pi/10+iv_{j}$ lies in the middle of the
analytical strip and has real part $3\pi/10$ and exist in the $(1,1)$
sector only. The two zeros of a short $2$-string $u_{j}=\pi/5+iw_{j}$,
$2\pi/5+iw_{j}$ have common imaginary parts and real parts $\pi/5$,
$2\pi/5$ respectively. The two zeros of a long $2$-string $u_{j}=-\pi/5+iw_{j}$,
$4\pi/5+iw_{j}$ have common imaginary parts and real parts $-\pi/5$,
$4\pi/5$ respectively so that these zeros sit at the edge of the
analytical strip. Lastly, a real 2-string consists of a pair of zeros
$u_{j}=\pi/10,\pi/2$ on the real axis. The string contents satisfy
the $(m,n)$ system 
\begin{eqnarray}
 & 2m+n+3-s=N\Leftrightarrow m+n=N-3+s,\qquad s=1,2\\
 & m=\text{\{number of short \ensuremath{2}-strings\}},\qquad n=\text{\{number of long \ensuremath{2}-strings\}}\label{stringdefs}
\end{eqnarray}
 There is always a real 2-string on the real axis and, in the $(r,s)=(1,1)$
sector, a single $1$-string furthest from the real axis. Each ``short
$2$-string\textquotedbl{} contributes two zeros and, by periodicity,
each ``long $2$-string contributes one zero. The 1-string contributes
one zero and so does the real 2-string since it is shared between
the upper and lower half planes. Consequently, the $(m,n)$ system
expresses the conservation of the total number of zeros in a periodicity
strip. The roles of $m$ and $n$ are interchanged under duality.
For the leading excitations $m$ is finite but $n\sim N$ as $N\to\infty$.

As explained in \cite{PCAI}, an excitation with string content $(m,n)$
is uniquely labeled by a set of quantum numbers 
\begin{equation}
I=(I_{1},I_{2},\ldots,I_{m})
\end{equation}
 where the integers $I_{j}\ge0$ give the number of long 2-strings
whose imaginary parts $w_{j}$ are greater than that of the given
short 2-string $v_{j}$. The short 2-strings $v_{j}$ and long 2-strings
$w_{j}$ labeled by $j=1$ are closest to the real axis. The quantum
numbers $I_{j}$ satisfy 
\begin{equation}
n\ge I_{1}\ge I_{2}\ge\dots\ge I_{m}\ge0.\label{iranges}
\end{equation}
 For given string content $(m,n)$, the lowest excitation occurs when
all of the short 2-strings are further out from the real axis than
all of the long 2-strings. In this case all of the quantum numbers
vanish $I_{j}=0$. Bringing the location of a short 2-string closer
to the real axis by interchanging the location of the short 2-string
with a long 2-string increments its quantum number by one unit and
increases the energy.

\subsubsection*{Finitized characters}

For $N=s-1$ (mod 2), the (fermionic) finitized characters are 
\begin{eqnarray}
(r,s)=(1,1): &  & \chi_{0}^{(N)}(q)=\sum_{m=0}^{(N-2)/2}q^{m^{2}+m}\left[\begin{array}{c}
N-2-m\\
m
\end{array}\right]_{q}\to\chi_{0}(q)\\
(r,s)=(1,2): &  & \chi_{-1/5}^{(N)}=\sum_{m=0}^{(N-1)/2}q^{m^{2}}\left[\begin{array}{c}
N-1-m\\
m
\end{array}\right]_{q}\to\chi_{1}(q)
\end{eqnarray}
where 
\[
\left[\begin{array}{c}
N\\
m
\end{array}\right]_{q}=\prod_{i=1}^{m}\frac{1-q^{N+1-i}}{1-q^{i}}\to\prod_{i=1}^{m}\frac{1}{1-q^{i}}\quad\mbox{for}\quad N\to\infty
\]
 These finitized characters can also be written in the form 
\begin{eqnarray}
\chi_{1,s}^{(N)}(q)=\sum q^{E}=\sum_{\sigma}q^{\sum_{j=1}^{N-1}jH(\sigma_{j-1},\sigma_{j},\sigma_{j+1})}
\end{eqnarray}
 where the sum is over all one-dimensional RSOS paths $\sigma=\{\sigma_{0},\sigma_{1},\ldots,\sigma_{N}\}$
on $A_{4}$ with $\sigma_{0}=s$ and $\sigma_{N}=1$. The energy function
$H$ is 
\begin{equation}
H(\sigma_{j-1},\sigma_{j},\sigma_{j+1})=\begin{cases}
1, & (\sigma_{j-1},\sigma_{j},\sigma_{j+1})=(2,1,2)\ \mbox{or}\ (3,4,3)\\
0, & \text{otherwise}
\end{cases}
\end{equation}
 Notice that this local energy function differs from the one introduced
by Forrester and Baxter.

\subsubsection*{Bijection between RSOS paths, strings and Virasoro modes}

\begin{figure}
\begin{centering}
\includegraphics[width=3.1cm,height=7.7cm,keepaspectratio]{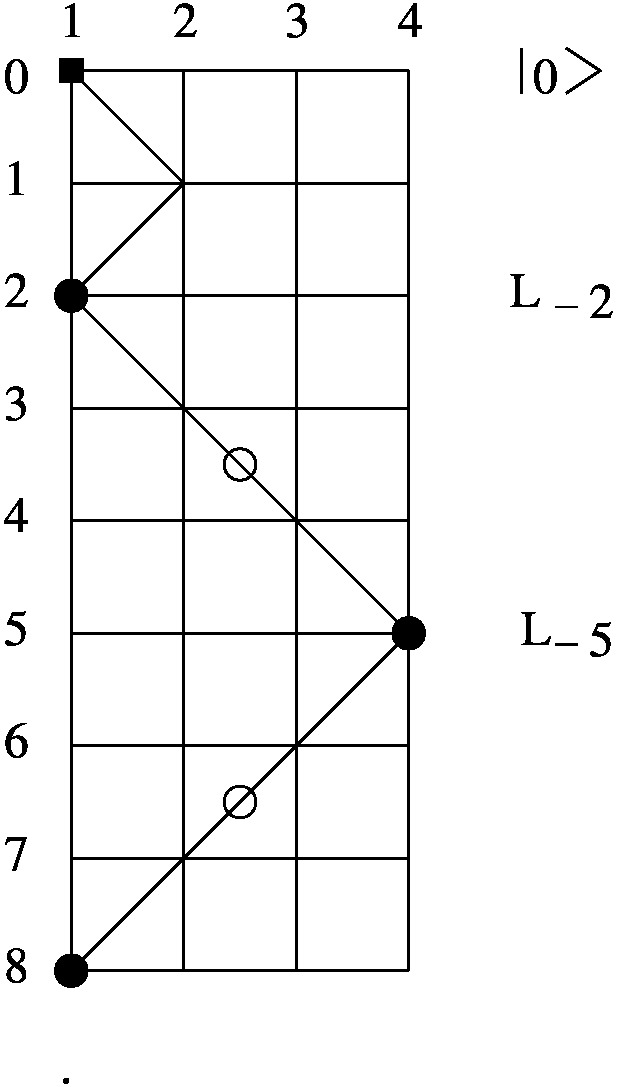}~~~\includegraphics[width=3.1cm,height=7.5cm,keepaspectratio]{boundzero1}~~~~~~~~~~~~\includegraphics[width=3.1cm,height=8cm,keepaspectratio]{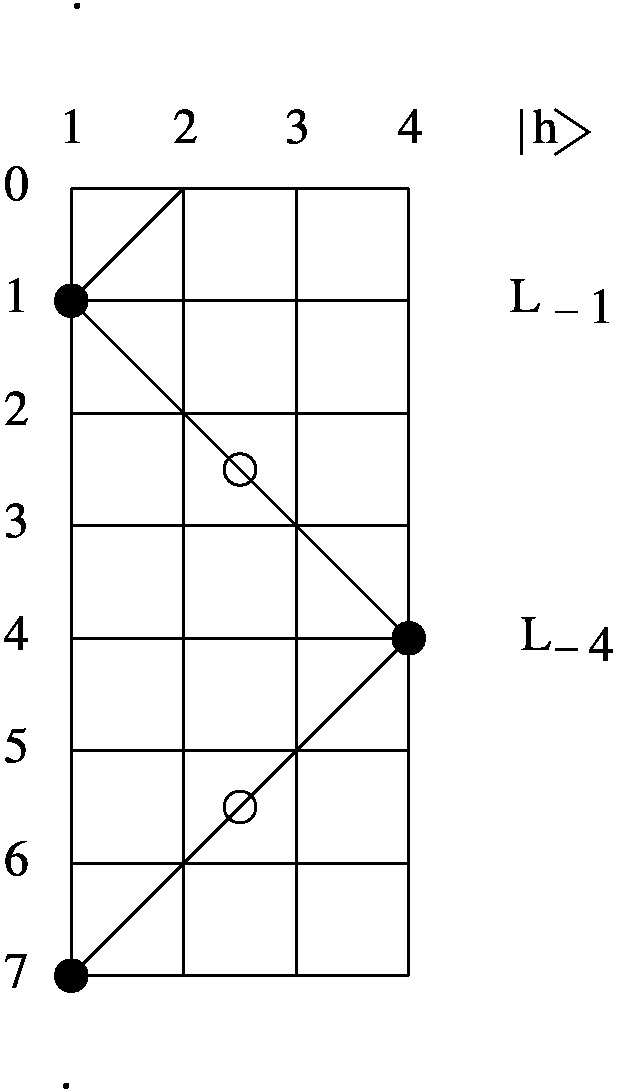}~~~\includegraphics[width=3.1cm,height=8cm,keepaspectratio]{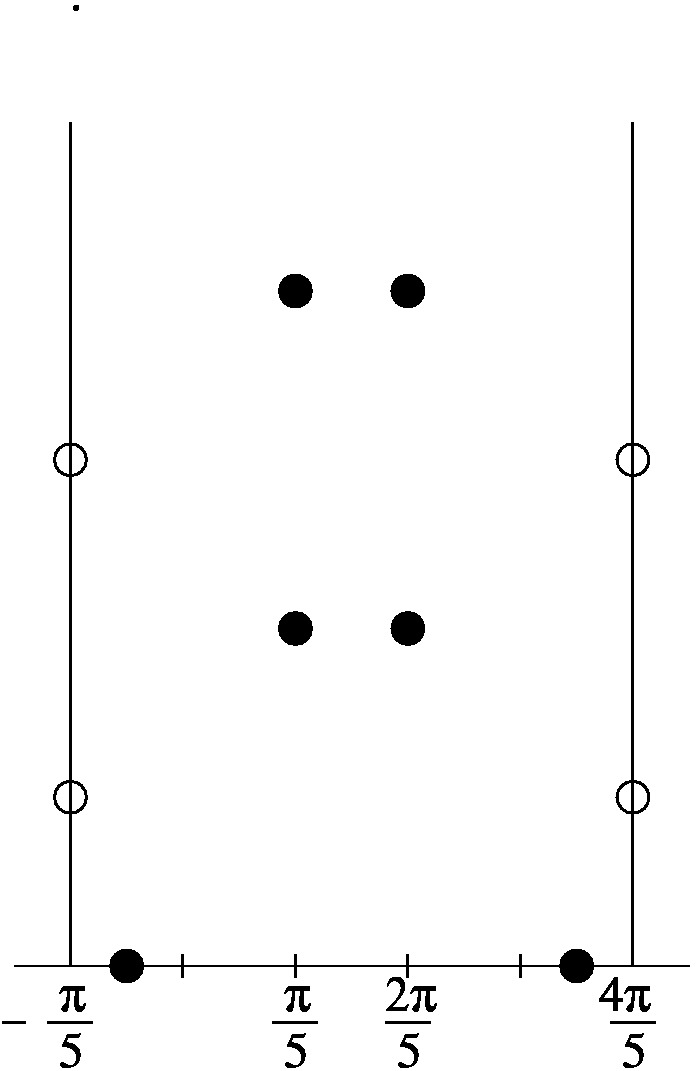}
\par\end{centering}

\vspace{-0in}
 \protect\caption{Bijection of one-dimensional configurational paths $\sigma$, strings
in the periodicity strip for the $(r,s)=(1,1)$ sector and the Virasoro
modes over $\vert0\rangle$ is shown on the left part. The right part
contains the analogous figures for the case $(r,s)=(1,2)$. The paths
$\sigma$ (rotated $90^{\circ}$ clockwise) are shown on the left
and the analytical strip containing a $1$-string (solid square),
short 2-strings (solid circles) and $2$-strings (open circles) in
the upper-half complex $u$-plane are shown on the right. For each
short two strings we associate a Virasoro mode where they are located,
$L_{-2}$ and $L_{-5}$ in the ($1,1$) and $L_{-1}$ , $L_{-4}$
in the $(1,2)$ cases. The string (particle) content is $m=n=2$ with
$N=8$ in the $(1,1)$ and with $N=7$ in the $(1,2)$ cases. The
energies of these excited states are $E=2+5=7$ ($E=1+4=5)$ since
there are two particles (short $2$-strings) at respective positions
$j=2,5$ ($1,4)$, respectively. The corresponding states in the Virasoro
modules are $L_{-5}L_{-2}\vert0\rangle$ and $L_{-4}L_{-1}\vert h\rangle$.\label{bijection}}
\end{figure}

There is in fact a bijection~\cite{FP02} between the one-dimensional
RSOS paths $\sigma$ that label the eigenstates (eigenvalues), the
allowed patterns of strings in the periodicity strip and the state
described in terms of the Virasoro modes. A triple $(\sigma_{j-1},\sigma_{j},\sigma_{j+1})=(2,1,2)$
or $(3,4,3)$ corresponds to a short $2$-string (particle) at position
$j$ and an insertion of a Virasoro mode $L_{-j}$ whereas a pair
segment $(\sigma_{j},\sigma_{j+1})=(2,3)$ or $(3,2)$ corresponds
to a long $2$-string (dual particle) at position $j+\frac{1}{2}$.
In addition, in the sector $(r,s)=(1,1)$, there is a $1$-string
at $j=0$ corresponding to the initial height $s=1$ at $j=0$. This
bijection is illustrated in Figure~\ref{bijection}. Notice that
only the relative positions of the long and short $2$-strings is
important. If $s=1$ the first and last ($3-s$) segments are inactive
whereas, if $s=2$, only the last ($3-s$) segments are inactive.
We see that the geometric constraint 
\begin{equation}
2m+n+3-s=N
\end{equation}
 agrees with the $(m,n)$ system.

\subsubsection*{Flow between boundary conditions}

We are in the position now to describe the boundary flows induced
by $\xi$ between the boundary conditions $(1,1)$ and $(1,2)$. This
flow is realized when $\xi$ goes from $0$ to $\infty$ and we can
describe it at the three different languages we already introduced. 

In terms of the zeros and $(n,m)$ system the flow is very simple:
the 1-string which exist only for the $(1,1)$ boundary condition
start to move to infinity in the imaginary direction as indicated
on the figure. There is no change in the 2-strings. 

\begin{figure}[H]
\begin{centering}
\includegraphics[height=6.5cm]{boundzero1} ~~~~~~~~\includegraphics[height=8.2cm]{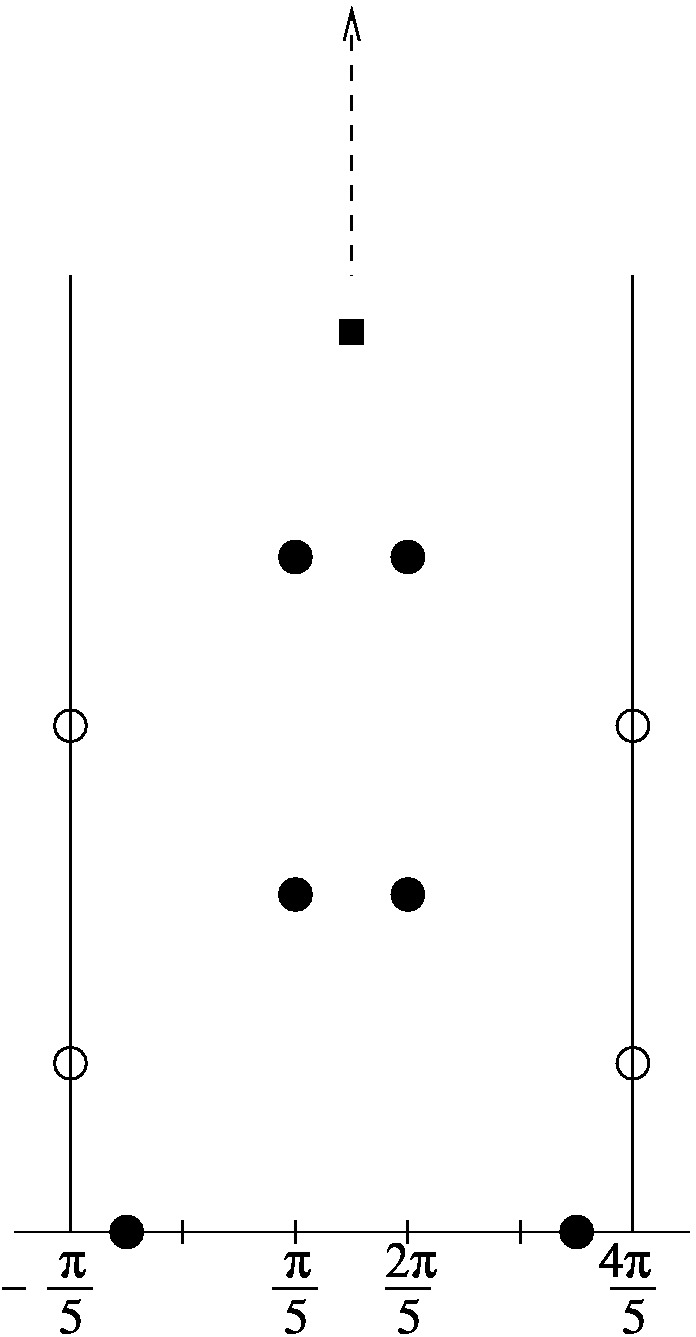}~~~~~~~~\includegraphics[height=7cm]{boundzero2a}
\par\end{centering}

\protect\caption{The boundary flow in terms of the zeros. On the left $\xi=0$ while
on the right $\xi=\infty$. In the $\xi\to\infty$ limit, shown in
the middle, nothing else happens than the imaginary location of the
1-string moves to infinity and disappears. The distribution of the
2-strings are not affected at all. }
\end{figure}

The flow in terms of the paths is also simple we merely have to remove
the first raw of the pathspace. 

Most enlightening is the flow in terms of the Virasoro modes. First
of all the highest weight state $\vert0\rangle$ flows to $\vert\Phi\rangle$,
and in the module the rule is very simple we have to increase the
index of every Virasoro mode by one $L_{-n}\to L_{-n+1}$:
\[
L_{-n_{1}}L_{-n_{2}}...L_{-n_{k}}\vert0\rangle\rightarrow L_{-n_{1}+1}L_{-n_{2}+1}...L_{-n_{k}+1}\vert h\rangle
\]
 This very simple flow is summarized for the first few excited states
in the following table 

\begin{center}
\begin{tabular}{|c|c|c|c|}
\hline 
Level & State in the $(1,1)$ module & State in the $(1,2)$ module & Level\tabularnewline
\hline 
\hline 
h.w. state & $|0\rangle$ & |$\Phi\rangle$ & h.w. state\tabularnewline
\hline 
2 & $L_{-2}|0\rangle$ & $L_{-1}|\Phi\rangle$ & 1\tabularnewline
\hline 
3 & $L_{-3}|0\rangle$ & $L_{-2}|\Phi\rangle$ & 2\tabularnewline
\hline 
4 & $L_{-4}|0\rangle$ & $L_{-3}|\Phi\rangle$ & 3\tabularnewline
\hline 
5 & $L_{-5}|0\rangle$ & $L_{-4}|\Phi\rangle$ & 4\tabularnewline
\cline{1-3} 
6 & $L_{-2}L_{-4}|0\rangle$ & $L_{-1}L_{-3}|\Phi\rangle$ & 4\tabularnewline
\cline{3-4} 
\multicolumn{1}{|c|}{6} & $L_{-6}|0\rangle$ & $L_{-5}|\Phi\rangle$ & 5\tabularnewline
\cline{1-2} 
7 & $L_{-2}L_{-5}|0\rangle$ & $L_{-1}L_{-4}|\Phi\rangle$ & 5\tabularnewline
\cline{3-4} 
7 & $L_{-7}|0\rangle$ & $L_{-6}|\Phi\rangle$ & 6\tabularnewline
\cline{1-2} 
8 & $L_{-2}L_{-6}|0\rangle$ & $L_{-1}L_{-5}|\Phi\rangle$ & 6\tabularnewline
8 & $L_{-3}L_{-5}|0\rangle$ & $L_{-2}L_{-4}|\Phi\rangle$ & 6\tabularnewline
\cline{3-4} 
8 & $L_{-8}|0\rangle$ & $L_{-7}|\Phi\rangle$ & 7\tabularnewline
\cline{1-2} 
9 & $L_{-2}L_{-7}|0\rangle$ & $L_{-1}L_{-6}|\Phi\rangle$ & 7\tabularnewline
9 & $L_{-3}L_{-6}|0\rangle$ & $L_{-2}L_{-5}|\Phi\rangle$ & 7\tabularnewline
\cline{3-4} 
9 & $L_{-9}|0\rangle$ & $L_{-8}|\Phi\rangle$ & 8\tabularnewline
\hline 
 &  &  & \tabularnewline
\hline 
\end{tabular}\smallskip{}

\par\end{center}

As expected and shown in the table above, the character will flow
from $X_{1,1}=1+q^{2}+q^{3}+q^{4}+q^{5}+2q^{6}+2q^{7}+3q^{8}+3q^{9}+...$
to $X_{1,2}=1+q+q^{2}+q^{3}+2q^{4}+2q^{5}+3q^{6}+3q^{7}+...$. 

The level by level flow agrees with the TCSA result of \cite{DPTW}
\smallskip{}

\subsubsection{Periodic case}

The best way to describe the periodic case is based on the previously
introduced boundary identification. In analyzing the zeros of the
transfer matrix we can distinguish two different appearance of zeros
as shown on the figure 4.4

\begin{figure}[H]
\begin{centering}
\includegraphics[width=4cm,height=8cm,keepaspectratio]{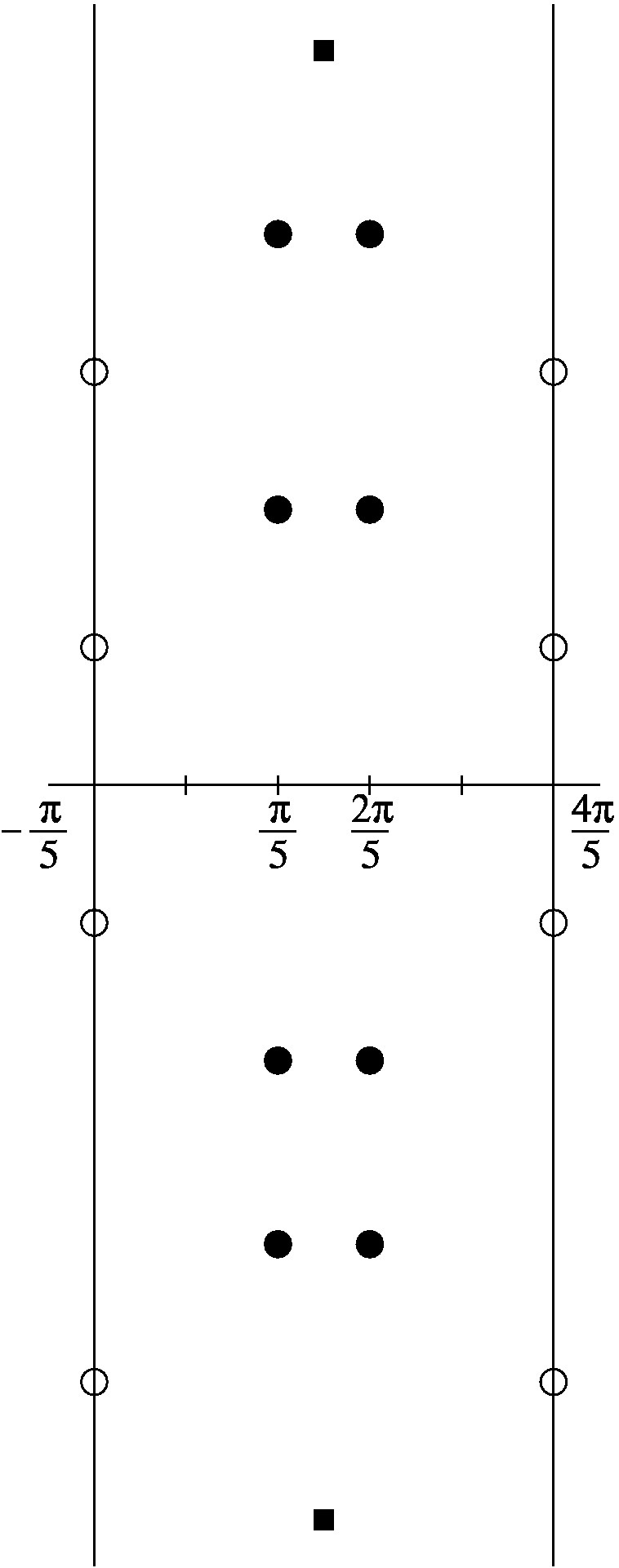}~~~~~~~\includegraphics[width=4cm,height=8cm,keepaspectratio]{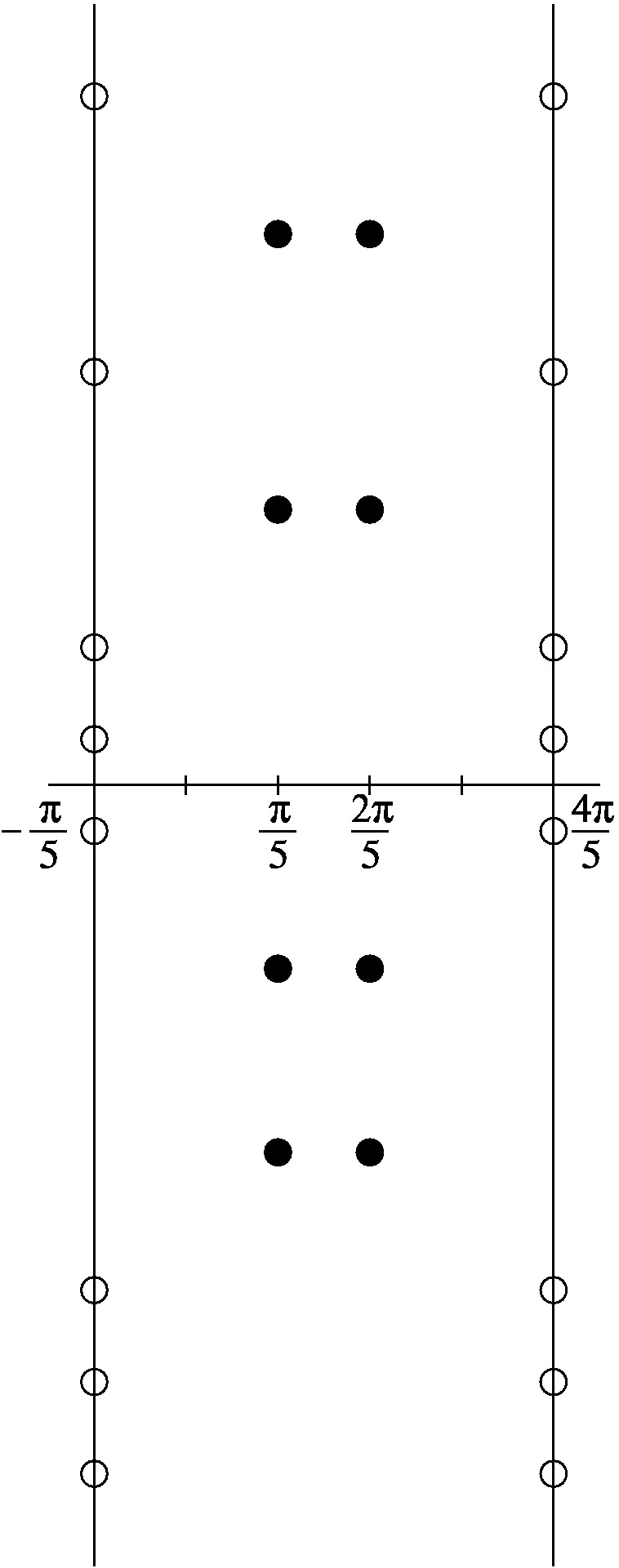}
\par\end{centering}

\protect\caption{The two appearing configurations of zeros of the transfer matrix eigenvalue
in the periodical case.}
\end{figure}

The first we can recognize is that we have similar short and long
2-strings and 1-strings as in the boundary case. What is different,
however, is that the zeros on the lower half plane are not necessarily
related to those on the upper half plane except for the 1-string.
So if we have a 1-string on the lower half we always have one the
upper half, too. In classifying the states we can use the already
developed classification for the boundary case, taking into account
that the lower and upper halves are independent. 

\begin{figure}
\begin{centering}
\includegraphics[width=3.5cm,height=8.5cm,keepaspectratio]{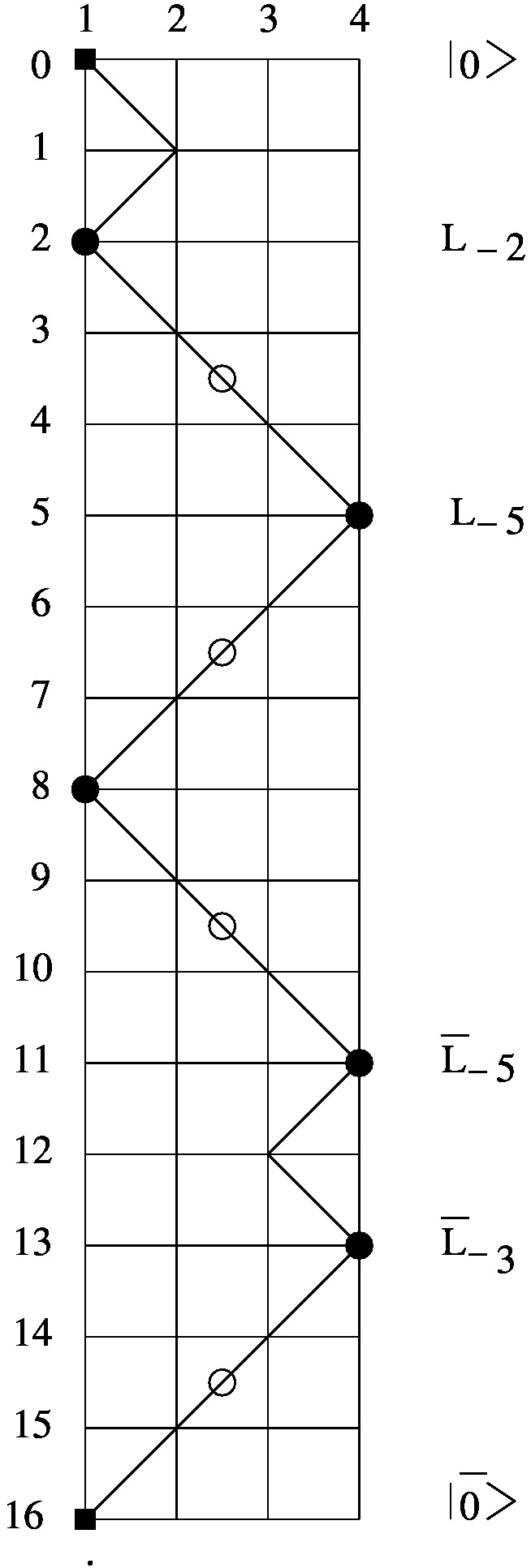}~~~\includegraphics[width=3.5cm,height=8.5cm,keepaspectratio]{periodic1}~~~~~~~~~~~~\includegraphics[width=3.5cm,height=8.5cm,keepaspectratio]{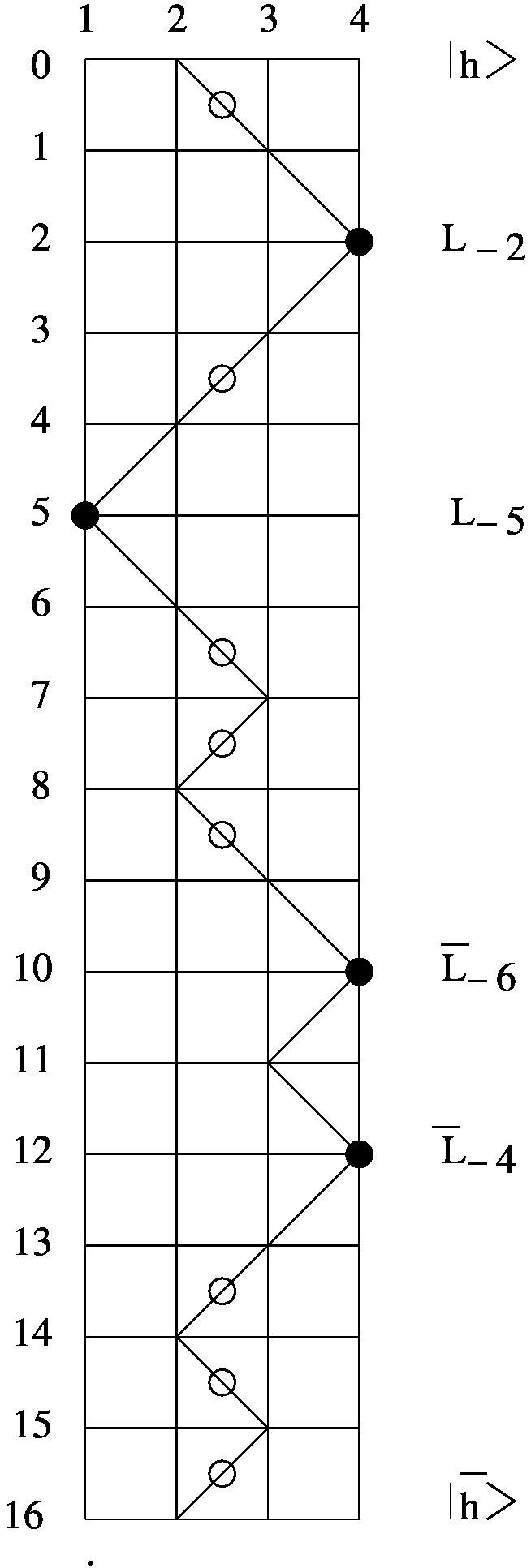}~~~\includegraphics[width=3.5cm,height=8.5cm,keepaspectratio]{periodic2}
\par\end{centering}

\vspace{-0in}
 \protect\caption{The identification between the zeros of the transfer matrix, paths
and Virasoro modes in the periodical case.}
\end{figure}

For the $(m,n)$ structure, we have to differentiate between the structures
on the 2 sides of the real axis. Now our lattice is $2N$ dimensional
with $N$ zeroes on each side. We define an $(m,n;\bar{m},\bar{n})$
system. On each side the eigenvalues with zeros at $\frac{3\pi}{10}$
correspond to the $s=1$ and have $N-2$ zeros, while the other eigenvalues
lie in the $s=2$ sector with $N$ zeros. Formulating this, we get
that $2m+n-2s+4=N$ or equivalently $2m+n=N+2s-4$, where $m$ is
the number of the short 2-strings and $n$ is the number of long 2-strings.
Similarly $2\bar{m}+\bar{n}=N+2s-4$.

\smallskip{}

Studying the zero structures, we find that they resemble the Hilbert
spaces of $\bar{V_{1}}\otimes V_{1}$ for the structures with zeroes
with the 1-string and $\bar{V_{2}}\otimes V_{2}$ for those without
the 1-string. In fact, the real axis separates the $\bar{V_{1}}$
and $\bar{V_{2}}$ parts of the tensored states from the $V_{1}$
and $V_{2}$ respectively. We define the vacuum $|0\rangle$ as the
state with $|0,0\rangle$ and the (1,2) ground state $|\Phi\rangle$
as the $|\frac{-1}{5},\frac{-1}{5}\rangle$ state. 

The state with one short string furthest from the real axis up below
the 1 string corresponds to $L_{-2}|0\rangle$ and moving the short
string downwards through the long strings increases the level by 1
for each permutation, thus creating the $L_{-n}|0\rangle$ states.
The mirror image on the zeroes below the real axis corresponds to
the $\bar{L}_{-n}|0\rangle$. A similar description also applies for
$L_{-n_{1}}...L_{-n_{N}}|0\rangle$.

For the (1,2) states, the lowest excitation appears with a short string
on the top of all long strings, with no 1-string above. This is $L_{-1}|\Phi\rangle$,
and every time we lower the short string below a long string we obtain
one extra unit of energy, hence we have all the $L_{-n}|\Phi\rangle$
and similarly for the mirror image of $\bar{L}$ and for combinations
of those.

Summarizing, in this model we find out that the first few states that
we obtain from the classification of the zeros of the eigenvalues
are:

\noindent \begin{center}
{\footnotesize{}}%
\begin{tabular}{|c|c|c|}
\hline 
\noalign{\vskip0.3mm}
{\footnotesize{}Level} & {\footnotesize{}$\bar{V_{1}}\otimes V_{1}$} & {\footnotesize{}$\bar{V_{2}}\otimes V_{2}$}\tabularnewline
\hline 
\hline 
\noalign{\vskip0.3mm}
{\footnotesize{}0} & {\footnotesize{}$|0\rangle$} & {\footnotesize{}$|\Phi\rangle$}\tabularnewline
\hline 
\noalign{\vskip0.3mm}
{\footnotesize{}1} &  & {\footnotesize{}$L_{-1}|\Phi\rangle$, $\bar{L}_{-1}|\Phi\rangle$}\tabularnewline
\hline 
\noalign{\vskip0.3mm}
{\footnotesize{}2} & {\footnotesize{}$L_{-2}|0\rangle$, $\bar{L}_{-2}|0\rangle$} & {\footnotesize{}$L_{-2}|\Phi\rangle$, $\bar{L}_{-2}|\Phi\rangle$,
$\bar{L}_{-1}L_{-1}|\Phi\rangle$}\tabularnewline
\hline 
\noalign{\vskip0.3mm}
{\footnotesize{}3} & {\footnotesize{}$L_{-3}|0\rangle$ , $\bar{L}_{-3}|0\rangle$} & {\footnotesize{}$L_{-3}|\Phi\rangle$, $\bar{L}_{-3}|\Phi\rangle$,
$\bar{L}_{-2}L_{-1}|\Phi\rangle$, $\bar{L}_{-1}L_{-2}|\Phi\rangle$}\tabularnewline
\hline 
\noalign{\vskip0.3mm}
\multicolumn{1}{|c}{{\footnotesize{}4}} & {\footnotesize{}$L_{-4}|0\rangle$ , $\bar{L}_{-4}|0\rangle$, $\bar{L}_{-2}L_{-2}|0\rangle$} & {\footnotesize{}$L_{-4}|\Phi\rangle$, $L_{-3}L_{-1}|\Phi\rangle$,
$L_{-3}\bar{L}_{-1}|\Phi\rangle$, $L_{-2}\bar{L}_{-2}|\Phi\rangle$,}\tabularnewline
\noalign{\vskip0.3mm}
\multicolumn{1}{|c}{} &  & {\footnotesize{} $\bar{L}_{-3}L_{-1}|\Phi\rangle$, $\bar{L}_{-3}\bar{L}_{-1}|\Phi\rangle$,
and $\bar{L}_{-4}|\Phi\rangle$}\tabularnewline
\hline 
\end{tabular}
\par\end{center}{\footnotesize \par}

\smallskip{}

Table 2 lists the first few states that we can see from the zero eigenvalues
of the periodic transfer matrix. In a lattice of $2N$ sites, one
can read all the states up to level $N$ completely. The same can
be done from the RSOS paths in the Hilbert space.

\smallskip{}

\subsubsection{The case of a seam}

Introducing a seam we can analyze the two limiting cases similarly
we analyzed in the boundary setting, namely $\xi$ going from $0$
to $\infty$. Clearly for $\xi=0$ the seam disappears (identity seam)
and we recover the results of the periodic boundary condition. In
the $\xi\to\infty$ limit we found the following identification between
the strings the paths and the Virasoro modes

\begin{figure}[H]
\begin{centering}
\includegraphics[width=1.9cm,height=8cm,keepaspectratio]{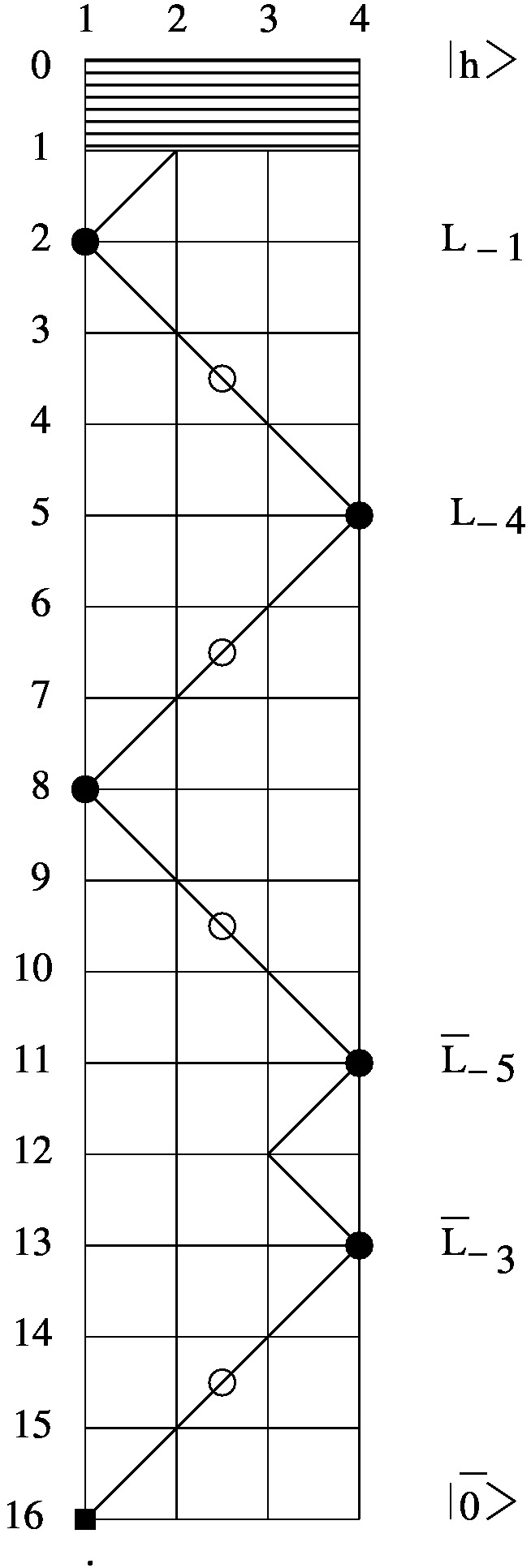}~~~\includegraphics[width=1.9cm,height=8cm,keepaspectratio]{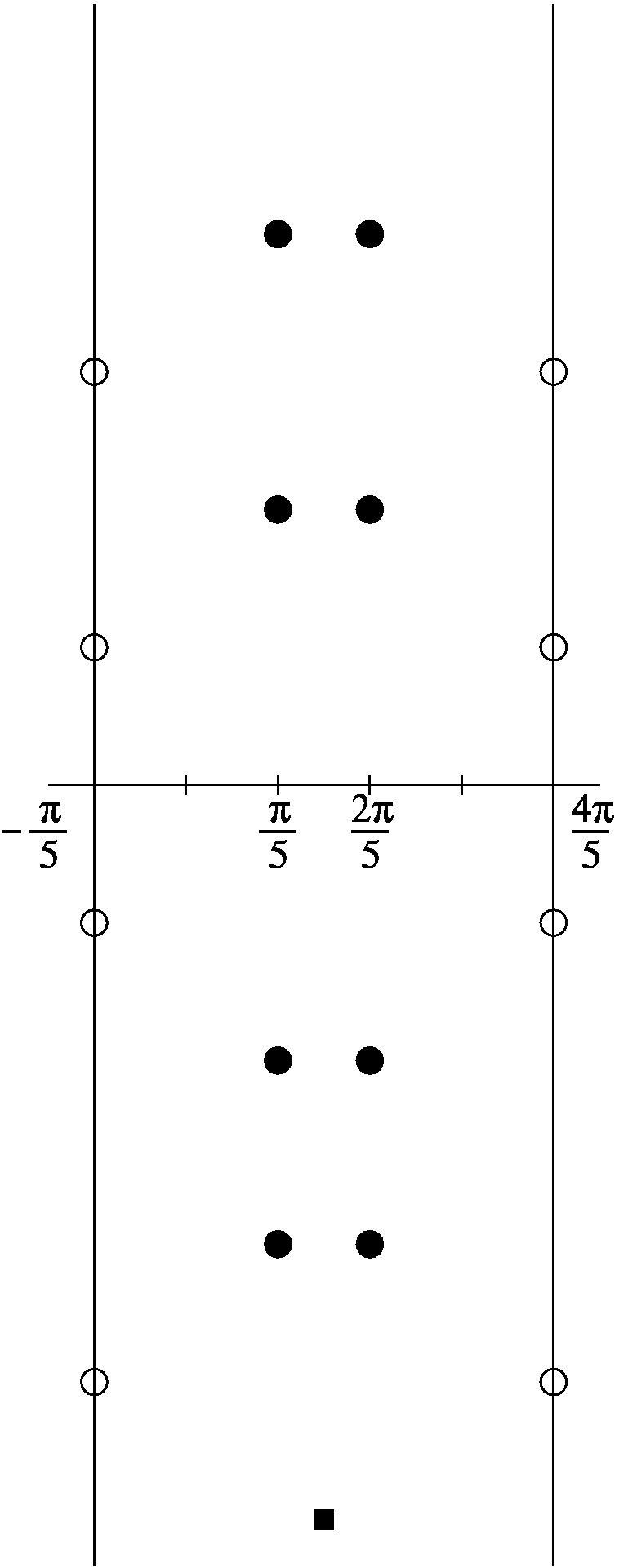}~~~~~~~~~\includegraphics[width=1.9cm,height=8cm,keepaspectratio]{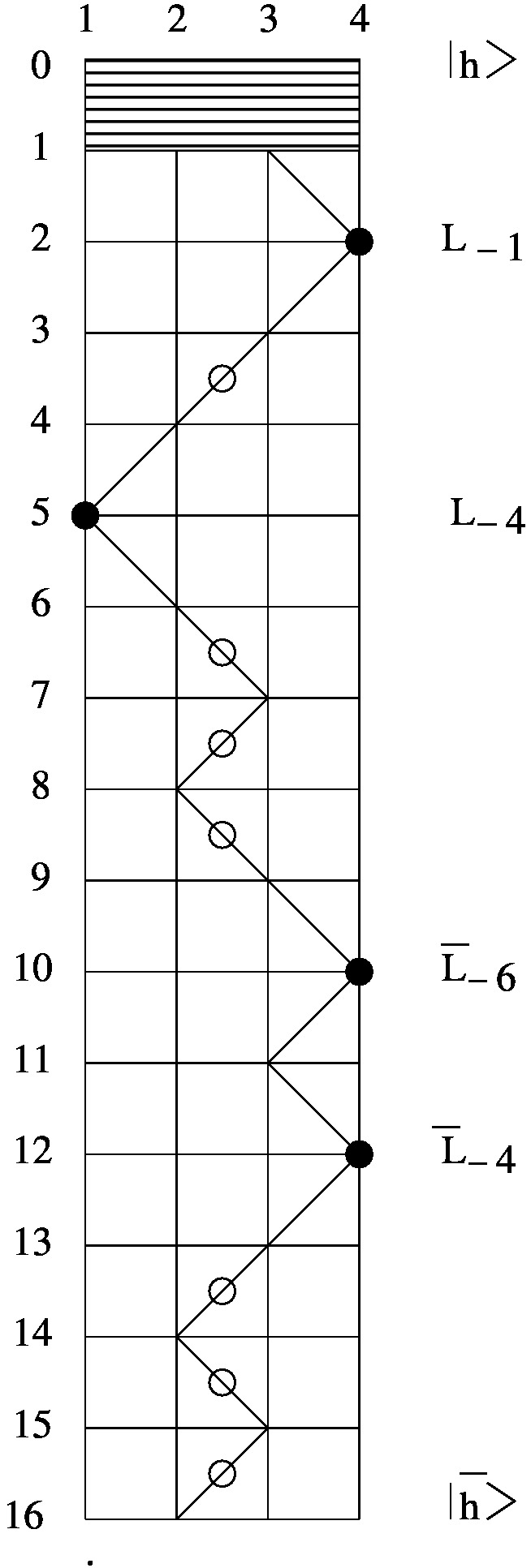}~~~\includegraphics[width=1.9cm,height=8cm,keepaspectratio]{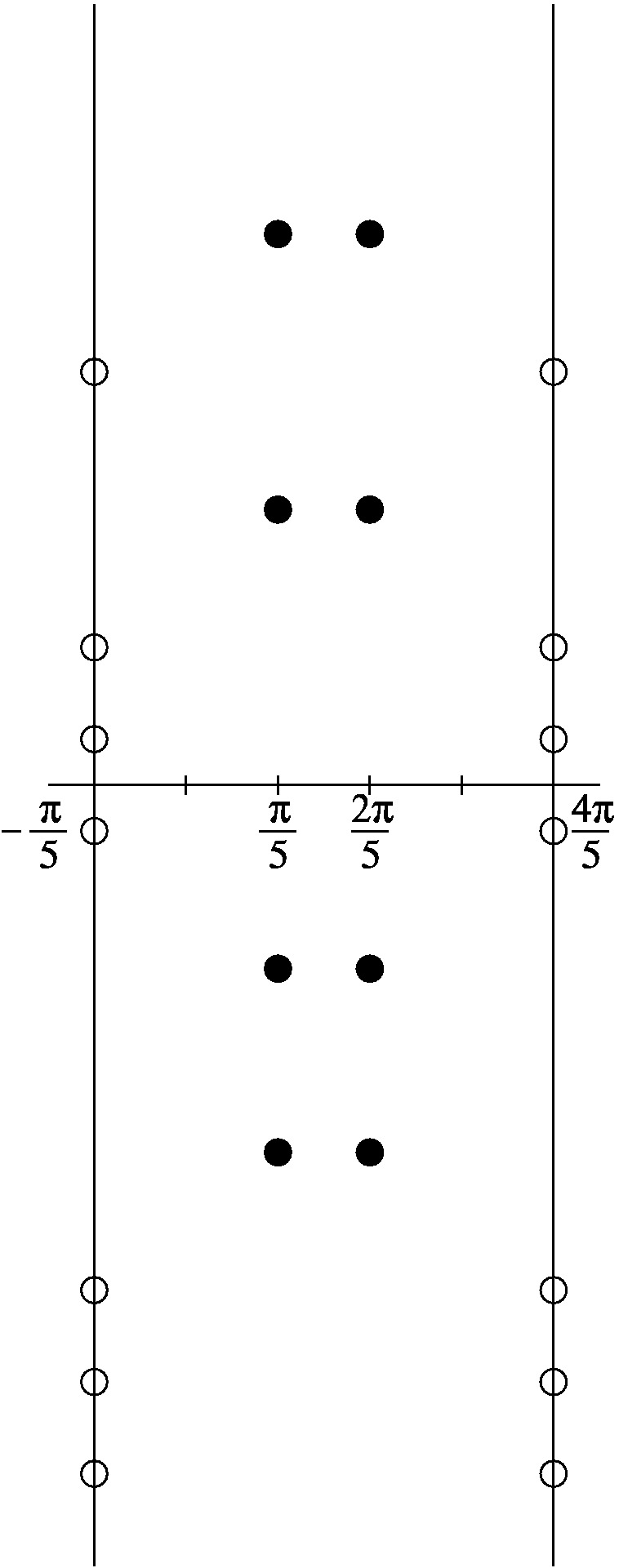}~~~~~~~~\includegraphics[width=1.9cm,height=8cm,keepaspectratio]{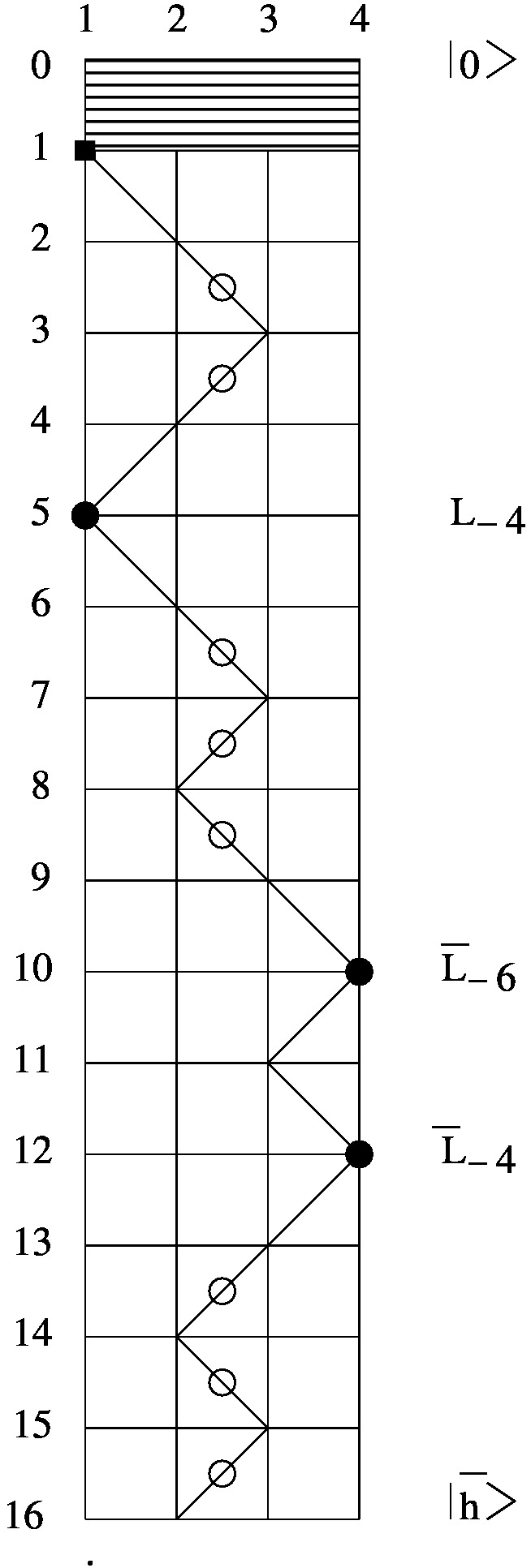}~~~-\includegraphics[width=1.9cm,height=8cm,keepaspectratio]{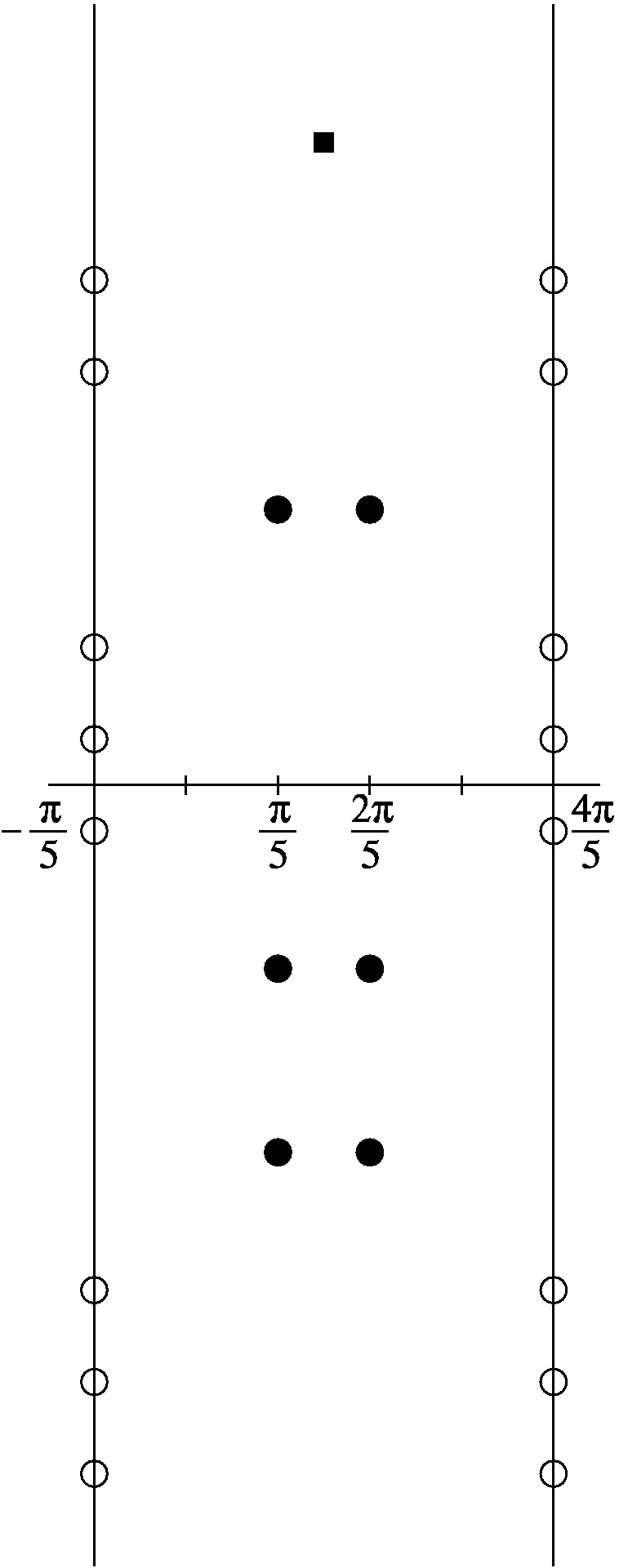}
\par\end{centering}

\vspace{0in}
 \protect\caption{The identification between the zeros of the transfer matrix, paths
and Virasoro modes in the case of a seam at the first site. The seam
shadows the periodicity of the paths.}
\end{figure}

The flows are very simple in terms of the zeros. We have the following
simple mechanism. The flows can be explained in three mechanisms:

A. If the outermost string is a 1-string, it flows towards infinity
with increasing $b$. (Plot on the left)

B. If the outermost string is a short 2-string, one of the zeroes
flows to infinity and the other goes to $\frac{3\pi}{10}$. (Plot
on the right)

C. If the outermost string is a long 2-string, it flows towards infinity.
(Plot in the middle)\smallskip{}

In terms of the states this is summarized as follows:

1. Due to type A flows: 
\[
L_{-N_{1}}...L_{-N_{n}}\bar{L}_{-\bar{N}_{1}}....\bar{L}_{-\bar{N}_{n}}|0\rangle\rightarrow L_{-N_{1}}...L_{-N_{n}}\bar{L}_{-\bar{N}_{1}+1}....\bar{L}_{-\bar{N}_{n}+1}|\bar{\phi}\rangle
\]
with $|0\rangle\rightarrow|\bar{\phi}\rangle$

2. Due to type B flows: 
\[
(L_{-N_{1}}...L_{-N_{n}}\bar{L}_{-\bar{N}_{1}}....\bar{L}_{-\bar{N}_{n}})\bar{L}_{-1}|\Phi\rangle\rightarrow(L_{-N_{1}}...L_{-N_{n}}\bar{L}_{-\bar{N}_{1}+1}....\bar{L}_{-\bar{N}_{n}+1})|\phi\rangle
\]

3. Due to type C flows: 
\[
L_{-N_{1}}...L_{-N_{n}}\bar{L}_{-\bar{N}_{1}}....\bar{L}_{-\bar{N}_{n}}|\Phi\rangle\rightarrow L_{-N_{1}}...L_{-N_{n}}\bar{L}_{-\bar{N}_{1}+1}....\bar{L}_{-\bar{N}_{n}+1}|\Phi\rangle
\]

Now using the first few states of the trivial defect case from Table
2, we will deduce their corresponding states using this mechanism,
and it is indeed what we can observe from the flows of the zero eigenvalues
as was shown for sample states above.

\noindent \begin{center}
\begin{tabular}{|c|c|c|c|}
\hline 
Level & Trivial Defect & Non-trivial Defect & Level\tabularnewline
\hline 
h.w. state & $|0\rangle$ & |$\bar{\phi}\rangle$ & h.w. state\tabularnewline
h.w. state & |$\Phi\rangle$ & |$\Phi\rangle$ & h.w. state\tabularnewline
\cline{1-2} 
1 & $\bar{L}_{-1}|\Phi\rangle$ & $|\phi\rangle$ & h.w. state\tabularnewline
\cline{3-4} 
1 & $L_{-1}|\Phi\rangle$ & $L_{-1}|\Phi\rangle$ & 1\tabularnewline
\cline{1-2} 
2 & $\bar{L}_{-2}|0\rangle$ & $\bar{L}_{-1}$$|\bar{\phi}\rangle$ & 1\tabularnewline
2 & $\bar{L}_{-2}|\Phi\rangle$ & $\bar{L}_{-1}|\Phi\rangle$ & 1\tabularnewline
2 & $\bar{L}_{-1}L_{-1}|\Phi\rangle$ & $L_{-1}|\phi\rangle$ & 1\tabularnewline
\cline{3-4} 
2 & $L_{-2}|\Phi\rangle$ & $L_{-2}|\Phi\rangle$ & 2\tabularnewline
\multicolumn{1}{|c|}{2} & $L_{-2}|0\rangle$ & $L_{-2}|\bar{\phi}\rangle$ & 2\tabularnewline
\cline{1-2} 
3 & $\bar{L}_{-3}|0\rangle$ & $\bar{L}_{-2}$$|\bar{\phi}\rangle$ & 2\tabularnewline
3 & $\bar{L}_{-3}|\Phi\rangle$ & $\bar{L}_{-2}|\Phi\rangle$ & 2\tabularnewline
3 & $\bar{L}_{-2}L_{-1}|\Phi\rangle$ & $\bar{L}_{-1}L_{-1}|\Phi\rangle$ & 2\tabularnewline
3 & $\bar{L}_{-1}L_{-2}|\Phi\rangle$ & $L_{-2}|\phi\rangle$ & 2\tabularnewline
\cline{3-4} 
3 & $L_{-3}|\Phi\rangle$ & $L_{-3}|\Phi\rangle$ & 3\tabularnewline
3 & $L_{-3}|0\rangle$ & $L_{-3}|0\rangle$ & 3\tabularnewline
\hline 
4 & $\bar{L}_{-1}\bar{L}_{-3}|\Phi\rangle$ & $\bar{L}_{-2}|\phi\rangle$ & 2\tabularnewline
\hline 
\end{tabular}
\par\end{center}

\smallskip{}

Table 3 shows the exact flow from each state in the trivial defect
Hilbert Space to its corresponding state in the non-trivial one up
to the second order descendent level in the defect Hilbert space.\medskip{}

It can be seen that the flow occurs from $\bar{V}_{1}\otimes V_{1}+\bar{V}_{2}\otimes V_{2}$
to $\bar{V}_{1}\otimes V_{2}+\bar{V}_{2}\otimes V_{1}+\bar{V}_{2}\otimes V_{2}$
as we expect from defect conformal field theory. Had we taken the
limit $b\rightarrow-\infty$, we would have got the similar total
outcome, but with $\phi$ and $\bar{\phi}$ exchanged, and $L_{-n}$
being the operator augmented to $L_{-n+1}$ instead of $\bar{L}$
as happens here.

\begin{figure}[H]
\begin{centering}
\includegraphics[width=2.02cm,height=9.1cm,keepaspectratio]{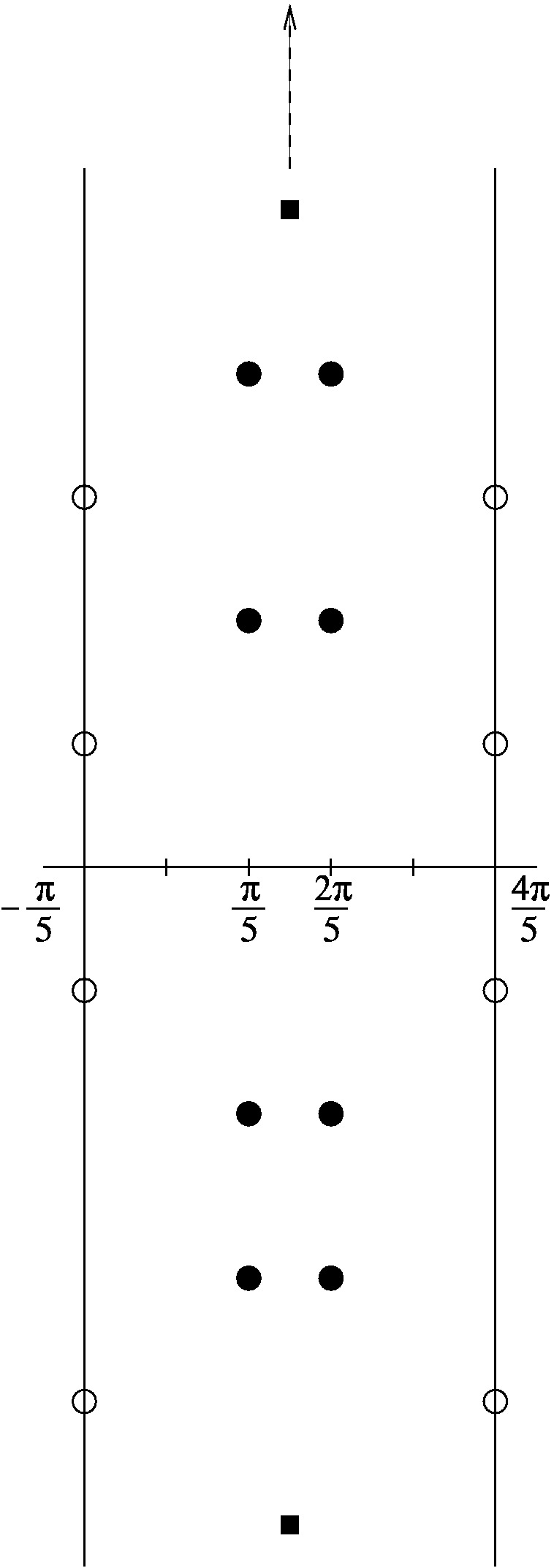}~~~\includegraphics[width=2.02cm,height=8.2cm,keepaspectratio]{seam1}~~~~~~~~~~~~~~~\includegraphics[width=2.02cm,height=8.8cm,keepaspectratio]{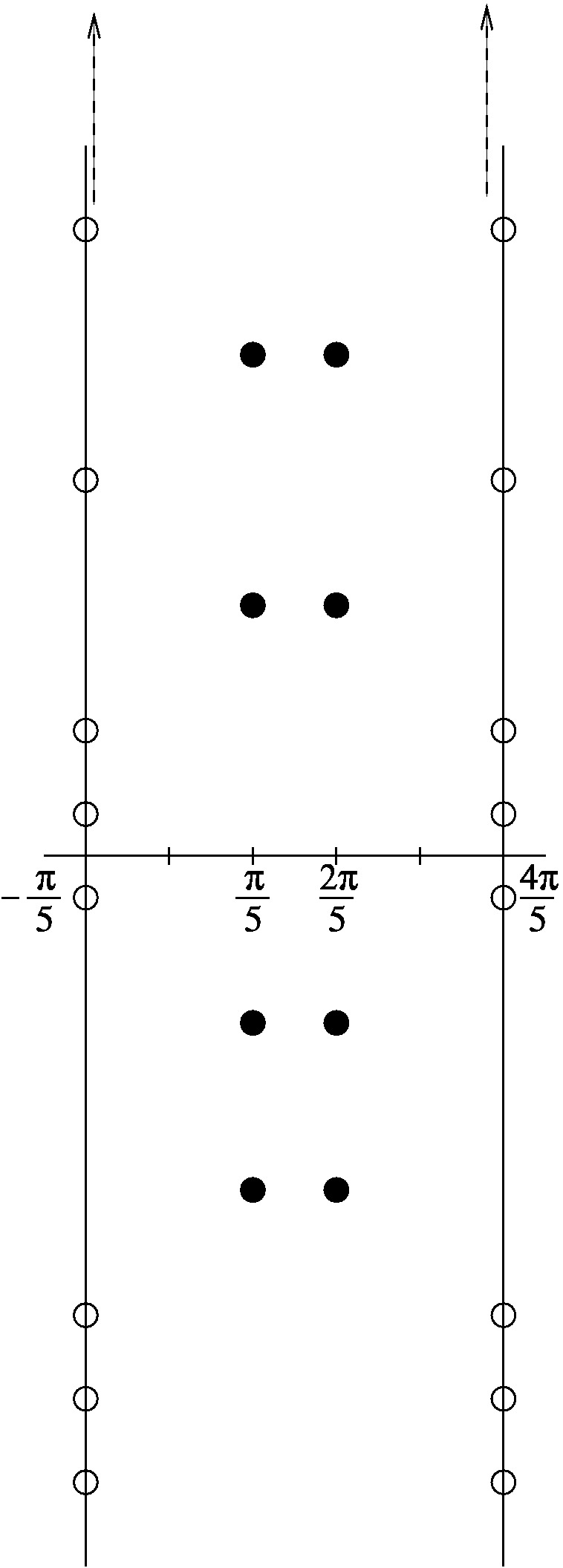}~~~\includegraphics[width=2.02cm,height=8cm,keepaspectratio]{seam2}~~~~~~~~~~~~\includegraphics[width=2.02cm,height=8.5cm,keepaspectratio]{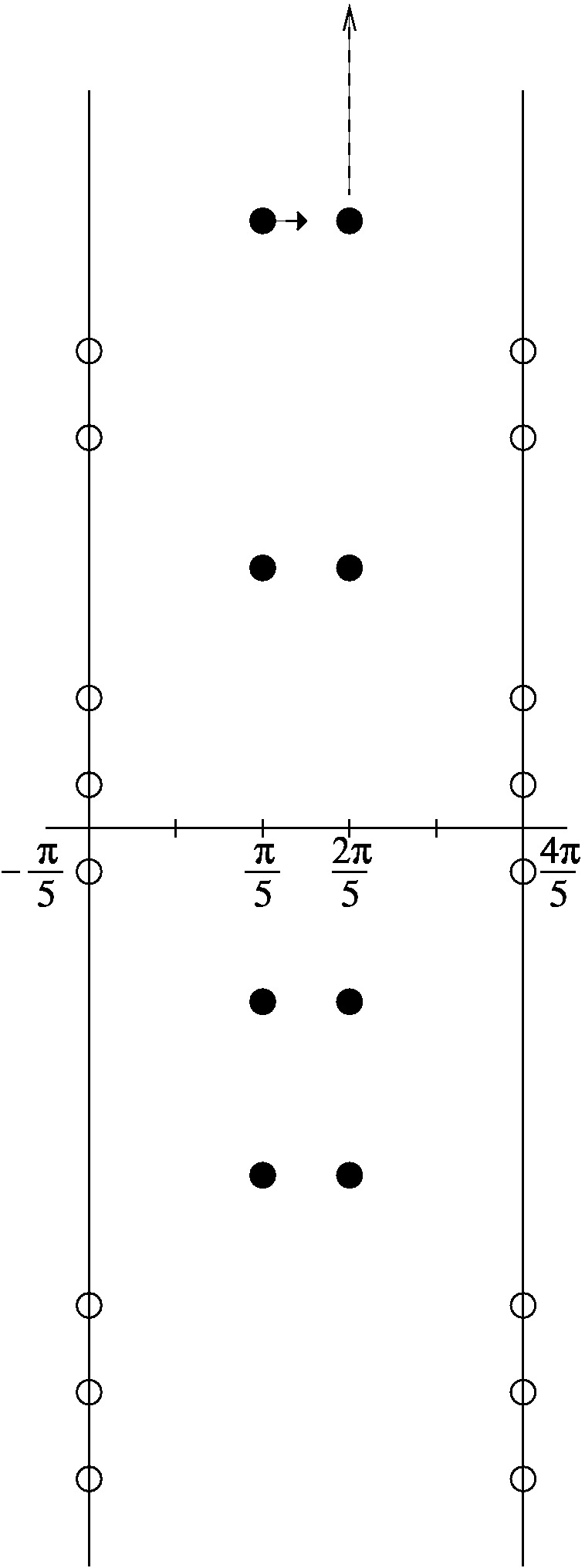}~~~\includegraphics[width=2.02cm,height=8cm,keepaspectratio]{seam3}
\par\end{centering}

\vspace{0.2in}
 \protect\caption{The identification between the zeros of the transfer matrix, paths
and Virasoro modes in the case of a seam at the first site. The seam
shadows the periodicity of the paths. The two figures on the left
correspond to type A flow, the outermost 1-string flows to infinity,
and the state flows from $L_{-3}L_{-5}\bar{L}_{-2}\bar{L}_{-5}|0\rangle\rightarrow L_{-3}L_{-5}\bar{L}_{-1}\bar{L}_{-4}|\bar{\phi}\rangle$.
The middle two figures are of type C, the long 2-string goes away
to infinity and the state flows from $L_{-4}L_{-6}\bar{L}_{-2}\bar{L}_{-5}|\Phi\rangle\rightarrow L_{-4}L_{-6}\bar{L}_{-1}\bar{L}_{-4}|\Phi\rangle$.
Finally the two figures on the right show how a short 2-string has
one of its zeroes flow to infinity and the other to the middle $u=\frac{3\pi}{10}$,
and the state flows from $L_{-4}L_{-6}\bar{L}_{-1}\bar{L}_{-5}|\Phi\rangle\rightarrow L_{-4}L_{-6}\bar{L}_{-4}|\phi\rangle$}
\end{figure}
\smallskip{}

\section{TBA Equations}

In this section we solve the TBA equations of the Lee-Yang model for
the periodic boundary conditions with and without a seam, and for
the boundary model in both the critical and the massive cases. We
derive those equations on the lattice, and after scaling we confirm
the results of the continuum limit equations which were derived in
\cite{LMSS,BS,DPTW}. Our approach is systematic since we know the
analytic structure and the zero eigenvalue locations in the analytic
strip, which allows us to solve the TBAs taking into account this
structure.

\subsection{Critical/Massless TBAs:}

\subsubsection{Periodic boundary conditions}

The transfer matrix satisfies the functional relation
\[
T(u)T(u+\lambda)=\left(\frac{\sin(\lambda+u)\sin(\lambda-u)}{\sin^{2}\lambda}\right)^{N}+\left(\frac{\sin u}{\sin\lambda}\right)^{N}T(u+3\lambda)
\]

We normalize the transfer matrix as:

\[
t(u)=\left(\frac{s(\lambda)s(u+2\lambda)}{s(u+\lambda)s(u+3\lambda)}\right)^{N}T_{1}(u)
\]
and we get that
\[
t(u)t(u+\lambda)=1+t(u+3\lambda)
\]
Using the periodicity $t(u)=t(u+\pi)$ we rewrite it as $t(u)t(u-\frac{2\pi}{5})=1+t(u-\frac{\pi}{5})$,
and after shifting $u$ we have
\[
t(u+\frac{\pi}{5})t(u-\frac{\pi}{5})=1+t(u)
\]

We decompose $t$ into two components $f$ and $l$:

\[
t(u)=f^{N}(u)l(u)
\]

where $f^{N}$ corresponds to the bulk free energy (order $N$ term)
and $l$ corresponds to the finite size corrections (order $\frac{1}{N}$
term) and $N$ is even. We want to kill the $N$th order zeros at
$-\frac{\pi}{5},\frac{4\pi}{5}$ and poles at $\frac{\pi}{5},\frac{2\pi}{5}$
in $f(u)$ which satisfies
\[
f(u+\frac{\pi}{5})f(u-\frac{\pi}{5})=f(u)
\]
The solution compatible with the analytical structure is 
\[
f(u)=\frac{\sin\frac{5u}{3}+\sin\frac{\pi}{3}}{\sin\frac{5u}{3}-\sin\frac{\pi}{3}}=-\tan(\frac{5u}{6}+\frac{\pi}{6})\tan(\frac{5u}{6}+2\frac{\pi}{6})
\]

which is basically the shifted S-matrix. $f(u)$ satisfies $f(u)f(u+\lambda)=1$.
Introducing the variable $\ u=\frac{3\pi}{10}+\frac{3ix}{5}\ $ we
write the functional equation as 
\[
t(x-i\frac{\pi}{3})t(x+i\frac{\pi}{3})=1+t(x)
\]
In this variable 
\begin{equation}
f(x)=-\tan(\frac{ix}{2}+\frac{5\pi}{12})\tan(\frac{ix}{2}+\frac{7\pi}{12})
\end{equation}

\subsubsection*{Vacuum state:}

The ground state has no zeroes inside the analytical strip, and since
it is analytic in the strip, we can use the functional relation to
write 
\[
\frac{t(x-i\frac{\pi}{3})t(x+i\frac{\pi}{3})}{t(x)}=\frac{l(x-i\frac{\pi}{3})l(x+i\frac{\pi}{3})}{l(x)}=1+t^{-1}(x)
\]

For the ground state, inside the physical strip $x\in[-\frac{i\pi}{3},\frac{i\pi}{3}]$,
both sides of the equation are ANZ, so we can take the log and solve
in Fourier space, then
\[
\log l(x)=-\varphi\star\log(1+t^{-1}(x))\qquad;\qquad\hat{\varphi}=\frac{1}{1-e^{k\frac{\pi}{3}}-e^{-k\frac{\pi}{3}}}
\]

where 
\[
\ \varphi(x)=\frac{1}{2\pi}\underset{-\infty}{\overset{\infty}{\int}}dk\hat{\varphi}e^{ikx}\ 
\]
 and 
\begin{equation}
\ \varphi\star f=\underset{-\infty}{\overset{\infty}{\int}}dy\varphi(x-y)f(y)\label{eq:convolution}
\end{equation}

Now we restore $t$: 
\begin{equation}
\log t(x)=N\log f(x)-\varphi\star\log(1+t^{-1}(x))
\end{equation}
This is the ground-state TBA on the lattice. In the thermodynamic
limit all interesting things happen around two domains: in the $u$
variable either on the upper half plane around $\frac{3}{5}i\log N$
or on the lower half plane around $-\frac{3}{5}i\log N$. For this
reason in the $x$ variable we focus on the behavior around $\pm\log N$.
Let us center the new functions around $x\pm\log N$ as $e^{\epsilon^{\pm}(x)}=\underset{N\to\infty}{\lim}t(x\pm\log N)$.
Taking the continuum limit ($N\to\infty$) on the source term we get
\[
N\lim_{N\to\infty}\log f(x+\log N)=N\lim_{N\to\infty}\log(1+4\sin\frac{\pi}{3}e^{\mp x}\frac{1}{N}+\dots)=4\sin\frac{\pi}{3}e^{\mp x}
\]
which leads to the massless ground-state TBA equations 
\begin{equation}
\epsilon^{\pm}(x)=4\sin\frac{\pi}{3}e^{\mp x}-\varphi\star\log(1+e^{-\epsilon^{\pm}(x)})
\end{equation}
where $4\sin\frac{\pi}{3}$ can be easily absorbed by shifting $x$.

\subsubsection*{Excited states:}

From our numerics, we know that in the excited states there are zero
eigenvalues in the analycity strip. We classified them before as 1-strings
and short 2-strings. The 1-strings occur at 
\[
u=\frac{3\pi}{10}+i\alpha
\]
 and additionally the short strings at 
\[
u=\begin{cases}
\frac{\pi}{5}+i\beta_{j}\\
\frac{2\pi}{5}+i\beta_{j}
\end{cases}
\]
For finite energy states in the continuum ($N\to\infty$) limit, they
go to infinity as $\alpha=\frac{3}{5}(\pm\log N+\tilde{\alpha}^{\pm})$
and $\beta_{j}=\frac{3}{5}(\pm\log N+\tilde{\beta}_{j}^{\pm})$ on
the upper/lower half plane, respectively as it can be analyzed numerically. 

In the $x=\frac{5}{3i}(u-\frac{3\pi}{10})$ variable they are located
at 
\[
\begin{array}{cc}
x_{0}^{\pm}=\frac{5\alpha}{3}=\log N+\tilde{\alpha}^{\pm}\\
x_{j}^{\pm}=\pm i\frac{\pi}{6}+\log N+\tilde{\beta}_{j}^{\pm}
\end{array}
\]
 As we would like to take logarithm we need functions free of zeros
and poles on the physical strip. The function which can eliminate
the single zero is 
\begin{equation}
\sigma_{0}(x)=\tanh\left(\frac{3}{4}x\right)
\end{equation}
while the one which eliminates the two zeros at $\pm\frac{\pi}{6}$
is 
\begin{equation}
\sigma_{1}(x)=f(x)^{-1}=\frac{\cosh x+\sin\frac{\pi}{3}}{\cosh x-\sin\frac{\pi}{3}}
\end{equation}
 These functions satisfy
\begin{equation}
\begin{array}{cc}
\sigma_{0}(x-\frac{i\pi}{3})\sigma_{0}(x+\frac{i\pi}{3})=1\\
\sigma_{1}(x-\frac{i\pi}{3})\sigma_{1}(x+\frac{i\pi}{3})=\sigma_{1}(x)
\end{array}
\end{equation}

With these functions we parametrize the normalized transfer matrix
eigenvalue as 
\begin{equation}
t(x)=f(x)^{N}\prod_{\pm}\sigma_{0}(x-x_{0}^{\pm})\prod_{j=1}^{M}\sigma_{1}(x-x_{j}^{\pm})l(x)
\end{equation}
which ensures that $l(x)$ is ANZ in the physical strip. The functional
equation then takes the form
\begin{equation}
\prod_{\pm}\sigma_{0}(x-x_{0}^{\pm})\frac{t(x-i\frac{\pi}{3})t(x+i\frac{\pi}{3})}{t(x)}=\frac{l(x-i\frac{\pi}{3})l(x+i\frac{\pi}{3})}{l(x)}=\prod_{\pm}\sigma_{0}(x-x_{0}^{\pm})(1+t^{-1}(x))
\end{equation}
Clearly both sides are ANZ in the interior of the physical strip:
the combination $\prod_{\pm}\sigma_{0}(x-x_{0}^{\pm})t^{-1}(x)$ is
regular at $x=x_{0}^{\pm}.$ Taking then logarithm and going to Fourier
space we find:
\[
\log l(x)=-\varphi\star\log(\prod_{\pm}\sigma_{0}(x-x_{0}^{\pm})(1+t^{-1}(x)))
\]
Restoring $t$ we obtain the excited-state massless TBA equation on
the lattice:

\begin{equation}
\begin{array}{cc}
\log t(x)=N\log f(x)+\sum_{\pm}\log(\sigma_{0}(x-x_{0}^{\pm}))+\sum_{j,\pm}\log(\sigma_{1}(x-x_{j}^{\pm}))\\
-\varphi\star\log(\prod_{\pm}\sigma_{0}(x-x_{0}^{\pm})(1+t^{-1}(x)))
\end{array}
\end{equation}
The parameters of the excited state $x_{i}=\{x_{0}^{\pm},x_{j}^{\pm}\}$
are determined self-consistently from the fact that it is a zero of
the transfer matrix: 
\[
t(x)\vert_{x=x_{i}\pm\frac{i\pi}{3}}=-1
\]
In the scaling limit we can focus on the two scaling domains at $\pm\log N$
by introducing 
\[
e^{\epsilon^{\pm}(x)}=\lim_{N\to\infty}\sigma_{0}(x\pm\log N-x_{0}^{\pm})^{-1}t(x+\log N)
\]
and in this limit, it satisfies the excited-state massless TBA equation:

\begin{equation}
\epsilon^{\pm}(x)=4\sin\frac{\pi}{3}e^{\mp x}+\sum_{j}\log\sigma_{1}(x-\tilde{\beta}_{j}^{\pm})-\varphi\star\log(\sigma_{0}(x-\tilde{\alpha}^{\pm})+e^{-\epsilon^{\pm}(x)})
\end{equation}
The parameters satisfy the following equations 
\[
e^{\epsilon^{\pm}(x)}\sigma_{0}(x-\tilde{\alpha}^{\pm})\vert_{x=\tilde{x}_{i}\pm\frac{i\pi}{3}}=-1
\]

\subsubsection{Seam}

The transfer matrix of the model with periodical boundary conditions
with a seam of parameter $\xi$ can be defined on a lattice of $N$
sites ($N$ is even) from the local face weights and satisfies the
relation:

{\footnotesize{}
\[
T_{1}(u)T_{1}(u+\lambda)=\frac{s(u+\lambda+\xi)s(\lambda-u-\xi)}{s^{2}(\lambda)}\left(\frac{s(u+\lambda)s(\lambda-u)}{s^{2}(\lambda)}\right)^{N-1}\mathbb{I}+\frac{s(u+\xi)}{s(\lambda)}\left(\frac{s(u)}{s(\lambda)}\right)^{N-1}T_{1}(u+3\lambda)
\]
}{\footnotesize \par}

Defining the normalization

\[
t(u)=\frac{s(\lambda)s(u+2\lambda+\xi)}{s(u+\lambda+\xi)s(u+3\lambda+\xi)}\left(\frac{s(\lambda)s(u+2\lambda)}{s(u+\lambda)s(u+3\lambda)}\right)^{N-1}T_{1}(u)
\]
We obtain the functional relations:
\[
t(u)t(u+\lambda)=1+t(u+3\lambda)
\]
Using the same relation assumed for the periodic boundary conditions,
we would like to kill the $(N-1)$th order zeros at $-\frac{\pi}{5},\frac{4\pi}{5}$
and poles at $\frac{\pi}{5},\frac{2\pi}{5}$, and the order one zeroes
at $-\frac{\pi}{5}-\xi=\frac{4\pi}{5}-\xi$ and poles $\frac{\pi}{5}-\xi$,
$\frac{2\pi}{5}-\xi$. Here $\xi=ib$, is a pure imaginary parameter.
We use the same 
\[
f(u)=-\tan(\frac{5u}{6}+\frac{\pi}{6})\tan(\frac{5u}{6}+2\frac{\pi}{6})
\]

and write the transfer matrix in the form
\[
t(u)=f(u)^{N-1}f(u+\xi)l(u)
\]
We introduce the variable $u=\frac{3\pi}{10}+\frac{3ix}{5}$ and solve
functional equation 
\[
t(x-i\frac{\pi}{3})t(x+i\frac{\pi}{3})=1+t(x)
\]

\subsubsection*{Vacuum state}

Since this state has no zeros in the physical strip we get that: 
\begin{equation}
\log t(x)=(N-1)\log f(x)+\log f(x-\frac{5i\xi}{3})-\varphi\star\log(1+t^{-1}(x))
\end{equation}
This is the ground-state TBA on the lattice with a seam. 

In the thermodynamic limit $u$ scales as $\pm i\frac{3}{5}\log N$
(upper or lower half plane), hence in the $x$ variable we focus on
the behavior around $\pm\log N$. We can also play with the parameter
$\xi$. If we do not scale it in the thermodynamic limit it simply
disappears from the equations. If we scale with $\mp i\frac{3}{5}\log N$
it will appear in the equation for $\epsilon^{\pm}$ only, respectively.
Let us focus on the plus sign: so $\xi=-i\frac{3}{5}(\tilde{\xi}+\log N$).
Now we center the new functions around $x\pm\log N$ as $e^{\epsilon^{\pm}(x)}=\underset{N\to\infty}{\lim}t(x\pm\log N)$.
Taking the continuum limit ($N\to\infty$) on the source term we get
the massless ground-state TBA equations in the presence of a seam:
\begin{equation}
\epsilon^{\pm}(x)=4\sin\frac{\pi}{3}e^{\mp x}+\log f(x-\tilde{\xi})-\varphi\star\log(1+e^{-\epsilon^{\pm}(x)})
\end{equation}
This agrees with the scaled bulk TBA equation as 
\[
f(x-\tilde{\xi})=T_{-}(\frac{i\pi}{2}+x)\quad;\quad b=3+\frac{6i\tilde{\xi}}{\pi}
\]

where $\ \ \ T_{-}(x)=[-1+b][1+b]\ \ \ $ with $\ \ \ [s]=i\frac{\sinh(\frac{x}{2}+i\pi\frac{s}{12})}{\sinh(\frac{x}{2}+i\pi\frac{s}{12}-i\frac{\pi}{2})}\ \ $

$T_{-}$ is the transmission matrix in the continuum theory with a
defect for $b=3+\frac{6i\tilde{\xi}}{\pi}$ as we expect from \cite{BS}.

\subsubsection*{Excited states}

For large $\xi$, where the flow of states has already occurred to
the (1,2) module, we might have 1-strings at 
\[
u=\frac{3\pi}{10}+i\alpha
\]
 and short 2-strings at 
\[
u=\begin{cases}
\frac{\pi}{5}+i\beta_{j}\\
\frac{2\pi}{5}+i\beta_{j}
\end{cases}
\]
For finite energy states in the continuum ($N\to\infty$) limit they
go to infinity as $\alpha=\frac{3}{5}(\pm\log N+\tilde{\alpha}^{\pm})$
and $\beta_{j}=\frac{3}{5}(\pm\log N+\tilde{\beta}_{j}^{\pm})$, but
here $\alpha$ occurs either in the upper or in the lower half plane,
and $\beta_{i}^{\pm}$ need not be symmetric with respect to the real
axis.

In the $x=\frac{5}{3i}(u-\frac{3\pi}{10})$ variable they are located
at 
\[
\begin{array}{cc}
x_{0}^{\pm}=\frac{5\alpha}{3}=\log N+\tilde{\alpha}^{\pm}\\
x_{j}^{\pm}=\pm i\frac{\pi}{6}+\log N+\tilde{\beta}_{j}^{\pm}
\end{array}
\]
Again, it is important to note that the $\pm$ sign here on $x$ is
to indicate whether the zero is occurring in the upper or the lower
half plane, and doesn't indicate symmetry with respect to the real
axis. Similar to the periodic case, to take logarithm we need functions
free of zeros and poles on the physical strip. The function which
can eliminate the single zero is $\ \sigma_{0}(x)=\tanh\left(\frac{3}{4}x\right)\ $
while the one which eliminates the two zeros at $\pm\frac{\pi}{6}$
is $\ \sigma_{1}(x)=f(x)^{-1}=\frac{\cosh x+\sin\frac{\pi}{3}}{\cosh x-\sin\frac{\pi}{3}}\ $
. 

With these functions we parametrize the normalized transfer matrix
eigenvalue as 
\begin{equation}
t(x)=f(x)^{N}\sigma_{0}(x-x_{0}^{\pm})\prod_{j=1}^{M}\sigma_{1}(x-x_{j}^{\pm})l(x)
\end{equation}

Here, there is no product over $\sigma_{0}$ as it only occurs once,
either in the upper or in the lower half planes. With this parametrization,
$l(x)$ is ANZ in the physical strip. Following the same derivation
of the excited periodic states, we find:
\[
\log l(x)=-\varphi\star\log(\sigma_{0}(x-x_{0}^{\pm})(1+t^{-1}(x)))
\]
Restoring $t$ we obtain
\begin{equation}
\begin{array}{cc}
\log t(x)=(N-1)\log f(x)+\log f(x-\frac{5i\xi}{3})+\log(\sigma_{0}(x-x_{0}^{\pm}))\\
+\sum_{j}\log(\sigma_{1}(x-x_{j}^{\pm}))-\varphi\star\log(\sigma_{0}(x-x_{0}^{\pm})(1+t^{-1}(x)))
\end{array}
\end{equation}
 In the scaling limit we, we repeat same scalings as before and we
get that $\ e^{\epsilon^{\pm}(x)}=\underset{N\rightarrow\infty}{\lim}\sigma_{0}(x\pm\log N-x_{0}^{\pm})^{-1}t(x+\log N)\ $
satisfies the equation:

\begin{equation}
\begin{array}{cc}
\epsilon^{\pm}(x)=4\sin\frac{\pi}{3}e^{\mp x}+\log f(x-\tilde{\xi})+\sum_{j}\log\sigma_{1}(x-\tilde{\beta}_{j}^{\pm})\\
-\varphi\star\log(\sigma_{0}(x-\tilde{\alpha}^{\pm})+e^{-\epsilon^{\pm}(x)})
\end{array}
\end{equation}

where, as mentioned before $\ f(x-\tilde{\xi})=T_{-}(\frac{i\pi}{2}+x)\quad;\quad b=3+\frac{6i\tilde{\xi}}{\pi}\ $and
the parameters satisfy the following equations $e^{\epsilon^{\pm}(x)}\sigma_{0}(x-\tilde{\alpha}^{\pm})\vert_{x=\tilde{x}_{i}\pm\frac{i\pi}{3}}=-1$.

\subsubsection{Boundary}

It is convenient to define the normalized transfer matrix for the
boundary double row transfer matrix as 
\[
t(u)=S_{s}(u)\,\frac{s^{2}(2u-\lambda)}{s(2u+\lambda)s(2u-3\lambda)}\Bigl[\frac{s(\lambda)s(u+2\lambda)}{s(u+\lambda)s(u+3\lambda)}\Bigr]^{N}D(u)
\]
 and 
\begin{eqnarray*}
S_{s}(u) & = & \begin{cases}
1, & s=1\\
\frac{s(\lambda)^{2}s(u+\xi+2\lambda)s(u-\xi+2\lambda)}{s(u+\xi+\lambda)s(u+\xi+3\lambda)s(u-\xi+3\lambda)s(u-\xi+\lambda)} & s=2
\end{cases}
\end{eqnarray*}
 where $N$ is even for $s=1$ and odd for $s=2$. It can then be
shown~\cite{BPO96} that the normalized transfer matrix satisfies
the universal TBA functional equation 
\begin{equation}
t(u)t(u+\lambda)=1+t(u+3\lambda)\label{e:functional-1-1}
\end{equation}

Using the periodicity $t(u)=t(u+\pi)$ we rewrite it as $t(u)t(u-\frac{2\pi}{5})=1+t(u-\frac{\pi}{5})$
and after shifting $u$ we have
\[
t(u+\frac{\pi}{5})t(u-\frac{\pi}{5})=1+t(u)
\]

We write 
\[
t(u)=f(u)^{2N}g(u)l(u)
\]

where $f^{2N}$ is the bulk free energy (order $N$ term), $g$ is
the boundary energy (order 1 term) and $l$ corresponds to the finite
size corrections (order $\frac{1}{N}$ term). Similarly as before
we would like to kill the $N$th order zeros at $-\frac{\pi}{5},\frac{4\pi}{5}$
and poles at $\frac{\pi}{5},\frac{2\pi}{5}$ by $f(u)$ which satisfies
$f(u+\frac{\pi}{5})f(u-\frac{\pi}{5})=f(u)$ where the solution $f(u)$
compatible with the analytical structure is 
\[
f(u)=\frac{\sin\frac{5u}{3}+\sin\frac{\pi}{3}}{\sin\frac{5u}{3}-\sin\frac{\pi}{3}}=-\tan(\frac{5u}{6}+\frac{\pi}{6})\tan(\frac{5u}{6}+\frac{2\pi}{6})
\]
which is basically the shifted S-matrix. 

Introducing the variable $u=\frac{3\pi}{10}+\frac{3ix}{5}$ we write
the functional equation as 
\begin{equation}
t(x-i\frac{\pi}{3})t(x+i\frac{\pi}{3})=1+t(x)\label{eq:frel}
\end{equation}

and in this variable, 
\[
f(x)=-\tan(\frac{ix}{2}+\frac{5\pi}{12})\tan(\frac{ix}{2}+\frac{7\pi}{12})
\]

\subsubsection*{(r,s)=(1,1) sector}

We also want to eliminate the zeroes of order 1 which originate from
the normalization and from the analytical strip.

For $s=1$, $S_{s}(u)=1$ , hence it has no contribution here. Hence
the only contributions come from a double zero at $u=\frac{\lambda}{2}$
and poles at $u=\frac{-\lambda}{2}=\frac{-3\pi}{10}=\frac{7\pi}{10}$
and $u=\frac{3\lambda}{2}=\frac{9\pi}{10}=\frac{-\pi}{10}$. Due to
$\frac{\pi}{2}$ periodicity, it also has poles at $\frac{\pi}{5}$
and $\frac{2\pi}{5}$. Also from the analytic structure there are
zeroes occurring at $\frac{\pi}{10}$ and $\frac{\pi}{2}$. The factor
that we need to insert and compatible with the analytical structure
is:

\[
g_{1,1}(u)=-\frac{\tan^{2}(\frac{5u}{6}-\frac{\pi}{4})\tan(\frac{5u}{6}-\frac{\pi}{12})\tan(\frac{5u}{6}-5\frac{\pi}{12})}{\tan(\frac{5u}{6}-\frac{\pi}{6})\tan(\frac{5u}{6}-\frac{\pi}{3})}
\]
In terms of the $x$ variable we get: 
\[
g_{1,1}(x)=-\frac{\tan^{2}(\frac{ix}{2})\tan(\frac{ix}{2}+\frac{\pi}{6})\tan(\frac{ix}{2}-\frac{\pi}{6})}{\tan(\frac{ix}{2}-\frac{\pi}{12})\tan(\frac{ix}{2}+\frac{\pi}{12})}
\]

where $g_{1,1}(x)$ satisfies the relation 
\begin{equation}
g_{1,1}(x-i\frac{\pi}{3})g_{1,1}(x+i\frac{\pi}{3})=g_{1,1}(x)
\end{equation}

Here it is important to mention that $g_{1,1}(x)=R_{1}^{-1}(\frac{i\pi}{2}+x)R_{1}^{-1}(\frac{i\pi}{2}-x)$
where $R_{1}$ is the reflection matrix of the continuum boundary
theory of the Lee-Yang model corresponding to the identity module,
defined by $R_{1}(x)=(\frac{1}{2})(\frac{3}{2})(-2)$ and $(s)=\frac{\sinh(\frac{x}{2}+i\frac{\pi s}{6})}{\sinh(\frac{x}{2}-i\frac{\pi s}{6})}$.
This will be important when we analyze the massive case.

In addition, in this sector we always have one strings at 
\[
u=\frac{3\pi}{10}+i\alpha
\]
 and short strings at 
\[
u=\begin{cases}
\frac{\pi}{5}+i\beta_{j}\\
\frac{2\pi}{5}+i\beta_{j}
\end{cases}
\]
For finite energy states in the continuum ($N\to\infty$) limit they
go to infinity as $\alpha=\frac{3}{5}(\pm\log N+\tilde{\alpha}^{\pm})$
and $\beta_{j}=\frac{3}{5}(\pm\log N+\tilde{\beta}_{j}^{\pm})$ on
the upper/lower half plane, respectively. In the boundary case $\tilde{\alpha}^{-}=-\tilde{\alpha}^{+}$,
and $\tilde{\beta}^{-}=\tilde{\beta}^{+}$, as we can see from their
symmetry with respect to the real axis.

In the $x=\frac{5}{3i}(u-\frac{3\pi}{10})$ variable they are located
at $x_{0}^{\pm}=\frac{5\alpha}{3}=\log N+\tilde{\alpha}^{\pm}$and
at $x_{j}^{\pm}=\pm i\frac{\pi}{6}+\log N+\tilde{\beta}_{j}^{\pm}$.
As we would like to take logarithm we need functions free of zeros
and poles on the physical strip. Those functions were defined before
as $\sigma_{0}(x)$ and $\sigma_{1}(x)$. With these functions we
parametrize the normalized transfer matrix eigenvalue as 
\begin{equation}
t(x)=f(x)^{2N}g_{1,1}(x)\prod_{\pm}\sigma_{0}(x-x_{0}^{\pm})\prod_{j=1}^{M}\sigma_{1}(x-x_{j}^{\pm})l(x)
\end{equation}
which ensures that $l(x)$ is ANZ in the physical strip. The functional
equation then takes the form
\[
\prod_{\pm}\sigma_{0}(x-x_{0}^{\pm})\frac{t(x-i\frac{\pi}{3})t(x+i\frac{\pi}{3})}{t(x)}=\frac{l(x-i\frac{\pi}{3})l(x+i\frac{\pi}{3})}{l(x)}=\prod_{\pm}\sigma_{0}(x-x_{0}^{\pm})(1+t^{-1}(x))
\]
Clearly both sides are ANZ in the interior of the physical strip:
the combination $\prod_{\pm}\sigma_{0}(x-x_{0}^{\pm})t^{-1}(x)$ is
regular and non-zero at $x=x_{0}^{\pm}.$ Taking the logarithm and
going to Fourier space we find:
\[
\log l(x)=-\varphi\star\log(\prod_{\pm}\sigma_{0}(x-x_{0}^{\pm})(1+t^{-1}(x)))
\]

where the convolution $\star$ was defined in equation (\ref{eq:convolution}).
Restoring $t$ we obtain

\begin{equation}
\begin{array}{cc}
\log t(x)=2N\log f(x)+\log g_{1,1}(x)+\sum_{\pm}\log(\sigma_{0}(x-x_{0}^{\pm}))+\sum_{j,\pm}\log(\sigma_{1}(x-x_{j}^{\pm}))\\
-\varphi\star\log(\prod_{\pm}\sigma_{0}(x-x_{0}^{\pm})(1+t^{-1}(x)))
\end{array}
\end{equation}
The parameters are determined by $t(x)\vert_{x=x_{i}\pm\frac{i\pi}{3}}=-1$.

In the scaling limit we can focus on the two scaling domains at $\pm\log N$
. We have calculated the scaling limits of $f,\ \sigma_{0}(x-x_{0}^{\pm})$and
$\sigma_{1}(x-x_{j}^{\pm})$ before, with the new function $g_{1,1}(x)$
scaling to 1 in the massless case, hence disappearing from the scaling
TBA. Using 
\begin{equation}
e^{\epsilon^{\pm}(x)}=\lim_{N\to\infty}\sigma_{0}(x\pm\log N-x_{0}^{\pm})^{-1}t(x+\log N)
\end{equation}

The parameters satisfy $e^{\epsilon^{\pm}(x)}\sigma_{0}(x-\tilde{\alpha}^{\pm})\vert_{x=\tilde{x}_{i}\pm\frac{i\pi}{3}}=-1$.

And $\epsilon(x)$ satisfies the equation
\begin{equation}
\epsilon^{\pm}(x)=8\sin\frac{\pi}{3}e^{\mp x}+\sum_{j}\log\sigma_{1}(x-\tilde{\beta}_{j}^{\pm})-\varphi\star\log(\sigma_{0}(x-\tilde{\alpha}^{\pm})+e^{-\epsilon^{\pm}(x)})
\end{equation}

This is the general massless TBA in the (1,1) sector. It is important
to mention that for the groundstate of this sector, there are no short
2-strings, hence $\sigma_{1}$ doesn't appear and the ground state
massless TBA of this sector on the lattice is:

\begin{equation}
\begin{array}{cc}
\log t(x)=2N\log f(x)+\log g_{1,1}(x)+\sum_{\pm}\log(\sigma_{0}(x-x_{0}^{\pm}))\\
-\varphi\star\log(\prod_{\pm}\sigma_{0}(x-x_{0}^{\pm})(1+t^{-1}(x)))
\end{array}
\end{equation}

while the scaled ground-state massless TBA is
\begin{equation}
\epsilon^{\pm}(x)=8\sin\frac{\pi}{3}e^{\mp x}+-\varphi\star\log(\sigma_{0}(x-\tilde{\alpha}^{\pm})+e^{-\epsilon^{\pm}(x)})
\end{equation}

\subsubsection*{(r,s)=(1,2) sector}

Similar analysis follows in this sector, where we have to take into
account the contribution of $S_{s}(u)$.

For $s=2$ , $S_{2}(u)$ is proportional to $f(u+\xi)f(u-\xi)$, therefore
we should account for their respective zeroes and poles.

The other contributions from the normalization and the analytic strip
are the same as before, and they were included in $g_{1,1}(u)$. Then
we need to define the new $g_{1,2}$ term as:

\[
g_{1,2}(u)=g_{1,1}(u)\times f(u+\xi)f(u-\xi)
\]
In terms of the $x$ variable we get: 
\begin{equation}
g_{1,2}(x)=-\frac{\tan^{2}(\frac{ix}{2})\tan(\frac{ix}{2}+\frac{\pi}{6})\tan(\frac{ix}{2}-\frac{\pi}{6})}{\tan(\frac{ix}{2}-\frac{\pi}{12})\tan(\frac{ix}{2}+\frac{\pi}{12})}f(x-\frac{5i\xi}{3})f(x+\frac{5i\xi}{3})
\end{equation}

where $g_{1,2}(x)$ satisfies the relation 
\[
g_{1,2}(x-i\frac{\pi}{3})g_{1,2}(x+i\frac{\pi}{3})=g_{1,2}(x)
\]

Again we note that $g_{1,2}(x)=R_{\phi}^{-1}(\frac{i\pi}{2}+x)R_{1}^{-1}(\frac{i\pi}{2}-x)$
where $R_{1}$ and $R_{\phi}$ are the reflection matrices of the
continuum boundary theory of the Lee-Yang model corresponding to the
$Id$ and $\phi$ modules of the Virasoro algebra with highest weights
$0$ and $\frac{-1}{5}$ respectively. They are defined by 

\[
\begin{array}{cc}
R_{1}(x)=(\frac{1}{2})(\frac{3}{2})(-2)\\
R_{\phi}=(\frac{1}{2})(\frac{3}{2})(-2)(\frac{b-1}{2})(\frac{b+1}{2})(\frac{5-b}{2})(\frac{-5-b}{2})
\end{array}
\]
 in the case where $b=3-\frac{10\xi}{\pi}$. This will also be important
in the analysis of the massive case.

In addition, in this sector, in general we will have short 2-strings
( no 1-strings) located at: 
\[
u=\begin{cases}
\frac{\pi}{5}+i\beta_{j}\\
\frac{2\pi}{5}+i\beta_{j}
\end{cases}
\]

In the $x=\frac{5}{3i}(u-\frac{3\pi}{10})$ variable they are located
at $x_{j}^{\pm}=\pm i\frac{\pi}{6}+\log N+\tilde{\beta}_{j}^{\pm}$. 

We parametrize the normalized transfer matrix eigenvalue as 
\begin{equation}
t(x)=f(x)^{2N}g_{1,2}(x)\prod_{j=1}^{M}\sigma_{1}(x-x_{j}^{\pm})l(x)
\end{equation}
which ensures that $l(x)$ is ANZ in the physical strip. The functional
equation then takes the form
\[
\frac{t(x-i\frac{\pi}{3})t(x+i\frac{\pi}{3})}{t(x)}=\frac{l(x-i\frac{\pi}{3})l(x+i\frac{\pi}{3})}{l(x)}=(1+t^{-1}(x))
\]
Both sides are ANZ in the interior of the physical strip. Taking then
logarithm and going to Fourier space we find:
\[
\log l(x)=-\varphi\star\log(1+t^{-1}(x))
\]
Restoring $t$ we obtain
\begin{equation}
\log t(x)=2N\log f(x)+\log g_{1,2}(x)+\sum_{j,\pm}\log(\sigma_{1}(x-x_{j}^{\pm}))-\varphi\star\log(1+t^{-1}(x))
\end{equation}
The parameters are determined by $t(x)\vert_{x=x_{i}\pm\frac{i\pi}{3}}=-1$.

In the scaling limit we again focus on the two scaling domains at
$\pm\log N$. We have calculated the scaling limits of those functions
before, with the new function $g_{1,2}(x)$ scaling to 1 in the massless
case for finite $\xi$. However, if we scale $\xi\rightarrow-i\frac{3}{5}(\tilde{\xi}+\log N)$,
and use $e^{\epsilon^{\pm}(x)}=\lim_{N\to\infty}t(x+\log N)$, it
satisfies the equation
\begin{equation}
\epsilon^{\pm}(x)=8\sin\frac{\pi}{3}e^{\mp x}+\log g_{1,2}(x-\tilde{\xi})+\sum_{j}\log\sigma_{1}(x-\tilde{\beta}_{j}^{\pm})-\varphi\star\log(1+e^{-\epsilon^{\pm}(x)})
\end{equation}

This is the general TBA in the (1,2) sector. For the groundstate of
this sector, there are no short 2-strings, hence $\sigma_{1}$ doesn't
appear and the ground state massless TBA of this sector on the lattice
is:

\begin{equation}
\log t(x)=2N\log f(x)+\log g_{1,2}(x)-\varphi\star\log(1+t^{-1}(x))
\end{equation}

while the scaled ground-state massless TBA is
\begin{equation}
\epsilon^{\pm}(x)=8\sin\frac{\pi}{3}e^{\mp x}+\log g_{1,2}(x-\tilde{\xi})-\varphi\star\log(1+e^{-\epsilon^{\pm}(x)})
\end{equation}

\subsection{Off-Critical/Massive TBAs:}

\subsubsection{Periodic boundary conditions}

In the massive description we have to make the following replacements
$\ \sin u\to\vartheta_{1}(u,q)$ and $\cos u\to\vartheta_{2}(u,q)$.
We solve the functional relations:
\[
t'(u)t'(u+\lambda)=1+t'(u+3\lambda)
\]
where the off-critical transfer matrix is double-periodic

\begin{equation}
t'(u)=t(u+\pi)\quad;\qquad t'(u+i\pi\epsilon)=C\cdot t'(u)\quad;\quad q=e^{-\pi\epsilon'}
\end{equation}

This means that $\partial_{u}\log t'(u)$ is also periodic. The prime
indicates that the expressions correspond to the massive case whose
functions are determined in terms of elliptic thetas to differentiate
them from the critical expressions used before which were trigonometric
functions.

Now we rewrite the functional relation as $t'(u)t'(u-\frac{2\pi}{5})=1+t'(u-\frac{\pi}{5})\ $,
and after shifting $u$ we get
\[
t'(u+\frac{\pi}{5})t'(u-\frac{\pi}{5})=1+t'(u)
\]
Similar to the critical case, we need to kill the $N$th order zeros
at $-\frac{\pi}{5},\frac{4\pi}{5}$ and poles at $\frac{\pi}{5},\frac{2\pi}{5}$
by $f'(u)$ which satisfies
\begin{equation}
f'(u+\frac{\pi}{5})f'(u-\frac{\pi}{5})=f'(u)\quad;\qquad f'(u+i\pi\epsilon)=f'(u)
\end{equation}
The resulting function $l'(u)$ will be analytical and nonzero in
the required domain. So we define 
\[
t'(u)=f'(u)^{N}l'(u)
\]
The solution for $f'(u)$ compatible with the analytical structure
is
\begin{equation}
f'(u,q)=-\frac{\vartheta_{1}(\frac{5u}{6}+\frac{\pi}{6},p)\vartheta_{1}(\frac{5u}{6}+\frac{2\pi}{6},p)}{\vartheta_{2}(\frac{5u}{6}+\frac{\pi}{6},p)\vartheta_{2}(\frac{5u}{6}+\frac{2\pi}{6},p)}
\end{equation}
where the periodicity requires $p=q^{\frac{6}{5}}.$ Introducing the
variable $u=\frac{3\pi}{10}+\frac{3ix}{5}$, we write the functional
equation as 
\begin{equation}
t'(x-i\frac{\pi}{3})t'(x+i\frac{\pi}{3})=1+t'(x)\label{eq:ttt}
\end{equation}
The periodicity box in the variable $x$ is $\Re e(x)\in(-\pi\epsilon\frac{5}{12},\pi\epsilon\frac{5}{12})$
and $\Im m(x)\in(-\frac{\pi}{3},\frac{\pi}{3})$, which is the analogue
of the physical strip.

\subsubsection*{Vacuum state}

We divide \ref{eq:ttt} by $t$ and use the functional relation to
write 
\[
\frac{t'(x-i\frac{\pi}{3})t'(x+i\frac{\pi}{3})}{t'(x)}=\frac{l'(x-i\frac{\pi}{3})l'(x+i\frac{\pi}{3})}{l'(x)}=1+t'^{-1}(x)
\]
After taking log (both sides are ANZ in the physical box) we solve
it in Fourier space. The functions are periodic with period $x\equiv x+V$,
where $V=\frac{5\pi\epsilon}{3}$. So we expand them as 
\[
h(x)=\sum_{k}e^{i\omega kx}h_{k}\quad;\qquad h_{k}=\frac{1}{V}\int_{-\frac{V}{2}}^{\frac{V}{2}}\, h(x)e^{-i\omega kx}dx\qquad;\qquad\omega=\frac{2\pi}{V}
\]
and solve the equation for $\log l'(x)$ as 
\[
\log l'_{k}=\frac{\log(1+t'^{-1})_{k}}{e^{\frac{\pi}{3}\omega k}+e^{-\frac{\pi}{3}\omega k}-1}
\]
our in real space 
\begin{equation}
\log l'(x)=-\varphi_{\epsilon}\star\log(1+t'^{-1}(x)):=\frac{1}{V}\int_{-\frac{V}{2}}^{\frac{V}{2}}dx\varphi_{\epsilon}(x-x')\log(1+t'^{-1}(x'))\label{eq:massiveconv}
\end{equation}
 where 
\begin{equation}
\varphi_{\epsilon}(x)=\sum_{k}\frac{e^{i\omega kx}}{1-e^{\frac{\pi}{3}\omega k}-e^{-\frac{\pi}{3}\omega k}}\label{eq:massivephi}
\end{equation}

Now we restore $t'$: 
\begin{equation}
\log t'(x)=N\log f'(x)-\varphi_{\epsilon}\star\log(1+t'^{-1}(x))
\end{equation}
This is the ground-state massive TBA on the lattice. 

We take the continuum limit such that we scale $x\to x+\log N$ and
$N\to\infty$:

\[
\lim_{N\to\infty}N\log f'(\frac{3\pi}{10}+\frac{3i(x+\log N)}{5},p)
\]
In order to have a finite limit we send $p\to0$ and obtain
\[
\lim_{N\to\infty}N\log f'(\frac{3\pi}{10}+\frac{3i(x+\log N)}{5},p)=2\sqrt{3}(e^{-x}+N^{2}p^{2}e^{x})+O(p^{4}N^{3})
\]
If we choose 
\begin{equation}
p=\frac{mL}{4\sqrt{3}}\frac{1}{N}=\frac{\mu}{N}
\end{equation}

This is equivalent to shifting $x$ as $x\rightarrow x-\log\mu.$Then
after shifting the $x$ variable we obtain the standard massive TBA
on the lattice: 
\begin{equation}
\log t'(x)=mL\cosh(x)-\varphi\star\log(1+t'^{-1}(x))
\end{equation}
as in this limit $\epsilon\to\infty$ and so $\varphi_{\epsilon}\to\varphi$. 

If we define 
\[
e^{\epsilon(x)}=\lim_{N\to\infty}t'(x+\log N-\log\mu)
\]

Then it satisfies the massive periodic TBA

\begin{equation}
\epsilon(x)=mL\cosh x-\varphi\star\log(1+e^{-\epsilon(x)})
\end{equation}

\subsubsection*{Excited states}

We repeat the same analysis as in the critical case and indicate the
important and new elements in the analysis. From the numerics, we
assume the existence of single real zeros and complex conjugated short
strings. However, contrary to the critical case we do not assume that
they are located exactly at $\pm\frac{i\pi}{6}$. The relevant functions
to eliminate these zeros with the required periodicity are

\begin{equation}
\begin{array}{cc}
\sigma'_{0}(x,r)=i\frac{\vartheta_{1}(\frac{3x}{4i},r)}{\vartheta_{2}(\frac{3x}{4i},r)}\quad;\qquad r=q^{\frac{4}{5}}\\
\sigma'_{1}(x,p)=-\frac{\vartheta_{2}(\frac{ix}{2}+\frac{5\pi}{12},p)\vartheta_{2}(\frac{ix}{2}+\frac{7\pi}{12},p)}{\vartheta_{1}(\frac{ix}{2}+\frac{5\pi}{12},p)\vartheta_{1}(\frac{ix}{2}+\frac{7\pi}{12},p)}
\end{array}
\end{equation}

$\sigma'_{0}(x,r)$ has a single zero at $x=0$ within the periodicity
box, while $\sigma'_{1}(x,q)$ has one at $i\frac{\pi}{6}$ and another
at $-i\frac{\pi}{6}$. So eliminating singularities on the upper half
plane at $\beta_{i}$ ($\Re e(\beta_{i})>0$) we use $\sigma'_{1}(x-\beta_{i}-\frac{i\pi}{6},p)$
while on the lower half plane at $\beta_{i}$ ($\Re e(\beta_{i})<0$)
we use $\sigma'_{1}(x-\beta_{i}+\frac{i\pi}{6},p)$. A short two string
with locations $\beta_{k}=u_{k}\pm iv_{k}$ can be eliminated by 
\begin{equation}
\sigma'_{1}(x-u_{k}-\frac{i\pi}{6}-iv_{k},p)\sigma'_{1}(x-u_{k}+iv_{k}+\frac{i\pi}{6},p)=\sigma'_{2}(x-u_{k},v_{v},p)
\end{equation}
For a state with one-strings at $x_{0}^{\pm}$ and short two-strings
at $x_{j}^{\pm}=(u_{j}^{\pm},v_{j}^{\pm})$, the calculation analogous
to the massless case leads to

\begin{equation}
\begin{array}{cc}
\log t'(x)=N\log f'(x)+\sum_{\pm}\log(\sigma'_{0}(x-x_{0}^{\pm},r))+\sum_{j,\pm}\log(\sigma'_{2}(x-u_{j}^{\pm},v_{j}^{\pm},p))\\
-\varphi_{\epsilon}\star\log(\prod_{\pm}\sigma'_{0}(x-x_{0}^{\pm},r)(1+t'^{-1}(x)))
\end{array}
\end{equation}

We do the continuum scaling as we did in the vacuum case while, in
addition, we suppose that the one string and the two strings scale
as $x_{k}\to x_{k}\mp\log\frac{N}{\mu}$ , then we have to take into
account that in the scaling limit

\begin{equation}
\begin{array}{cc}
\underset{N\to\infty}{\lim}\sigma'_{0}(x\pm\log\frac{N}{\mu}-x_{0}^{\pm},r)^{-1}=\sigma_{0}(x-\tilde{\alpha}_{j})\\
\underset{N\to\infty}{\lim}\sigma'_{2}(x\pm\log\frac{N}{\mu}-u_{k},v_{v},p)=\sigma_{2}(x-u_{k},v_{v})
\end{array}
\end{equation}

where we retain the trigonometric expressions $\sigma_{0}(x)=\tanh(\frac{3x}{4})$
and $\sigma_{1}(x)=-\tan(\frac{ix}{2}-\frac{\pi}{12})\tan(\frac{ix}{2}+\frac{\pi}{12})$,
and define 

\[
\sigma_{2}(x-u_{k},v_{v})=\sigma_{1}(x-u_{k}-\frac{i\pi}{6}-iv_{k})\sigma_{1}(x-u_{k}+iv_{k}+\frac{i\pi}{6})
\]

we can define the function
\[
e^{\epsilon(x)}=\lim_{N\to\infty}\sigma'_{0}(x+\log\frac{N}{\mu}-x_{0}^{\pm})^{-1}t'(x+\log\frac{N}{\mu})
\]

and it satisfies the equation
\begin{equation}
\epsilon(x)=mL\cosh x+\sum_{j}\log\sigma_{2}(x-\tilde{u}_{j},\tilde{v}_{j})-\varphi\star\log(\sigma_{0}(x-\tilde{\alpha}_{j})+e^{-\epsilon(x)})
\end{equation}
The parameters as before, satisfy the following equations $e^{\epsilon(x)}\sigma_{0}(x)\vert_{x=\tilde{x}_{i}\pm\frac{i\pi}{3}}=-1$.

This is the massive excited state for the Lee-Yang model with periodic
boundary conditions.

\subsubsection{Seam}

In a similar derivation to the critical periodic model with a seam,
we derive the massive one here. The periodic transfer matrix in presence
of a seam satisfies the following relation:

{\footnotesize{}
\[
T_{1}(u)T_{1}(u+\lambda)=\frac{s(u+\lambda+\xi)s(\lambda-u-\xi)}{s^{2}(\lambda)}\left(\frac{s(u+\lambda)s(\lambda-u)}{s^{2}(\lambda)}\right)^{N-1}\mathbb{I}+\frac{s(u+\xi)}{s(\lambda)}\left(\frac{s(u)}{s(\lambda)}\right)^{N-1}T_{2}(u)
\]
}{\footnotesize \par}

with

\[
t'(u)=\frac{s(\lambda)s(u+2\lambda+\xi)}{s(u+\lambda+\xi)s(u+3\lambda+\xi)}\left(\frac{s(\lambda)s(u+2\lambda)}{s(u+\lambda)s(u+3\lambda)}\right)^{N-1}T_{1}(u)
\]

we obtain the relation

\[
t'(u)t'(u+\lambda)=1+t'(u+3\lambda)
\]
Then, writing $t'(u)=f'^{N-1}(u)f'(u+\xi)l'(u)$, and killing the
zeroes of the normalization, exactly as we did before and making the
change of variables $u=\frac{3\pi}{10}+\frac{3ix}{5}$, we obtain
$ $ 
\begin{equation}
f'(x)=-\frac{\vartheta_{1}(\frac{ix}{2}+\frac{5\pi}{12},p)\vartheta_{1}(\frac{ix}{2}+\frac{7\pi}{12},p)}{\vartheta_{2}(\frac{ix}{2}+\frac{5\pi}{12},p)\vartheta_{2}(\frac{ix}{2}+\frac{7\pi}{12},p)}
\end{equation}

and the periodicity requires that $p=q^{\frac{6}{5}}$ as before.

\subsubsection*{Vacuum state}

Since this state has no zeros in the physical strip we get that: 
\begin{equation}
\log t'(x)=(N-1)\log f'(x)+\log f'(x-\frac{5i\xi}{3})-\varphi_{\epsilon}\star\log(1+t'^{-1}(x))
\end{equation}

where the convolution $\star$ and $\varphi_{\epsilon}$ were defined
in (\ref{eq:massiveconv}) and (\ref{eq:massivephi}) respectively.
This is the massive ground-state TBA on the lattice with a seam. 

We take now the scaling limit. We calculated the scalings of those
functions before in the previous section. Only $\log f'(x-\frac{5i\xi}{3})$
was not considered. As $x\to x+\log\frac{N}{\mu}$ and taking $\xi=-i\frac{3}{5}(\tilde{\xi}+\log\frac{N}{\mu}$),
we get that the continuum limit of $f'(x-\frac{5i\xi}{3})$ is exactly
its trigonometric form $f(x-\tilde{\xi)}$ where

\[
f(x)=-\tan(\frac{ix}{2}+\frac{5\pi}{12})\tan(\frac{ix}{2}+\frac{7\pi}{12})
\]

Defining $e^{\epsilon(x)}=\underset{N\rightarrow\infty}{\lim}t'(x+\log\frac{N}{\mu})$
we get

\begin{equation}
\epsilon(x)=mL\cosh x+\log f(x-\tilde{\xi})-\varphi\star\log(1e^{-\epsilon(x)})
\end{equation}

In other words, the ground-state massive seam TBA becomes:

\begin{equation}
\epsilon(x)=mL\cosh x+\log T_{-}(\frac{i\pi}{2}+x)-\varphi\star\log(1+e^{-\epsilon(x)})
\end{equation}

\subsubsection*{Excited states:}

To eliminate the zeros of the 1-string and short 2-strings with the
required periodicity, we use $\sigma'_{0}(x,r)$ and $\sigma'_{1}(x,p)$.
As before a short two string with locations $\beta_{k}=u_{k}\pm iv_{k}$
can be eliminated by 
\begin{equation}
\sigma'_{1}(x-u_{k}-\frac{i\pi}{6}-iv_{1k},q)\sigma'_{1}(x-u_{k}+iv_{2k}+\frac{i\pi}{6},q)=\sigma'_{2}(x-u_{k},v_{v},q)
\end{equation}
where $v_{1}$ and $v_{2}$ are not necessarily equal. 

Now suppose in the state we have one string at $x_{0}^{\pm}$ (either
in the upper or in the lower half plane) and short two strings at
$x_{j}^{\pm}=(u_{j}^{\pm},v_{j}^{\pm})$. A calculation analogous
to the massless case leads to 

\begin{equation}
\begin{array}{cc}
\log t'(x)=(N-1)\log f'(x)+\log f'(x-\tilde{\xi})+\sum_{\pm}\log(\sigma'_{0}(x-x_{0}^{\pm},r))+\\
\sum_{j,\pm}\log(\sigma'_{2}(x-u_{j}^{\pm},v_{j}^{\pm},p))-\varphi\star\log(\prod_{\pm}\sigma'_{0}(x-x_{0}^{\pm},r)(1+t'^{-1}(x)))
\end{array}
\end{equation}
We already know the scaling limit of this relation. Defining the function
\begin{equation}
e^{\epsilon(x)}=\lim_{N\to\infty}\sigma'_{0}(x\pm\log\frac{N}{\mu}-x_{0}^{\pm})^{-1}t'(x+\log\frac{N}{\mu})
\end{equation}
It satisfies the equation
\[
\epsilon(x)=mL\cosh x+\log f_{T}(x-\tilde{\xi})+\sum_{j}\log\sigma_{2}(x-\tilde{u}_{j},\tilde{v}_{j})-\varphi\star\log(\sigma_{0}(x-\tilde{\alpha}_{j})+e^{-\epsilon(x)})
\]
or using the relation derived before $f(x-\tilde{\xi})=T_{-}(\frac{i\pi}{2}+x)\quad;\quad b=3+\frac{6i\tilde{\xi}}{\pi}$
then

\begin{equation}
\epsilon(x)=mL\cosh x+\log T_{-}(\frac{i\pi}{2}+x)+\sum_{j}\log\sigma_{2}(x-\tilde{u}_{j},\tilde{v}_{j})-\varphi\star\log(\sigma_{0}(x-\alpha)+e^{-\epsilon(x)})
\end{equation}
which is the massive excited-state TBA of the Lee-Yang model with
a seam.

\subsubsection{Boundary}

Similar to the massless case we define: 
\[
t(u)=S_{s}(u)\,\frac{\vartheta_{1}^{2}(2u-\lambda,q)}{\vartheta_{1}(2u+\lambda,q)\vartheta_{1}(2u-3\lambda,q)}\Bigl[\frac{\vartheta_{1}(\lambda,q)\vartheta_{1}(u+2\lambda,q)}{\vartheta_{1}(u+\lambda,q)\vartheta_{1}(u+3\lambda,q)}\Bigr]^{N}D(u)
\]

with 
\begin{eqnarray*}
S_{s}(u) & = & \begin{cases}
1, & s=1\\
\frac{\vartheta_{1}(\lambda,q)^{2}\vartheta_{1}(u+\xi+2\lambda,q)\vartheta_{1}(u-\xi+2\lambda,q)}{\vartheta_{1}(u+\xi+\lambda,q)\vartheta_{1}(u+\xi+3\lambda,q)\vartheta_{1}(u-\xi+3\lambda,q)\vartheta_{1}(u-\xi+\lambda,q)} & s=2
\end{cases}
\end{eqnarray*}
 Then the normalized transfer matrix satisfies the universal TBA functional
equation 
\[
t(u)t(u+\lambda)=1+t(u+3\lambda)
\]

We write 
\[
t'(u)=f'(u)^{2N}g'(u)l'(u)
\]

To kill the $N$th order zeroes and poles, and after shifting $u$,
and with $p=q^{\frac{6}{5}}$we have

\[
f'(x)=-\frac{\vartheta_{1}(\frac{ix}{2}+\frac{5\pi}{12},\ p)}{\vartheta_{2}(\frac{ix}{2}+\frac{5\pi}{12},\ p)}\frac{\vartheta_{1}(\frac{ix}{2}+\frac{7\pi}{12},\ p)}{\vartheta_{2}(\frac{ix}{2}+\frac{7\pi}{12},\ p)}
\]

\subsubsection*{(r,s)=(1,1) sector:}

We also want to eliminate the zeroes of order 1 which originate from
the normalization and from the analytical strip. In a similar analysis
of the critical case, and taking into account the periodicity implied
by the normalization, we deduce that the factor that we need to insert
and compatible with the analytical structure is:

\[
g'_{1,1}(u)=-\frac{\vartheta_{1}^{2}(\frac{5u}{6}-\frac{\pi}{4},p^{\frac{1}{2}})\vartheta_{1}(\frac{5u}{6}-\frac{\pi}{12},p^{\frac{1}{2}})\vartheta_{1}(\frac{5u}{6}-\frac{5\pi}{12},p^{\frac{1}{2}})\vartheta_{2}(\frac{5u}{6}-\frac{\pi}{6},p^{\frac{1}{2}})\vartheta_{2}(\frac{5u}{6}-\frac{\pi}{3},p^{\frac{1}{2}})}{\vartheta_{2}^{2}(\frac{5u}{6}-\frac{\pi}{4},p^{\frac{1}{2}})\vartheta_{2}(\frac{5u}{6}-\frac{\pi}{12},p^{\frac{1}{2}})\vartheta_{2}(\frac{5u}{6}-\frac{5\pi}{12},p^{\frac{1}{2}})\vartheta_{1}(\frac{5u}{6}-\frac{\pi}{6},p^{\frac{1}{2}})\vartheta_{1}(\frac{5u}{6}-\frac{\pi}{3},p^{\frac{1}{2}})}
\]
In terms of the $x$ variable we get: 
\[
g'_{1,1}(x)=-\frac{\vartheta_{1}^{2}(\frac{ix}{2},p^{\frac{1}{2}})\vartheta_{1}(\frac{ix}{2}+\frac{\pi}{6},p^{\frac{1}{2}})\vartheta_{1}(\frac{ix}{2}-\frac{\pi}{6},p^{\frac{1}{2}})\vartheta_{2}(\frac{ix}{2}-\frac{\pi}{12},p^{\frac{1}{2}})\vartheta_{2}(\frac{ix}{2}+\frac{\pi}{12},p^{\frac{1}{2}})}{\vartheta_{2}^{2}(\frac{ix}{2},p^{\frac{1}{2}})\vartheta_{2}(\frac{ix}{2}+\frac{\pi}{6},p^{\frac{1}{2}})\vartheta_{2}(\frac{ix}{2}-\frac{\pi}{6},p^{\frac{1}{2}})\vartheta_{1}(\frac{ix}{2}-\frac{\pi}{12},p^{\frac{1}{2}})\vartheta_{1}(\frac{ix}{2}+\frac{\pi}{12},p^{\frac{1}{2}})}
\]

The priodicity of $g$ is half that of $f$, due to the fact that
the order 1 normalization is a function of $2u$, rather than $u$,
hence the period is divided by two, or equivalently the nome of $g$
is $p^{\frac{1}{2}}.$ 

$g'_{1,1}(x)$ satisfies the relation 
\begin{equation}
g'_{1,1}(x-i\frac{\pi}{3})g'_{1,1}(x+i\frac{\pi}{3})=g'_{1,1}(x)
\end{equation}

We eliminate the zeros of the 1-string and short 2-strings with the
required periodicity, we use $\sigma'_{0}(x,r)$ and $\sigma'_{2}(x-u_{k},v_{v},p)$.
As before a short two string with locations $\beta_{k}=u_{k}\pm iv_{k}$
can be eliminated by 
\begin{equation}
\sigma'_{1}(x-u_{k}-\frac{i\pi}{6}-iv_{1k},q)\sigma'_{1}(x-u_{k}+iv_{2k}+\frac{i\pi}{6},q)=\sigma'_{2}(x-u_{k},v_{v},q)
\end{equation}
where $v_{1}$ and $v_{2}$ are not necessarily equal. 

In the $x=\frac{5}{3i}(u-\frac{3\pi}{10})$ variable they are located
at $x_{0}^{\pm}=\frac{5\alpha}{3}=\log N+\tilde{\alpha}^{\pm}$and
at $x_{j}^{\pm}=\pm i\frac{\pi}{6}+\log N+\tilde{\beta}_{j}^{\pm}$.
As we would like to take logarithm we need functions free of zeros
and poles on the physical strip. Those functions were defined before
as $\sigma'_{0}(x,r)$ and $\sigma'_{2}(x,p)$. 

With these functions we parametrize the normalized transfer matrix
eigenvalue as 
\begin{equation}
t'(x)=f'(x)^{2N}g'_{1,1}(x)\prod_{\pm}\sigma'_{0}(x-x_{0}^{\pm},r)\prod_{j=1}^{M}\sigma'_{2}(x-u_{k},v_{v},q)l'(x)
\end{equation}
which ensures that $l'(x)$ is ANZ in the physical strip. 

We find:
\[
\log l'(x)=-\varphi_{\epsilon}\star\log(\prod_{\pm}\sigma'_{0}(x-x_{0}^{\pm})(1+t'^{-1}(x)))
\]

where the convolution $\star$ was defined in equation (\ref{eq:massiveconv}).
Restoring $t$ we obtain

\begin{equation}
\begin{array}{cc}
\log t'(x)=2N\log f'(x)+\log g'_{1,1}(x)+\sum_{\pm}\log(\sigma'_{0}(x-x_{0}^{\pm}))+\sum_{j,\pm}\log(\sigma'_{2}(x-u_{k},v_{v},q))\\
-\varphi_{\epsilon}\star\log(\prod_{\pm}\sigma'_{0}(x-x_{0}^{\pm})(1+t'^{-1}(x)))
\end{array}
\end{equation}
The parameters are determined by $t(x)\vert_{x=x_{i}\pm\frac{i\pi}{3}}=-1$.

In the scaling limit, $g'_{1,1}(x)\rightarrow g_{1,1}(x)=R_{1}^{-1}(\frac{i\pi}{2}+x)R_{1}^{-1}(\frac{i\pi}{2}-x)$. 

Using 
\begin{equation}
e^{\epsilon(x)}=\lim_{N\to\infty}\sigma'_{0}(x+\log\frac{N}{\mu}-x_{0}^{\pm})^{-1}t'(x+\log\frac{N}{\mu})
\end{equation}

The parameters satisfy $e^{\epsilon(x)}\sigma'_{0}(x-\tilde{\alpha}^{\pm})\vert_{x=x_{i}\pm\frac{i\pi}{3}}=-1$. 

And $\epsilon(x)$ satisfies the equation

\begin{equation}
\begin{array}{cc}
\epsilon(x)=2mL\cosh x-\log R_{1}(\frac{i\pi}{2}+x)R_{1}(\frac{i\pi}{2}-x)+\sum_{j}\log\sigma'_{2}(x-u_{k},v_{v},q)\\
-\varphi\star\log(\sigma_{0}(x-\tilde{\alpha}^{\pm})+e^{-\epsilon(x)})
\end{array}
\end{equation}

This is the general massive TBA in the (1,1) sector. It is important
to mention that for the groundstate of this sector, there are no short
2-strings, hence $\sigma_{1}$ doesn't appear and the ground state
massive TBA of this sector on the lattice is:

\begin{equation}
\begin{array}{cc}
\log t'(x)=2N\log f'(x)+\log g'_{1,1}(x)+\sum_{\pm}\log(\sigma'_{0}(x-x_{0}^{\pm}))\\
-\varphi_{\epsilon}\star\log(\prod_{\pm}\sigma'_{0}(x-x_{0}^{\pm})(1+t'^{-1}(x)))
\end{array}
\end{equation}

while the scaled ground-state massive TBA is
\begin{equation}
\epsilon(x)=2mL\cosh x-\log R_{1}(\frac{i\pi}{2}+x)R_{1}(\frac{i\pi}{2}-x)+-\varphi\star\log(\sigma_{0}(x-\tilde{\alpha}^{\pm})+e^{-\epsilon(x)})
\end{equation}

\subsubsection*{(r,s)=(1,2) sector}

Similar analysis follows in this sector, where we have to take into
account the contribution of $S_{s}(u)$.

For $s=2$ , $S_{2}(u)$ is proportional to $f'(u+\xi)f'(u-\xi)$,
therefore we should account for their respective zeroes and poles.

The other contributions from the normalization and the analytic strip
are the same as before, and they were included in $g'_{1,1}(u)$.
Then we need to define the new $g'_{1,2}$ term as:

\[
g'_{1,2}(u)=g'_{1,1}(u)\times f'(u+\xi)f'(u-\xi)
\]
In terms of the $x$ variable we get:

\begin{equation}
\begin{array}{cc}
g'_{1,2}(x)=-\frac{\vartheta_{1}^{2}(\frac{ix}{2},p^{\frac{1}{2}})\vartheta_{1}(\frac{ix}{2}+\frac{\pi}{6},p^{\frac{1}{2}})\vartheta_{1}(\frac{ix}{2}-\frac{\pi}{6},p^{\frac{1}{2}})\vartheta_{2}(\frac{ix}{2}-\frac{\pi}{12},p^{\frac{1}{2}})\vartheta_{2}(\frac{ix}{2}+\frac{\pi}{12},p^{\frac{1}{2}})}{\vartheta_{2}^{2}(\frac{ix}{2},p^{\frac{1}{2}})\vartheta_{2}(\frac{ix}{2}+\frac{\pi}{6},p^{\frac{1}{2}})\vartheta_{2}(\frac{ix}{2}-\frac{\pi}{6},p^{\frac{1}{2}})\vartheta_{1}(\frac{ix}{2}-\frac{\pi}{12},p^{\frac{1}{2}})\vartheta_{1}(\frac{ix}{2}+\frac{\pi}{12},p^{\frac{1}{2}})}\\
\times f'(x-\frac{5i\xi}{3})f'(x+\frac{5i\xi}{3})
\end{array}
\end{equation}

$g_{1,2}(x)$ satisfies the relation 
\[
g'_{1,2}(x-i\frac{\pi}{3})g'_{1,2}(x+i\frac{\pi}{3})=g'_{1,2}(x)
\]

The zeros short 2-strings are eliminated by $\sigma'_{2}(x-u_{k},v_{v},p)$

In the $x=\frac{5}{3i}(u-\frac{3\pi}{10})$ variable they are located
at $x_{j}^{\pm}=\pm i\frac{\pi}{6}+\log N+\tilde{\beta}_{j}^{\pm}$.
As we would like to take logarithm we need functions free of zeros
and poles on the physical strip. We take $\sigma'_{1}(x,p)$ to accommodate
for the zeroes of the short strings. With these functions we parametrize
the normalized transfer matrix eigenvalue as 
\[
t'(x)=f'(x)^{2N}g'_{1,2}(x)\prod_{j=1}^{M}\sigma'_{2}(x-u_{k},v_{v},p)l'(x)
\]
then we obtain
\begin{equation}
\log t'(x)=2N\log f'(x)+\log g'_{1,2}(x)+\sum_{j,\pm}\log(\sigma'_{2}(x-u_{k},v_{v},p))-\varphi_{\epsilon}\star\log(1+t'^{-1}(x))
\end{equation}
The parameters are determined by $t'(x)\vert_{x=x_{i}\pm\frac{i\pi}{3}}=-1$.

In the scaling ($N\rightarrow\infty)$ limit, as $x\rightarrow x+\log\frac{N}{\mu}$
and $\xi\rightarrow-i\frac{3}{5}(\tilde{\xi}+\log\frac{N}{\mu})$
then

\[
g'_{1,2}(x)\rightarrow g_{1,2}(x)=R_{\phi}^{-1}(\frac{i\pi}{2}+x)R_{1}^{-1}(\frac{i\pi}{2}-x)
\]

for $b=3+\frac{6\tilde{\xi}}{\pi}$ in the expression of $R_{\phi}$. 

Using $e^{\epsilon(x)}=\lim_{N\to\infty}t(x+\log\frac{N}{\mu})$,
it satisfies the equation

\begin{equation}
\begin{array}{cc}
\epsilon(x)=2mL\cosh x-\log R_{\phi}^{-1}(\frac{i\pi}{2}+x)R_{1}^{-1}(\frac{i\pi}{2}-x)+\sum_{j}\log\sigma'_{2}(x-u_{k},v_{v},p)\\
-\varphi\star\log(1+e^{-\epsilon(x)})
\end{array}
\end{equation}

This is the general massive TBA in the (1,2) sector. For the groundstate
of this sector, there are no short 2-strings, hence $\sigma_{1}$
doesn't appear and the ground state massless TBA of this sector on
the lattice is:

\begin{equation}
\log t'(x)=2N\log f'(x)+\log g'_{1,2}(x)+-\varphi\star\log(1+t'^{-1}(x))
\end{equation}

while the scaled ground-state massive boundary TBA in the (1,2) sector
is
\begin{equation}
\epsilon(x)=2mL\cosh x-\log R_{\phi}^{-1}(\frac{i\pi}{2}+x)R_{1}^{-1}(\frac{i\pi}{2}-x)+-\varphi\star\log(1+e^{-\epsilon(x)})
\end{equation}

We have derived all the TBAs for the periodical boundary model with
and without a seam, and for the boundary model in both critical and
massive regimes for ground and excited states.

\pagebreak{}

\chapter{Energy corrections: from Lee-Yang to SYM}

As mentioned in the introduction, AdS/CFT correspondence \cite{MGKPW}
relates the string energies on the $AdS_{5}\times S^{5}$ background
to the anomalous dimensions of gauge invariant operators in maximally
supersymmetric four dimensional $SU(N)$ gauge theory. The correspondence
is particularly useful in the large $N$ limit when it can be described
by a two dimensional integrable field theory \cite{AFro}. The advantage
of the integrable two dimensional point of view lies in its non-perturbative
nature, which nevertheless, can be used to calculate perturbative
quantities, too. Indeed, the leading finite size correction of a two
particle state can be described in terms of the asymptotic Bethe ansatz
and the generalized Lüscher formulas \cite{BJa}, which, when expanded
in the coupling constant, provides the exact perturbative anomalous
dimension of the Konishi operator up to seven loops.

To show this resemblance, we start by calculating higher order corrections
for the energy of a one particle state in the Lee-Yang model. We do
this because it turns out that the $F-term$ corrections can be calculated
in an analogous way for several other models. We deduce a general
result that can be applied to integrable models including the $AdS$
superstring sigma model. 

In the bulk Lee-Yang model, we know that the multiparticle state energy
is given by

\[
E_{\{n_{j}\}}(L)=m{\displaystyle \sum_{j=1}^{N}\cosh\theta_{j}-m{\displaystyle \int_{-\infty}^{+\infty}\frac{d\theta}{2\pi}\cosh\theta\log(1+e^{-\epsilon(\theta)})}}
\]

The ground state TBA is given by

\begin{equation}
\epsilon(\theta)=mL\cosh\theta-\frac{1}{2\pi}\intop_{-\infty}^{+\infty}d\theta'\phi(\theta-\theta')\log(1+e^{-\epsilon(\theta)})
\end{equation}

where $\phi(\theta)=-i\partial_{\theta}\log S(\theta)$ and $S(\theta)$
is the scattering matrix of the model \cite{Dorey}

In the large volume limit, it reduces to:

\[
\epsilon(\theta)=mL\cosh\theta
\]

Substituting this into the energy equation we get that the leading
finite size correction is given by:

\[
E_{0}(L)=-m{\displaystyle \int_{-\infty}^{+\infty}\frac{d\theta}{2\pi}\cosh\theta\log(1+e^{-mL\cosh\theta})}
\]

\begin{equation}
=-m{\displaystyle \int_{-\infty}^{+\infty}\frac{d\theta}{2\pi}\cosh\theta e^{-mL\cosh\theta}}
\end{equation}

while for a moving one particle state TBA we need to do one more iteration.
The one particle TBA equation is

\begin{equation}
\epsilon(\theta)=mL\cosh\theta+\log\frac{S(\theta-\theta_{0})}{S(\theta+\theta_{0})}-\frac{1}{2\pi}\intop_{-\infty}^{+\infty}d\theta'\phi(\theta-\theta')\frac{S(\theta+\theta_{0})}{S(\theta-\theta_{0})}e^{-mL\cosh\theta})
\end{equation}

The $\theta_{0}$ is not arbitrary. It is determined from the relation
$\epsilon(\theta_{0})=i\pi(2n+1)$ therefore (neglecting the third
term for large $L)$ we have

\[
mL\cosh\theta_{0}+i\pi-\log S(2\theta_{0})=i\pi(2n+1)
\]
 or 

\[
mL\cosh\theta_{0}-\log S(2\theta_{0})=2i\pi n
\]

For a standing particle, $n=0$ thus the first two terms need to cancel
each other. For large $L$ this can only happen if $\theta_{0}$ causes
a singularity in $S(2\theta_{0}),$ hence for $\theta_{0}=\frac{i\pi}{6}$.
Continuing the analysis for a standing particle, the contribution
to the $F-terms$ will come from the energy correction

\[
E_{1}(L)=-m\intop_{-\infty}^{+\infty}d\theta\cosh\theta\frac{S(\theta+\theta_{0})}{S(\theta-\theta_{0})}e^{-mL\cosh\theta}
\]

Hence the first order correction for a standing particle state if
given by:

\[
E_{1}(L)-E_{0}(L)=-m\intop_{-\infty}^{+\infty}d\theta\cosh\theta(\frac{S(\theta+\theta_{0})}{S(\theta-\theta_{0})}-1)e^{-mL\cosh\theta}
\]

Now using that $\theta_{0}=\frac{i\pi}{6}+i\delta$, where $\delta$
is a small correction , and the fusion relation of the Lee-Yang $S(\theta+i\pi/3)S(\theta-i\pi/3)=S(\theta)$
we get the energy correction for the one-particle state. $\delta$
will lead to $\mu-term$ corrections whereas the $F-term$ contribution
will be:

\begin{equation}
E_{1}(L)-E_{0}(L)=-m\intop_{-\infty}^{+\infty}d\theta\cosh\theta(S(\frac{i\pi}{2}-\theta)-1)e^{-mL\cosh\theta}
\end{equation}

This final result is general. It applies to the $F-term$ contribution
of the Lee-Yang \cite{Dorey}, the sinh-Gordon \cite{BJa}, some sectors
of the sine-Gordon and several other models. It is conjectured to
be the relation that gives the higher order $F$ correction terms
for integrable model TBAs. It is generalized to the $AdS$ model.

With the $AdS/CFT$ conjecture and the identical 4-loop corrections
that we find on both sides of the duality, we can calculate higher
order corrections for the superstring model and assume that those
results will hold on the other side of the duality.The direct perturbative
gauge theory calculations are very cumbersome and have been evaluated
for the Konishi operator \emph{up to 4 and 5 loops} \emph{only} \cite{FSSZ4,FSSZ3}
, where the wrapping part of the correction matches exactly the Lüscher
type finite size correction \cite{BJa}. Thus the integrability based
``string theory'' techniques provide a way to go beyond the available
perturbative calculations and collect information about higher order
results. 

The $\beta$-deformed theory is an $\mathcal{N}=1$ supersymmetric,
exactly marginal deformation of the $\mathcal{N}=4$ SYM and has been
always the testing ground of the AdS/CFT duality. The dispersion relation
of the excitations was calculated exactly in \cite{MPSZ}, which together
with wrapping corrections \cite{FSSZ1,FSSZ2} provides the exact anomalous
dimension of a single impurity operator $\mbox{Tr}(XZ^{L})$. On the
string theory side this operator corresponds to a one particle state
in the $su(2)$ sector in finite volume. The leading finite size effect
follows from the momentum quantization, or asymptotic Bethe Ansatz
equation \cite{BRo}. As the volume decreases Lüscher/wrapping type
correction becomes important \cite{BJa,BJL,ABBN} and for an exact
description they have to be summed up. An educated way to take all
finite size correction into account would be to extend the Thermodynamic
Bethe Ansatz of the $\mathcal{N}=4$ theory \cite{GKV,AFro,BFTa,AfroTBA,GKKV,AFroS,CFTat}
to the $\beta$-deformed case. This has been recently done in \cite{ABBN-2}.

We are interested in the $\beta$ -deformed theory at a particular
value of $\beta$, namely for the simplest nontrivial $\beta=\frac{1}{2}$
and evaluate the Lüscher correction at NNLO in the coupling. This
value of $\beta$ is interesting as it correspond to a particular
orbifold theory, too. As it was conjectured in \cite{Gunn} the anomalous
dimension of the $su(2)$ particle ($\mbox{Tr}(XZ)$) in the $\beta=\frac{1}{2}$-deformed
theory coincides with the anomalous dimension of the $sl(2)$ particle
($\mbox{Tr}(DZZ)$) of the orbifold theory. 

In the results, multiple zeta functions appear. Multiple zeta values
(MZVs) are the generalization of zeta functions: 
\[
\zeta(a_{1},a_{2},\dots,a_{n})=\sum_{j_{1}>j_{2}>\dots>j_{n}>0}\frac{1}{j_{1}^{a_{1}}j_{2}^{a_{2}}\dots\, j_{n}^{a_{n}}}
\]
and are trivially related to the values of nested harmonic sums taken
at infinity. A recent review is presented in \cite{BBBZ}.

We calculate the anomalous dimension of the operator $\mbox{Tr}(XZ)$
in the $\beta=\frac{1}{2}$ deformed theory. It is a very special
state as it has a vanishing rapidity $u=0$, ($p=\pi$), which is
not effected by finite size corrections. In contrast, the energy is
shifted by vacuum polarization effects as 
\begin{equation}
E=E_{ABA}+\Delta E\label{eq:E}
\end{equation}
The asymptotic Bethe Ansatz energy $E_{ABA}$ is simply the dispersion
relation of a standing particle:
\begin{eqnarray*}
E_{ABA} & = & \sqrt{1+16g^{2}\sin^{2}(\frac{p}{2})}=\sqrt{1+16g^{2}}\\
 & = & 1+8g^{2}-32g^{4}+256g^{6}-2560g^{8}+28672g^{10}-344064g^{12}+O\left(g^{13}\right)
\end{eqnarray*}
while $\Delta E$ corresponds to the wrapping interactions and has
the expansion 
\begin{equation}
\Delta E=\Delta E_{4}g^{8}+\Delta E_{5}g^{10}+\Delta E_{6}g^{12}+\dots\label{eq:DeltaE}
\end{equation}
In \cite{BJL}the LO correction was calculated 
\begin{equation}
\Delta E_{4}=128(4\zeta(3)-5\zeta(5))
\end{equation}
The NLO correction turned out to be \cite{BHJL} : 
\begin{equation}
\Delta E_{5}=-128(12\zeta(3)^{2}+32\zeta(3)+40\zeta(5)-105\zeta(7))
\end{equation}
 And in \cite{BaOm} we evaluated the NNLO correction and obtained
\begin{equation}
\Delta E_{6}=-128(48\zeta(3)^{2}-592\zeta(5)-24\zeta(3)(8+15\zeta(5))-322\zeta(7)+1701\zeta(9))
\end{equation}
which completes the anomalous dimension of the operator up to six
loops. \pagebreak{}

\chapter{Conclusions}

In the first part of this thesis I presented an introductory review
of the Lee Yang model. The in chapter 3, I initiated the form factor
program for purely transmitting integrable defect theories. I restricted
for a single particle type, but the extension of the program for diagonal
bulk scatterings and diagonal transmissions is straightforward \cite{Ol1}.
We laid down axioms for the form factors of operators localized both
in the bulk and also on the defect. We determined the solutions of
the consistency requirements for bulk operators in terms of the bulk
form factors together with the transmission matrix. These form factors
determine the correlation functions of bulk operators, which we elaborated
in details for the two point functions. In the case of defect operators
we gave the general form of the solutions and explicitly calculated
for the free boson and for some operator in the Lee-Yang model. We
also described how the fusion method can be used to generate new form
factor solutions from old ones.

In the analysis of the Lee-Yang model we observed relations between
the defect operators which should have the origin in defect conditions.
The lack of the Lagrangian definition of the model prevented us to
analyze this question. In order to achieve this aim one has to analyze
the simultaneous integrable defect and bulk perturbations of the defect
Lee-Yang model using conformal perturbation theory and establish the
relation between the bulk and defect couplings, which maintains integrability.
This approach then can be used to derive defect conditions which will
provide relations between fields living on the defect. In pushing
forward this program one has to solve the defect Lee-Yang model first.
The explicit knowledge of the correlation functions together with
the structure constants will make it possible to bridge the operators
appearing in the form factor program to their UV counterparts by analyzing
the short distance behavior of the two point functions obtained from
the explicit form factor solutions.

Another interesting problem is to see that we have as many polynomial
solution of the form factor equation as many local operators existing
in the theory. We have seen this coincidence in the case of the free
boson. In the case of the Lee-Yang model the nontrivial mixing between
the left and right degrees of freedom and the various cancellations
between the leading order scaling terms prevented us from performing
this analysis. Possibly a more careful analysis along the line of
\cite{SZT} would clear up this point as well.

We have analyzed the free boson and the Lee-Yang model, the method,
however, has a straightforward application for the sinh-Gordon model
adopting ideas from the boundary form factor solutions \cite{Ol2,Gabor1}.

The defect form factors in the Lee-Yang model can be tested by extending
them for finite volume and comparing to direct TCSA data. They also
can be used to build up finite temperature defect correlation functions.
These are direct generalizations of the related boundary analysis
developed in \cite{Gabor2,Gabor3}.

Whereas in the second part, in chapter 4, we defined the Lee Yang
model on the lattice as a special case of the $A_{L}$ Forrester-Baxter
model. We defined the face weights, for both massless and massive
models and then defined the transfer matrices for the periodic boundary
conditions, with and without a seam and also the double row transfer
matrices for the boundary case. Analyzing their analytic structure
by finding the zero eigenvalues of those matrices using numerical
methods we were able to provide a systematic classification and deduce
their finitized characters. We were also able to deduce a direct bijection
between the paths, the zero structure and the Virasoro modes. Given
the analytic structure of the transfer matrix for any energy eigenvalue,
we can associate it in a one-to-one correspondence with a state in
the Virasoro module. For the boundary case, I classified the boundary
flows between the states of the two modules of the model exactly.
More interestingly, we applied this to the case of a seam where the
operator content and the flow patterns are much richer, and the method
accurately describes all the flows in terms of the states, the zero
structure and the path diagrams.

With the knowledge of the analytic structure, and using the functional
relations of the model in the three different boundary conditions,
one can solve the TBA equations for all cases in the massless and
the massive regimes. The seam TBA is also consistent with the transmission
matrix suggested in the bootstrap. The results also confirm what was
found before through the TBA approach. However, before, the analytic
structure was guessed. On the lattice, in contrast, it is exactly
determined.

The methods of this part are general in the sense that they can be
applied for a variety of models of more complexity. A future line
of research can work on solving other integrable models.

\begin{doublespace}
In the large volume limit, I calculated the Luscher correction terms
of the TBA energy of a one-particle state to higher orders. This generalization
is important as similar techniques are used in the analysis of energy
corrections due to wrappings for superstrings in the sigma model.
I display those ideas in the end of the dissertation as this is an
important direction of work in string theory models, particularly
in the AdS/CFT correspondence.
\end{doublespace}

\pagebreak{}

\chapter*{Appendix}

\addcontentsline{toc}{chapter}{Appendix}

\section*{Calculation of the anomalous dimension from AdS/CFT}

In this appendix we explain how to obtain the leading Lüscher correction
for a one particle state in the $\beta=\frac{1}{2}$ deformed theory
by recalling the available formulas from the literature.

As the anomalous dimension of the single impurity operator corresponds
to a standing particle state, we have to calculate the standard Lüscher
correction. It describes how the one particle energy (dispersion relation)
\begin{equation}
E=E(p=\pi)=\sqrt{1+16g^{2}\sin^{2}(\frac{p}{2})}=\sqrt{1+16g^{2}}=\epsilon
\end{equation}
is modified due to vacuum polarization effects. The $p=\pi$ momentum
corresponds to vanishing rapidity $u=0$ which is protected by symmetry.
Consequently the leading finite size correction is the energy correction
which originates from virtual particles propagating around the circle
and can be written as
\begin{equation}
\Delta E=-\sum_{Q=1}^{\infty}\int\frac{dq}{2\pi}\mbox{sTr}(S_{Q1}^{Q1}(q,0))e^{-\tilde{\epsilon}_{Q}(q)L}+O(g^{16})\label{eq:Luscher}
\end{equation}
Here we sum up for all bound-states of charge $Q$ of the mirror model,
whose momenta are $q$, $S_{Q1}^{Q1}$ describes how they scatter
on the fundamental particle and $\tilde{\epsilon}(q)$ denotes their
mirror energy. We have to expand this expression in NNLO in $g$.
The above form exactly describes the energy correction up to the order
$g^{14}$ only since at the order $g^{16}$ double wrapping effects
will contribute, too. Let us analyze the $g$ dependence of the various
terms.

\subsubsection*{Exponential factor}

The mirror energy has the following parametrization
\begin{equation}
e^{-\tilde{\epsilon}_{Q}(q)}=\frac{z^{-}(q,Q)}{z^{+}(q,Q)}\quad;\qquad z^{\pm}(q,Q)=\frac{q+iQ}{4g}\left(\sqrt{1+\frac{16g^{2}}{q^{2}+Q^{2}}}\pm1\right)
\end{equation}

\subsubsection*{Matrix part of the scattering matrix}

The contribution of the scattering matrix can be factored as 
\begin{equation}
\mbox{sTr}(S_{Q1}^{Q1}(q,0))=S_{scalar}(q,0)\mbox{sTr}(S_{matrix}^{su(2)}(q,0))^{2}
\end{equation}
where the super-trace of the matrix part contain the contributions
of all polarizations. The various polarizations of the mirror bound-states
can be labeled in the super-space formalism as $(w_{3}^{j}w_{4}^{Q-j}$,
$w_{3}^{j}w_{4}^{Q-2-j}\theta_{1}\theta_{2}$, $w_{3}^{j}w_{4}^{Q-1-j}\theta_{3}$,
$w_{3}^{j}w_{4}^{Q-1-j}\theta_{4})$ see \cite{BJa,AFroSmSB} for
the details. The super-trace in the $\beta$- deformed theory evaluates
in the previous basis as 
\begin{equation}
\mbox{sTr}(S_{matrix}^{su(2)}(z,x))=\sum_{j=0}^{Q}SB1_{j}(z,x)+\sum_{j=0}^{Q-2}SB2_{j}(z,x)+i\sum_{j=0}^{Q-1}SF1_{j}(z,q)-i\sum_{j=0}^{Q-1}SF2_{j}(z,q)
\end{equation}
The S-matrix elements can be extracted from \cite{BJa}. In calculating
the corrections for the $su(2)$ representative neither of the S-matrix
contributions depends on $j$ and they read explicitly as 
\[
SB1_{j}(z,x)=\frac{z^{+}-x^{+}}{z^{-}-x^{+}}\frac{\tilde{\eta}_{1}}{\eta_{1}}\quad;\qquad SB2_{j}(z,x)=\frac{z^{+}-x^{-}}{z^{-}-x^{+}}\frac{(1-x^{+}z^{-})}{(1-x^{-}z^{-})}\frac{x^{-}}{x^{+}}\frac{\tilde{\eta}_{1}}{\eta_{1}}\left(\frac{\tilde{\eta}_{2}}{\eta_{2}}\right)^{2}
\]
\[
SF1_{j}(z,x)=\frac{z^{+}-x^{-}}{z^{-}-x^{+}}\frac{\tilde{\eta}_{1}}{\eta_{1}}\frac{\tilde{\eta}_{2}}{\eta_{2}}\quad;\qquad SF2_{j}(z,x)=\frac{z^{+}-x^{+}}{z^{-}-x^{+}}\frac{(1-x^{+}z^{-})}{(1-x^{-}z^{-})}\frac{x^{-}}{x^{+}}\frac{\tilde{\eta}_{1}}{\eta_{1}}\frac{\tilde{\eta}_{2}}{\eta_{2}}
\]
The appearing string frame factors can be written as $\frac{\tilde{\eta}_{1}}{\eta_{1}}=\sqrt{\frac{z^{-}}{z^{+}}}$
and $\left(\frac{\tilde{\eta}_{2}}{\eta_{2}}\right)^{2}=\frac{x^{-}}{x^{+}}$.
The $x^{\pm}$ parameters depend on the momentum the usual way 
\begin{equation}
x^{\pm}(p)=\frac{\cot\frac{p}{2}\pm i}{4g}\left(1+\sqrt{1+16g^{2}\sin^{2}\frac{p}{2}}\right)
\end{equation}
which in our case results in 
\[
x^{\pm}=\pm x=\pm\frac{i}{4g}(1+\epsilon)\quad;\qquad\frac{x^{-}}{x^{+}}=-1
\]
With these variables the super-trace of the matrix part takes a particularly
simple form
\begin{equation}
\mbox{sTr}(S_{matrix}^{su(2)}(z,x))=\frac{2x(1+2Qz^{-}(x-z^{+}))}{(x-z^{-})(1+xz^{-})}e^{-\tilde{\epsilon}_{Q}(q)/2}
\end{equation}
We have checked that the matrix part of the $sl(2)$ representative
of the orbifold model gives the same result. There, the deformation
is such that the fermions do not contribute as the two undeformed
fermionic S matrix elements are the same \cite{LuRe,BJL}.

\subsubsection*{Scalar part of the scattering matrix}

The scalar part of the scattering matrix of a charge $Q$ bound-state
can be obtained by multiplying the scalar factors of its individual
scattering constituents. The charge $Q$ bound-state composed of elementary
magnons as $z=(z_{1},\dots,z_{Q})$, such that $z^{-}=z_{1}^{-}$
and $z_{Q}^{+}=z^{+}$and the bound-state condition is also satisfied
$z_{i}^{+}=z_{i+1}^{-}$. Thus the full scalar factor as the product
of the elementary scalar factors turns out to be \cite{BJa}: 
\begin{equation}
S_{scalar}(z,x)=\prod_{i=1}^{Q}S^{sl(2)}(z_{i},x)=\prod_{i=1}^{Q}e^{-2i\sigma(z_{i},x)}\frac{z_{i}^{-}-x^{+}}{z_{i}^{+}-x^{-}}\frac{1-\frac{1}{z_{i}^{+}x^{-}}}{1-\frac{1}{z_{i}^{-}x^{+}}}
\end{equation}
In calculating the Lüscher correction we have to evaluate this expression
when $z$ is in the mirror kinematics ($\vert z^{-}\vert<1,\vert z^{+}\vert>1$).
The analytical continuation has been carefully elaborated in \cite{AFroDr}:

\begin{equation}
S_{scalar}^{-1}(z,x)=\Sigma_{Q,1}^{2}(z,x)S^{su(2)}(z,x)
\end{equation}
A particularly good feature of the formula is that both expressions
depend on $z^{\pm}$ only and not on the individual $z_{i}^{\pm}$.
Explicitly the $su(2)$ scalar factors read as: 

\begin{equation}
S^{su(2)}(z,x)=\frac{(z^{+}-x^{-})(z^{+}-x^{+})}{(z^{-}-x^{+})(z^{-}-x^{-})}\frac{(1-\frac{1}{z^{+}x^{-}})(1-\frac{1}{z^{+}x^{+}})}{(1-\frac{1}{z^{-}x^{+}})(1-\frac{1}{z^{-}x^{-}})}
\end{equation}
while for the case $\vert x^{\pm}\vert>1$ following \cite{AFroDr}
we can write 
\begin{eqnarray*}
-i\log\Sigma_{Q,1}(z,x) & = & \Phi(z^{+},x^{+})-\Phi(z^{+},x^{-})-\Phi(z^{-},x^{+})+\Phi(z^{-},x^{-})\\
 &  & +\frac{1}{2}\left[-\Psi(z^{+},x^{+})+\Psi(z^{+},x^{-})-\Psi(z^{-},x^{+})+\Psi(z^{-},x^{-})\right]\\
 &  & +\frac{1}{2i}\log\left[\frac{(z^{+}-x^{+})(x^{-}-\frac{1}{z^{+}})^{2}}{(z^{+}-x^{-})(x^{-}-\frac{1}{z^{-}})(x^{+}-\frac{1}{z^{-}})}\right]
\end{eqnarray*}
We have to be careful as the conventions of \cite{AFroDr} are different
from ours. To turn into our conventions one has to replace $z^{\pm}\to z^{\mp}$
and $x^{\pm}\to x^{\mp}.$ There are integral representations for
$\Phi$ and$\Psi$ as 

\begin{equation}
\Phi(x_{1},x_{2})=i\oint_{C_{1}}\frac{dw_{1}}{2\pi i}\oint_{C_{1}}\frac{dw_{2}}{2\pi i}\frac{1}{w_{1}-x_{1}}\frac{1}{w_{2}-x_{2}}\log\frac{\Gamma(1+ig(w_{1}+w_{1}^{-1}-w_{2}-w_{2}^{-1}))}{\Gamma(1-ig(w_{1}+w_{1}^{-1}-w_{2}-w_{2}^{-1}))}
\end{equation}

\begin{equation}
\Psi(x_{1},x_{2})=i\oint_{C_{1}}\frac{dw_{2}}{2\pi i}\frac{1}{w_{2}-x_{2}}\log\frac{\Gamma(1+ig(x_{1}+x_{1}^{-1}-w_{2}-w_{2}^{-1}))}{\Gamma(1-ig(x_{1}+x_{1}^{-1}-w_{2}-w_{2}^{-1}))}
\end{equation}
where the integrations are for the unit circle. They are well-defined
provided none of the $x_{i}$ lies on the unit circle. 

As by now we have collected all the necessary formulas we turn to
analyze their weak coupling expansions.

\subsubsection*{Weak coupling expansion}

Our aim is to calculate the weak coupling expansion of $\Delta E$
for $L=1$. In doing so we decompose the integrand of the Lüscher
correction 
\begin{equation}
\Delta E=-\sum_{Q=1}^{\infty}\int\frac{dq}{2\pi}P(q,Q)\Sigma(q,Q)
\end{equation}
into a simpler rational part 
\begin{equation}
P(q,Q)=\frac{4x^{2}(1+2Qz^{-}(x-z^{+}))^{2}}{((xz^{-})^{2}-1)(x^{2}-(z^{+})^{2})}\left(\frac{z^{-}}{z^{+}}\right)^{2}
\end{equation}
which contains both the matrix part and the rational part of the scalar
factor, and into the more complicated $\Sigma$ part: 
\begin{eqnarray*}
i\log\Sigma(q,Q) & = & 2(\Phi(z^{+},x^{+})-\Phi(z^{+},x^{-})-\Phi(z^{-},x^{+})+\Phi(z^{-},x^{-}))\\
 &  & -\Psi(z^{+},x^{+})+\Psi(z^{+},x^{-})-\Psi(z^{-},x^{+})+\Psi(z^{-},x^{-})
\end{eqnarray*}
We expand these functions in $g^{2}$ as 
\begin{equation}
P(q,Q)=P_{8}(q,Q)g^{8}+P_{10}(q,Q)g^{10}+P_{12}(q,Q)g^{12}+\dots
\end{equation}
\begin{equation}
\Sigma(q,Q)=1+\Sigma_{2}(q,Q)g^{2}+\Sigma_{4}(q,Q)g^{4}+\dots
\end{equation}
The expansion of the rational part is quite straightforward and we
obtain 
\[
P_{8}(q,Q)=\frac{4096Q^{2}(-1+q^{2}+Q^{2})^{2}}{(q^{2}+Q^{2})^{4}(q^{4}+(-1+Q^{2})^{2}+2q^{2}(1+Q^{2}))}
\]

\begin{eqnarray*}
\frac{P_{10}(q,Q)}{P_{8}(q,Q)} & = & -\frac{8(7q^{4}+3(Q^{2}-1)^{2}+10q^{2}(1+Q^{2}))}{(q^{2}+Q^{2})(q^{4}+(-1+Q^{2})^{2}+2q^{2}(1+Q^{2}))}
\end{eqnarray*}
{\footnotesize{}
\begin{eqnarray*}
\frac{P_{12}(q,Q)}{P_{10}(q,Q)} & = & -\frac{2(27+268q^{2}+704q^{4})}{(3+28q^{2}+64q^{4})(q^{2}+Q^{2})}+\frac{16(-3q^{2}+2q^{2}Q)}{(1+4q^{2})(1+q^{2}-2Q+Q^{2})}\\
 &  & -\frac{16(3q^{2}+2q^{2}Q)}{(1+4q^{2})(1+q^{2}+2Q+Q^{2})}-\frac{8(-81q^{2}-117q^{4}+16q^{6}+27q^{2}Q^{2}+16q^{4}Q^{2})}{(3+16q^{2})(3+10q^{2}+7q^{4}-6Q^{2}+10q^{2}Q^{2}+3Q^{4})}
\end{eqnarray*}
}In expanding the $\Psi$ and $\Phi$ functions we use the same method
we used in (7). The expansion of the $\Psi(x_{1},x_{2})$ functions
for $\vert x_{2}\vert>1$ (string region) reads as follows 
\begin{eqnarray*}
\Psi(x_{1},x_{2}) & = & -\frac{g}{x_{2}}(\Psi(1-ig(x_{1}+x_{1}^{-1}))+\Psi(1+ig(x_{1}+x_{1}^{-1})))\\
 &  & -\frac{ig^{2}}{2x_{2}^{2}}(\Psi_{1}(1-ig(x_{1}+x_{1}^{-1}))-\Psi_{1}(1+ig(x_{1}+x_{1}^{-1})))\\
 &  & +\frac{g^{3}}{2x_{2}}(\Psi_{2}(1-ig(x_{1}+x_{1}^{-1}))+\Psi_{2}(1+ig(x_{1}+x_{1}^{-1})))+\dots
\end{eqnarray*}
where $\Psi_{n}(x)=(\frac{d}{dx})^{n}(\log(\Gamma(x))$ are the standard
polygamma functions. If $\vert x_{1}\vert>1$ then $\Phi(x_{1},x_{2})$
starts at $g^{6}$. In the opposite case using the identity $\Phi(x_{1},x_{2})=\Phi(0,x_{2})-\Phi(x_{1}^{-1},x_{2})$,
being valid if $\vert x_{1}\vert\neq1$, we can calculate the leading
expansion of $\Phi$ as 
\[
\Phi(0,x)=\frac{2}{x}(\gamma_{E}g-3\zeta(3)g^{3}+\dots)
\]
Using functional identities valid for integer $Q$ we obtained
\[
\Sigma_{2}(q,Q)=-\frac{16Q}{q^{2}+Q^{2}}-8\Bigr(S_{1}\Bigl(\frac{Q-iq-2}{2}\Bigl)+S_{1}\Bigl(\frac{Q+iq-2}{2}\Bigl)\Bigl)
\]
and {\footnotesize{}
\begin{eqnarray*}
\Sigma_{4}(q,Q) & = & \frac{1}{2}\Sigma_{2}(q,Q)^{2}+\frac{64Q(1+q^{2}+Q^{2})}{(q^{2}+Q^{2})^{2}}-\frac{32iq}{q^{2}+Q^{2}}\Bigr(S_{2}\Bigl(\frac{Q-iq-2}{2}\Bigl)-S_{2}\Bigl(\frac{Q+iq-2}{2}\Bigl)\Bigl)+\\
 &  & 32\Bigr(S_{1}\Bigl(\frac{Q-iq-2}{2}\Bigl)+S_{1}\Bigl(\frac{Q+iq-2}{2}\Bigl)\Bigl)+8\Bigr(S_{3}\Bigl(\frac{Q-iq-2}{2}\Bigl)+S_{3}\Bigl(\frac{Q+iq-2}{2}\Bigl)\Bigl)+18\zeta(3)
\end{eqnarray*}
}where $S_{n}(x)$ are the analytical continuation of the harmonic
sums $S_{n}(N)=\sum_{k=1}^{N}\frac{1}{k^{n}}$. They are related to
the polygamma functions as%
\footnote{For $n=0$ one has to replace $\zeta(1)$ with $\gamma_{E}$. %
} 
\[
\Psi_{n}(Q)=(-1)^{n+1}n!(\zeta(n+1)-S_{n+1}(Q-1)
\]

Further details are given in \cite{BaOm}.

\pagebreak{}


\begin{thebibliography}{100}
\bibitem{MGKPW} J. M. Maldacena, \emph{\textquotedblleft The large
N limit of superconformal field theories and supergravity\textquotedblright ,}
Adv. Theor. Math. Phys. 2 (1998) 231 {[}Int. J. Theor. Phys. 38 (1999)
1113{]}, {[}hep-th/9711200{]}; S. S. Gubser, I. R. Klebanov and A.
M. Polyakov, \emph{\textquotedblleft Gauge theory correlators from
non-critical string theory\textquotedblright ,} Phys. Lett. B 428
(1998) 105, {[}hep-th/9802109{]}; E. Witten, \emph{\textquotedblleft Anti-de
Sitter space and holography\textquotedblright ,} Adv. Theor. Math.
Phys. 2 (1998) 253, {[}hep-th/9802150{]}.

\bibitem{MinZa} J. A. Minahan and K. Zarembo,\emph{ The Bethe-ansatz
for N = 4 super Yang-Mills,} JHEP 0303, 013 (2003), {[}hep-th/0212208{]}.

\bibitem{BKS} N. Beisert, C. Kristjansen and M. Staudacher, \emph{The
dilatation operator of N = 4 super Yang-Mills theory,} Nucl. Phys.
B 664 (2003) 131, {[}hepth/0303060{]}.

\bibitem{BStau} N. Beisert and M. Staudacher, \emph{The N = 4 SYM
integrable super spin chain,} Nucl. Phys. B 670 (2003) 439, {[}hep-th/0307042{]}.

\bibitem{BDKM} A. V. Belitsky, S. E. Derkachov, G. P. Korchemsky
and A. N. Manashov, \emph{Quantum integrability in (super) Yang-Mills
theory on the light-cone}, Phys. Lett. B 594 (2004) 385 {[}hep-th/0403085{]}.

\bibitem{BPR} I. Bena, J. Polchinski and R. Roiban, \emph{Hidden
symmetries of the $AdS_{5}\times S^{5}$ superstring,} Phys. Rev.
D 69 (2004) 046002, {[}hep-th/0305116{]}.

\bibitem{DNW} L. Dolan, C. R. Nappi and E. Witten,\emph{ A relation
between approaches to integrability in superconformal Yang-Mills theory,
}JHEP 0310 (2003) 017, {[}hep-th/0308089{]}.

\bibitem{KW}F.A. Smirnov, \emph{Form factors in completely integrable
models of quantum field theory}, Advanced Series in Mathematical Physics
Vol. 14 World Scientific, Singapore.M. Karowski and P. Weisz, \emph{Nucl.
Phys.} \textbf{B139} (1978) 455-476. 

\bibitem{FFBootstrap}Hratchya M. Babujian, Angela Foerster, Michael
Karowski,\emph{ The Form Factor Program: a Review and New Results
- the Nested SU(N) Off-Shell Bethe Ansatz}, arXiv:hep-th/0609130.

\bibitem{chew} G. Chew, \emph{the S-matrix theory of strong interactions},
New York, W.A. Benjamin Inc,. 1961

\bibitem{STW} B. Schroer, T.T. Troung and P.Weisz, \emph{Towards
an explicit construction of the sine-Gordon theory}, Phys. Lett. B,
1976, V.63, 422\textendash 424.

\bibitem{KarT} Karowski M., Thun H.J., Truong T.T., Weisz P.,\emph{
On the uniqueness of a purely elastic S matrix in 1 + 1 dimensions},
Phys. Lett. B, 1977, V.67, 321\textendash 322.

\bibitem{KarT2}{]} Karowski M., Thun H.J., \emph{Complete S matrix
of the massive Thirring model}, Nuclear Phys. B, 1977, V.130, 295\textendash 308.

\bibitem{BKW} Berg B., Karowski M., Weisz, P.,\emph{ Construction
of Green functions from an exact S matrix}, Phys. Rev. D, 1979, V.19,
2477\textendash 2479.

\bibitem{KarW} Karowski M., Weisz P., \emph{Exact form factors in
(1+1)-dimensional field theoretic models with soliton behavior}, Nuclear
Phys. B, 1978, V.139, 455\textendash 476.

\bibitem{BFK1} Babujian H., Foerster A., Karowski M., \emph{The nested
SU(N) off-shell Bethe ansatz and exact form factors}, J. Stat. Mech.,
2006, to appear, hep-th/0611012

\bibitem{BK1} Babujian H., Karowski M., \emph{Exact form factors
for the scaling Z(N)-Ising and the affine A(N \textminus{} 1) Toda
quantum field theories}, Phys. Lett. B, 2003, V.575, 144\textendash 150,
hep-th/0309018

\bibitem{BFK2} Babujian H., Foerster A., \emph{Karowski M., Exact
form factors in integrable quantum field theories: the scaling Z(N)-Ising
model}, Nuclear Phys. B, 2006, V.736, 169\textendash 198, hep-th/0510062.

\bibitem{Bab1} Babujian H.M., \emph{Correlation function in WZNW
model as a Bethe wave function for the Gaudin magnetics, in Gosen
1990, Proceedings}, Theory of Elementary Particles, 1990, 12\textendash 23
(see High Energy Physics Index, 1991, V.29, N 12257).

\bibitem{Bab2} Babujian H.M., \emph{Off-shell Bethe ansatz equation
and N point correlators in SU(2) WZNW theory}, J. Phys. A: Math. Gen.,
1993, V.26, 6981\textendash 6990.

\bibitem{BF1} Babujian H.M., Flume R., \emph{Off-shell Bethe ansatz
equation for Gaudin magnets and solutions of Knizhnik\textendash{}
Zamolodchikov equations}, Modern Phys. Lett. A, 1994, V.9, 2029\textendash 2040,
hep-th/9310110

\bibitem{FR1} Frenkel I.B., Reshetikhin N.Y., \emph{Quantum affine
algebras and holonomic difference equations}, Comm. Math. Phys., 1992,
V.146, 1\textendash 60

\bibitem{SV1} Schechtman V., Varchenko A., \emph{Arrangements of
hyperplanes and Lie algebra homology}, Invent. Math., 1991, V.106,
139\textendash 194.

\bibitem{exactS}Patrick Dorey, \emph{Exact S-matrices}, arXiv:hep-th/9810026. 

\bibitem{GZ}S. Ghoshal and A. Zamolodchikov, \emph{Boundary S-Matrix
and Boundary State in Two-Dimensional Integrable Quantum Field Theory},
Int.J.Mod.Phys. A9 (1994) 3841-3886; Erratum-ibid. A9 (1994) 4353.

\bibitem{BCT}Patrick Dorey, Roberto Tateo, Gerard Watts, \emph{ Generalizations
of the Coleman-Thun mechanism and boundary reflection factors}, Phys.
Lett. \textbf{B448} (1999) 249-256.

\bibitem{BRF}Z. Bajnok, G. B\"ohm, G. Tak\'acs, \emph{Boundary
reduction formula}, J.Phys. \textbf{A35} (2002) 9333-9342. 

\bibitem{BBT}Z. Bajnok, G. B\"ohm, G. Tak\'acs, \emph{On perturbative
quantum field theory with boundary,} Nucl.Phys. \textbf{B682} (2004)
585-617.

\bibitem{BFF}Z. Bajnok, L. Palla, G. Takacs, \emph{On the boundary
form factor program,} Nucl.Phys. \textbf{B750} (2006) 179-212 

\bibitem{DMS1}G. Delfino, G. Mussardo and P. Simonetti, \emph{Statistical
Models with a Line of Defect}, Phys.Lett. B \textbf{328} (1994) 123. 

\bibitem{DMS2}G. Delfino, G. Mussardo and P. Simonetti, \emph{Scattering
Theory and Correlation Functions in Statistical Models with a Line
of Defect}, \emph{Nucl. Phys. B} \textbf{432} (1994) 518.

\bibitem{Defboot}O.A. Castro-Alvaredo, A. Fring and F. Göhmann, \emph{On
the absence of simultaneous reflection and transmission in integrable
impurity systems}, hep-th/0201142. 

\bibitem{BaOmFF} Z. Bajnok, O. El Deeb: \emph{Form factors in the
presence of integrable defects}, Nuclear Physics B 832, p. 500-519,
2010, arXiv:0909.3200 {[}hep-th{]}

\bibitem{DefBound}Z. Bajnok, A. George, \emph{From Defects to Boundaries,}
Int.J.Mod.Phys. \textbf{A21} (2006) 1063-1078. 

\bibitem{ClDef} P. Bowcock, E. Corrigan, C. Zambon, \emph{Classically
integrable field theories with defects}, Int.J.Mod.Phys. \textbf{A19S2}
(2004) 82-91. 

\bibitem{DAT}P. Bowcock, E. Corrigan, C. Zambon, \emph{Affine Toda
field theories with defects,} JHEP \textbf{0401} (2004) 056. 

\bibitem{BS}Z. Bajnok, Zs. Simon, \emph{Solving topological defects
via fusion}, Nucl.Phys.\textbf{B802} (2008) 307-329. 

\bibitem{XXZ}Z. Bajnok\emph{, Equivalences between spin models induced
by defects}, J.Stat.Mech. \textbf{0606} (2006) P010. 

\bibitem{KL} Robert Konik, Andre' LeClair\emph{, Purely Transmitting
Defect Field Theories,} Nucl.Phys. \textbf{B538} (1999) 587-611. 

\bibitem{sgdef}P. Bowcock, E.Corrigan, C. Zambon, \emph{Some aspects
of jump-defects in the quantum sine-Gordon model}, JHEP \textbf{0508}
(2005) 023. 

\bibitem{QAT} E. Corrigan, C. Zambon, \emph{ On purely transmitting
defects in affine Toda field theory}, JHEP \textbf{07} (2007) 001. 

\bibitem{SYLbulk}J. L. Cardy and G. Mussardo, \emph{S matrix of the
Lee-Yang edge singularity in two dimensions}, Phys. Lett. \textbf{B225}
(1989) 275-278.

\bibitem{1ptf}P. Dorey, M. Pillin, R. Tateo and G.M.T. Watts, \emph{One-point
functions in perturbed boundary conformal field theories}, Nucl. Phys.
\textbf{B594} (2001) 625.

\bibitem{KRW} M. Kormos, I. Runkel, G.M.T. Watts,\emph{ Defect flows
in minimal models, }arXiv:0907.1497 

\bibitem{LY2pt} A.B. Zamolodchikov, \emph{Two point correlation function
in scaling Lee-Yang model}, Nucl. Phys. \textbf{B348} (1991) 619-641. 

\bibitem{FPR} G. Feverati, P. A. Pearce and F. Ravanini, \textit{Exact
$\varphi_{1,3}$ boundary flows of the tricritical Ising model}, hep-th/0308075,
Nucl. Phys. B 675, 469--515 (2003).

\bibitem{PCAI} P. A. Pearce, L. Chim and C. Ahn, \textit{Excited
TBA equations I: massive tricritical Ising model}, hep-th/0012223,
Nucl. Phys. \textbf{B601}, 539-568 (2001).

\bibitem{PCAII} P. A. Pearce, L. Chim and C. Ahn, \textit{Excited
TBA equations II: massless flow from tricritical to critical Ising
model}, Nucl. Phys. \textbf{B660}, 579--606 (2003).

\bibitem{FB85} P. J. Forrester and R. J. Baxter, \emph{Further exact
solutions of the eight-vertex SOS model and generalizations of the
Rogers-Ramanujan identities}, J. Stat. Phys. \textbf{38} (1985) 435--472.

\bibitem{FP02} G. Feverati and P. A. Pearce, \emph{Critical RSOS
and Minimal Models I: Paths, Fermionic Algebras and Virasoro Modules},
hep-th/0211185 (2002).

\bibitem{PearN} Pearce P. and Nienhuis B., \emph{Scaling limit of
RSOS lattice models and TBA equations}, Nucl.Phys. B519 (1998) 579-596,
hep-th/9711185

\bibitem{Riggs89} H. Riggs, \emph{Solvable lattice models with minimal
and nonunitary critical behavior in two dimensions}, Nucl. Phys. \textbf{B326}
(1989) 673--688.

\bibitem{YangYang} C.N. Yang and C.P. Yang,\emph{ Thermodynamics
of a one-dimensional system of bosons with repulsive delta-function
interaction}, J. Math. Phys. \textbf{10} (1969) 1115.

\bibitem{Zam90} Al.B. Zamolodchikov, \emph{Thermodynamic Bethe ansatz
in relativistic models: Scaling 3-state potts and Lee-Yang models},
Nucl. Phys. \textbf{B342} (1990) 695-720.

\bibitem{Zam91a} Al.B. Zamolodchikov, \emph{Thermodynamic Bethe ansatz
for RSOS scattering theories}, Nucl. Phys. \textbf{B358} (1991) 497-523.

\bibitem{Zam91b} Al.B. Zamolodchikov,\emph{ From tricritical Ising
to critical Ising by thermodynamic Bethe ansatz}, Nucl. Phys. \textbf{B358}
(1991) 524-546.

\bibitem{Zam91c} Al.B. Zamolodchikov,\emph{ TBA equations for integrable
perturbed coset models}, Nucl. Phys. \textbf{B366} (1991) 122-132.

\bibitem{Martins} M. J. Martins, \emph{Complex excitations in the
thermodynamic Bethe-anzatz approach}, Phys. Rev. Lett. \textbf{67}
(1991) 419-421.

\bibitem{Fendley} P. Fendley, \emph{Excited-state thermodynamics},
Nucl. Phys. \textbf{B374} (1992) 667-691.

\bibitem{Dorey} P. Dorey and R. Tateo, \emph{Excited states by analytic
continuation of TBA equations}, Nucl. Phys. \textbf{B482} (1996) 639-659.

\bibitem{BLZ} V.V. Bazhanov, S.L. Lukyanov, A.B. Zamoldchikov, \emph{Quantum
field theories in finite volume: Excited state energies}, Nucl. Phys.
\textbf{B489} (1997) 487-531.

\bibitem{Rav} D. Fioravanti, A. Mariottini, E. Quattrini, and F.
Ravanini,\emph{ Excited state Destri-De Vega equation for sine-Gordon
and restricted sine-Gordon models}, Phys. Lett. \textbf{B390} (1997)
243-251.

\bibitem{Bax80} R. J. Baxter, \emph{Hard hexagons: exact solution},
J. Phys. \textbf{A13} (1980) L61.

\bibitem{Bax82} R. J. Baxter, \emph{``Exactly Solved Models in Statistical
Mechanics''}. Academic Press, London, 1982.

\bibitem{ABF84} G. E. Andrews, R. J. Baxter and P. J. Forrester,
\emph{Eight-vertex SOS model and generalized Rogers-Ramanujan-type
identities}, J. Stat.\ Phys.\ \textbf{35} (1984) 193-266.

\bibitem{Huse84} D.A. Huse, \emph{Exact exponents for infinitely
many new multicritical points}, Phys.\ Rev.\ \textbf{B30} (1984)
3908-3915.

\bibitem{BPZ84} A. A. Belavin, A. M. Polyakov and A. B. Zamolodchikov,\emph{
Infinite conformal symmetry in two-dimensional quantum field theory},
Nucl.\ Phys.\ \textbf{B241} (1984) 333-380.

\bibitem{BaxP82} R. J. Baxter and P. A. Pearce, \emph{Hard hexagons:
interfacial tension and correlation length}, J. Phys.\ \textbf{A15}
(1982) 897.

\bibitem{BaxP83} R. J. Baxter and P. A. Pearce, \emph{Hard squares
with diagonal attractions}, J. Phys.\ \textbf{A16} (1983) 2239.

\bibitem{PearK91} P. A. Pearce and A. Klümper,\emph{ Finite-size
corrections and scaling dimensions of solvable lattice models: An
analytic method}, Phys. Rev. Lett. \textbf{66} (1991) 974-977.

\bibitem{KlumP91} A. Klümper and P. A. Pearce, \emph{Analytic calculation
of scaling dimensions: Tricritical hard squares and critical hard
hexagons}, J. Stat.\ Phys.\ \textbf{64} (1991) 13-76.

\bibitem{KlumP92} A. Klümper and P. A. Pearce,\emph{ Conformal weights
of RSOS lattice models and their fusion hierarchies}, Physica \textbf{A183}
(1992) 304-350.

\bibitem{BPO96} R.E. Behrend, P.A. Pearce and D.L. O'Brien, \emph{Interaction-round-a-face
models with fixed boundary conditions: The ABF fusion hierarchy},
J. Stat. Phys. \textbf{84}, 1-48 (1996).

\bibitem{BaOmL} Z. Bajnok, O. El Deeb, P. Pearce: Lattice, In preparation.

\bibitem{Polyakov} A.M. Polyakov, \emph{Conformal symmetry of critical
fluctuations}, JEPT Lett. 12 (1970), 381-383

\bibitem{Ginsparg} P.H. Ginsparg, \emph{Applied Conformal Field Theory},
1988, hep-th/9108028

\bibitem{Cardy} J. Cardy, \emph{Conformal Field Theory and Statistical
Mechanics}, 2008, arxiv:0807.3472{[}cond-mat.stat-mech{]}

\bibitem{PBZstat} A.M. Polyakov, A.A. Belavin and A.B. Zamolodchikov,
\emph{Infinite Conformal Symmetry of Critical Fluctuations in Two
Dimensions}, J. Stat.Phys. 34 (1984) 763

\bibitem{Zam89} A.B. Zamolodchikov, \emph{Integrable field theory
from conformal field theory}, Adv. Stud. Pure Math. 19 (1989) 641-674

\bibitem{Zam87} A.B. Zamolodchikov, \emph{Higher order integrals
of motion in two dimensional models of the field theory with a broken
conformal symmetry}, JETP Lett 46 (1987) 160-164

\bibitem{Zam87R} A.B. Zamolodchikov, \emph{Renormalization group
and perturbation theory near fixed points in two dimensional field
theory}, Sov. J.Nucl. Phys. 46 (1987), 1090

\bibitem{Zam90tba} A.B. Zamolodchikov, \emph{Thermodynamic Bethe
Ansatz in relativistic models. Scaling three Potts and Lee-Yang models},
Nucl. Phys. B342 (1990) 695-720

\bibitem{KlMe} T.R. Klassen and E. Melzer, \emph{The thermodynamics
of purely elastic scattering theories and conformal perturbation theory},
Nucl. Phys. B350 (1991) 635-689

\bibitem{CaMu} J.L. Cardy and G. Mussardo, \emph{S matrix of the
Yang-Lee edge singularity in two dimensions}, Phys. Lett. B225 (1989)
275

\bibitem{PearW00} P.A. Pearce and S.O. Warnaar, in preparation (2000).

\bibitem{Mus92}G. Mussardo, \emph{Off-critical statistical models:
Factorized scattering theories and bootstrap program}, Phys. Rept
\textbf{218} (1992) 215-379.

\bibitem{Yang67} C.N. Yang, \emph{Some Exact Results for the Many-Body
Problem in One Dimension with Repulsive Delta-Function Interaction},
Phys. Rev. Lett. \textbf{19} (1967) 1312-1314.

\bibitem{Bax72} R. J. Baxter, \emph{Partition function of the Eight-Vertex
lattice model}, Annals of Phys. \textbf{70} (1972) 193-228.

\bibitem{ZamZ} A.B. Zamolodchikov and A.B. Zamolodchikov, \emph{Factorized
S-matrices in two dimensions as the exact solutions of certain relativistic
quantum field theory models}, Annals of Phys. \textbf{120} (1979)
253-291.

\bibitem{DPTW}P. Dorey, A. Pocklington, R. Tateo and G. Watts, \emph{TBA
and TCSA with boundaries and excited states}, Nucl. Phys. B525 (1998)
641-663

\bibitem{LMSS} A. LeClair, G. Mussardo. H. Saleur and S.Skorkik,
\emph{Boundary energy and boundary states in integrable quantum field
theories}, Nucl. Phys. B453:581-618, 1995

\bibitem{Ol1} Olalla A. Castro-Alvaredo\emph{, Form factors of boundary
fields for A(2)-affine Toda field theory}, J. Phys. \textbf{A41}:
(2008) 19400

\bibitem{Ol2}5 Olalla A. Castro-Alvaredo, \emph{Boundary form factors
of the sinh-Gordon model with Dirichlet boundary conditions at the
self-dual point,} J. Phys. \textbf{A39}: (2006) 11901-11914

\bibitem{Gabor1} G. Takacs\emph{, Form factors of boundary exponential
operators in the sinh-Gordon model,} Nucl. Phys. \textbf{B805:} (2008)
391-417.

\bibitem{Gabor2}G. Takacs, \emph{Finite temperature expectation values
of boundary operators,} Nucl. Phys. \textbf{B805}: (2008) 391-417.

\bibitem{Gabor3}M. Kormos, G. Takacs, \emph{Boundary form factors
in finite volume,} Nucl. Phys. B803: (2008) 277-298. 

\bibitem{SZT} M. Szots, G. Takacs\emph{, Spectrum of local boundary
operators from boundary form factor bootstrap}, Nucl. Phys. \textbf{B785}:
(2007) 211-233

\bibitem{AFro} G. Arutyunov, S. Frolov, \emph{Foundations of the
$AdS_{5}\times S^{5}$ Superstring. Part I}, J. Phys. A42 (2009) 254003.
arXiv:0901.4937, doi:10.1088/1751-8113/42/25/254003.

\bibitem{BJa} Z. Bajnok, R. A. Janik, \emph{Four-loop perturbative
Konishi from strings and finite size effects for multiparticle states},
Nucl. Phys. B807 (2009) 625\textendash 650. arXiv:0807.0399, doi:10.1016/j.nuclphysb.2008.08.020.

\bibitem{FSSZ3} F. Fiamberti, A. Santambrogio, C. Sieg, D. Zanon,
\emph{Wrapping at four loops in N=4 SYM}, Phys. Lett. B666 (2008)
100\textendash 105. arXiv:0712.3522, doi:10.1016/j.physletb.2008.06.061.

\bibitem{FSSZ4} F. Fiamberti, A. Santambrogio, C. Sieg, D. Zanon,
\emph{Anomalous dimension with wrapping at four loops in N=4 SYM},
Nucl. Phys. B805 (2008) 231\textendash 266. arXiv:0806.2095, doi:10.1016/j.nuclphysb.2008.07.014.

\bibitem{BJL} Z. Bajnok, R. A. Janik, T. Lukowski, \emph{Four loop
twist two, BFKL, wrapping and strings}, Nucl. Phys. B816 (2009) 376\textendash 398.
arXiv:0811.4448, doi:10.1016/j.nuclphysb.2009.02.005.

\bibitem{BHJL} Z. Bajnok, A. Hegedus, R. A. Janik, T. Lukowski, \emph{Five
loop Konishi from AdS/CFT}, arXiv:0906.4062.

\bibitem{MPSZ} A. Mauri, S. Penati, A. Santambrogio, D. Zanon, \emph{Exact
results in planar N = 1 superconformal Yang-Mills theory}, JHEP 11
(2005) 024. arXiv:hep-th/0507282.

\bibitem{FSSZ1} F. Fiamberti, A. Santambrogio, C. Sieg, D. Zanon,
\emph{Finite-size effects in the superconformal beta-deformed N=4
SYM}, JHEP 08 (2008) 057. arXiv:0806.2103, doi:10.1088/1126-6708/2008/08/057.

\bibitem{FSSZ2} F. Fiamberti, A. Santambrogio, C. Sieg, D. Zanon,
\emph{Single impurity operators at critical wrapping order in the
beta-deformed N=4 SYM}, JHEP 08 (2009) 034. arXiv:0811.4594, doi:10.1088/1126-6708/2009/08/034.

\bibitem{BRo} N. Beisert, R. Roiban, \emph{Beauty and the twist:
The Bethe ansatz for twisted N = 4 SYM}, JHEP 08 (2005) 039. arXiv:hep-th/0505187.

\bibitem{ABBN} C. Ahn, Z. Bajnok, D. Bombardelli, R. I. Nepomechie,\emph{
Finite-size effect for four-loop Konishi of the beta- deformed N=4
SYM}, arXiv:1006.2209

\bibitem{GKV} N. Gromov, V. Kazakov, P. Vieira,\emph{ Exact Spectrum
of Anomalous Dimensions of Planar N=4 Supersymmetric Yang-Mills Theory},
Phys. Rev. Lett. 103 (2009) 131601. arXiv:0901.3753, doi:10.1103/PhysRevLett.103.131601.

\bibitem{Afro2} G. Arutyunov, S. Frolov, \emph{String hypothesis
for the $AdS_{5}\times S^{5}$ mirror}, JHEP 03 (2009) 152. arXiv:0901.1417,
doi:10.1088/1126-6708/2009/03/152.

\bibitem{BFTa} D. Bombardelli, D. Fioravanti, R. Tateo, \emph{Thermodynamic
Bethe Ansatz for planar AdS/CFT: a proposal}, J. Phys. A42 (2009)
375401. arXiv:0902.3930, doi:10.1088/1751-8113/42/37/375401.

\bibitem{AfroTBA} G. Arutyunov, S. Frolov, \emph{Thermodynamic Bethe
Ansatz for the $AdS_{5}\times S^{5}$ Mirror Model}, JHEP 05 (2009)
068. arXiv:0903.0141, doi:10.1088/1126-6708/2009/05/068.

\bibitem{GKKV} N. Gromov, V. Kazakov, A. Kozak, P. Vieira, \emph{Exact
Spectrum of Anomalous Dimensions of Planar N = 4 Supersymmetric Yang-Mills
Theory: TBA and excited states}, Lett. Math. Phys. 91 (2010) 265\textendash 287.
arXiv:0902.4458, doi:10.1007/s11005-010-0374-8.

\bibitem{AFroS} G. Arutyunov, S. Frolov, R. Suzuki, \emph{Exploring
the mirror TBA}, JHEP 05 (2010) 031. arXiv:0911.2224, doi:10.1007/JHEP05(2010)031.

\bibitem{CFTat} A. Cavaglia, D. Fioravanti, R. Tateo,\emph{ Extended
Y-system for the $AdS_{5}/CFT_{4}$ correspondence}, arXiv:1005.3016.

\bibitem{GLeMa} N. Gromov, F. Levkovich-Maslyuk, \emph{Y-system and
beta-deformed N=4 Super-Yang-Mills}, arXiv:1006.5438.

\bibitem{ABBN-1} C. Ahn, Z. Bajnok, D. Bombardelli, R. I. Nepomechie,
\emph{Twisted Bethe equations from a twisted S-matrix}, arXiv:1010.3229.

\bibitem{Gunn} J. Gunnesson, \emph{Wrapping in maximally supersymmetric
and marginally deformed N=4 Yang-Mills}, JHEP 04 (2009) 130. arXiv:0902.1427,
doi:10.1088/1126-6708/2009/04/130.

\bibitem{ArLT} G. Arutyunov, M. de Leeuw, S. J. van Tongeren, \emph{Twisting
the Mirror TBA}, arXiv:1009.4118.

\bibitem{AFroSm} G. Arutyunov, S. Frolov, R. Suzuki, \emph{Five-loop
Konishi from the Mirror TBA}, JHEP 04 (2010) 069. arXiv:1002.1711,
doi:10.1007/JHEP04(2010)069.

\bibitem{Bahe1} J. Balog, A. Hegedus, \emph{5-loop Konishi from linearized
TBA and the XXX magnet}, JHEP 06 (2010) 080. arXiv:1002.4142, doi:10.1007/JHEP06(2010)080.

\bibitem{BaHe} J. Balog, A. Hegedus, \emph{The Bajnok-Janik formula
and wrapping corrections}, arXiv:1003.4303.

\bibitem{BBBZ} D. H. Bailey, J. M. Borwein, D. Broadhurst, W. Zudilin,
\emph{Experimental Mathematics and Mathematical Physics}, arXiv:1005.0414.

\bibitem{AFroSmSB} G. Arutyunov, S. Frolov, \emph{The S-matrix of
String Bound States}, Nucl. Phys. B804 (2008) 90\textendash 143. arXiv:0803.4323,
doi:10.1016/j.nuclphysb.2008.06.005.

\bibitem{LuRe} T. Lukowski, A. Rej, V. N. Velizhanin, \emph{Five-Loop
Anomalous Dimension of Twist-Two Operators}, arXiv:0912.1624.

\bibitem{AFroDr} G. Arutyunov, S. Frolov, \emph{The Dressing Factor
and Crossing Equations}, J. Phys. A42 (2009) 425401. arXiv:0904.4575,
doi:10.1088/1751-8113/42/42/425401.

\bibitem{ABBN-2} C. Ahn, Z. Bajnok, D. Bombardelli, R. Nepomechie,
\emph{``TBA, NLO Luscher correction, and double wrapping in twisted
AdS/CFT'', }arXiv:1108.4914v3 {[}hep-th{]}

\bibitem{BaOm} Z. Bajnok, O. El Deeb: \emph{6-loop anomalous dimension
of a single impurity operator from AdS/CFT and multiple zeta values},
JHEP 1101:054,2011, arXiv:1010.5606 {[}hep-th{]}\pagebreak{}\end{thebibliography}
\end{document}